\documentclass[a4paper,fleqn,usenatbib]{mnras}

\usepackage{newtxtext,newtxmath}

\usepackage[T1]{fontenc}
\usepackage{ae,aecompl}

\usepackage{graphicx}	
\usepackage{amsmath}	
\usepackage{amssymb}	

\newcommand\blue{\color{blue}}

\newcommand\T{\rule{0pt}{2.6ex}}       
\newcommand\B{\rule[-1.2ex]{0pt}{0pt}} 

\defcitealias{heesen_16a}{HD16}

\title[Radio haloes in nearby galaxies]{Radio
  haloes in nearby galaxies modelled with 1D cosmic-ray transport using {\LARGE SPINNAKER}} 

\author[V.~Heesen et al.]{V.~Heesen,$^{1,2}$\thanks{E-mail: {\blue
      volker.heesen@hs.uni-hamburg.de}} M.~Krause,$^{3}$ R.~Beck,$^{3}$ B.~Adebahr,$^{3,4,5}$ 
  D.~J. Bomans,$^{5}$ \newauthor E.~Carretti,$^{6,7}$ M.~Dumke,$^{3,8}$ G.~Heald,$^{4,6}$ J.~Irwin,$^{9}$
  B.~S.~Koribalski,$^{6}$ \newauthor D.~D.~Mulcahy,$^{10}$ T.~Westmeier$^{11}$ and R.-J.~Dettmar$^{5}$\\
$^{1}$School of Physics and Astronomy, University of Southampton, Southampton SO17 1BJ, UK\\
$^{2}$Universit\"at Hamburg, Hamburger Sternwarte, Gojenbergsweg 112, D-21029, Hamburg, Germany\\
$^{3}$Max-Planck-Institut f\"ur Radioastronomie, Auf dem H\"ugel 69, D-53121 Bonn, Germany\\
$^{4}$ASTRON, the Netherlands Institute for Radio Astronomy, Postbus 2, 7990 AA, Dwingeloo, The Netherlands\\
$^{5}$Astronomisches Institut der Ruhr-Universit\"at Bochum,
Universit\"atsstr. 150, D-44780 Bochum, Germany\\
$^{6}$CSIRO Astronomy and Space Science, Australia Telescope National
Facility, P.O.~Box 76, Epping, NSW 1710, Australia\\
$^{7}$INAF Osservatorio Astronomico di Cagliari, Via della Scienza 5, 09047 Selargius (CA), Italy\\
$^{8}$European Southern Observatory, Alonso de Cordova 3107, Vitacura, Santiago, Chile\\
$^{9}$Department of Physics, Engineering Physics, \& Astronomy, Queen's University, Kingston, Ontario, K7L 3N6, Canada\\
$^{10}$Jodrell Bank Centre for Astrophysics, Alan Turing Building, Oxford Road, Manchester M13 9PL, UK\\
$^{11}$ICRAR, M468, The University of Western Australia, 35 Stirling Highway, Crawley, WA 6009, Australia}

\date{Accepted 2018 January 4. Received 2017 December 14; in
  original form 2016 September 6}

\pubyear {2017}

\begin{document}
\label{firstpage}
\pagerange{\pageref{firstpage}--\pageref{lastpage}}

\maketitle

\begin{abstract}
We present radio continuum maps of 12 nearby ($D\leq 27$~Mpc), edge-on
($i\geq 76\degr$), late-type spiral galaxies mostly at $1.4$ and 5~GHz, observed
with the Australia Telescope Compact Array, Very Large Array, Westerbork
Synthesis Radio Telescope, Effelsberg 100-m and
Parkes 64-m telescopes. All galaxies show clear evidence of radio haloes,
including the first detection in the Magellanic-type galaxy NGC~55. In 11 galaxies,
we find a thin and a thick disc that can be better fitted by exponential
rather than Gaussian functions. We fit our {\small SPINNAKER} (SPectral
INdex Numerical Analysis of K(c)osmic-ray Electron Radio-emission) 1D
cosmic-ray transport models to the vertical model profiles of the
non-thermal intensity and to the non-thermal radio spectral
index in the halo. We simultaneously fit for the advection
speed (or diffusion coefficient) and magnetic field scale height. In the thick
disc, the magnetic field scale heights range from 2 to 8 kpc with an average across the
sample of $3.0\pm 1.7$~kpc; they show no correlation with either star-formation rate (SFR), SFR surface density ($\Sigma_{\rm SFR}$) or rotation speed ($V_{\rm rot}$). The advection speeds
range from 100 to 700~$\rm km\,s^{-1}$ and display correlations of $V\propto \rm
SFR^{0.36\pm 0.06}$ and $V\propto \Sigma_{\rm SFR}^{0.39\pm 0.09}$; they agree
remarkably well with the escape velocities ($0.5\leq V/V_{\rm
  esc}\leq 2$), which can be explained by cosmic-ray driven winds. Radio haloes show the presence of disc winds in galaxies with $\Sigma_{\rm SFR} > 10^{-3}~\rm M_{\sun}\,yr^{-1}\,kpc^{-2}$ that extend over several kpc and are driven by processes related to the distributed star formation in the disc.
\end{abstract}


\section{Introduction}
\label{intro}


This paper is a follow-up study of our recent paper about the `Advective and
diffusive cosmic-ray transport in galactic haloes' \citep[][hereafter
\citetalias{heesen_16a}]{heesen_16a}, where we presented 1D cosmic ray transport models now
implemented in the software {\scriptsize SPINNAKER}  (SPectral
INdex Numerical Analysis of K(c)osmic-ray Electron Radio-emission).\footnote{\href{www.github.com/vheesen/Spinnaker}{www.github.com/vheesen/Spinnaker}}
We have extended our previous radio continuum observations 
of two edge-on
galaxies, adding another 10 galaxies, to study advection speed,
magnetic field scale height and diffusion coefficient across a wider range of
galactic parameters. We then study
the influence of the star-formation rate (SFR), SFR surface density ($\Sigma_{\rm
  SFR}$)
and rotation speed ($V_{\rm rot}$) on these results. We use a
combination of so far unpublished and archival data from the Australia Telescope Compact Array
(ATCA), the Very Large Array (VLA) and the
Westerbork Synthesis Radio Telescope (WSRT). For a
sub-sample of four galaxies we use single-dish observations with both the 
Effelsberg 100-m and Parkes 64-m telescopes in order to complement the
interferometric data. The majority of the interferometric observations were
taken before the recent upgrade of the correlators, so that we mostly use 100-MHz-bandwidths, rather than the GHz-bandwidths which are available
nowadays. While this leads to reduced sensitivities of tens rather than a few $\umu\rm Jy\,beam^{-1}$, a
sample like this is a good reference against which to compare the new wide-band data as
observed with the Karl G.\ Jansky VLA \citep{wiegert_15a}. 

Our sample contains 12 nearby edge-on galaxies (11 late-type spiral galaxies
and 1 Magellanic-type galaxy), which lie in
the SFR range of $0.1\leq {\rm
  SFR/(M_{\sun}\,yr^{-1})}\leq 16$. Where possible, we produce
maps that have a spatial resolution of 1~kpc or better. The reason for this is
to allow the separation of the thin and thick radio disc, which are present
in most galaxies. The thin disc (with a scale height of a few 100~pc) represents a region of current/recent star formation populated by supernova remnants and \ion{H}{ii} regions \citep{ferriere_01a}; the thick disc (with a scale height of a few kpc) is what we refer to in this work as the `halo' and represents the region of extra-planar gas, magnetic fields and cosmic rays with no ongoing star formation \citep{rossa_03a,krause_17a}. As we will show in
this paper, we observe a transition between thin and thick disc at a height of
$|z|\approx 1~\rm kpc$, so that we will refer to the halo as emission at
$|z|\gtrapprox 1$~kpc. We are aware that the transition between the disc and halo is
gradual (the `disc--halo' interface), and the observed transition will also
depend on the spatial resolution.\footnote{In this paper, we are using `thick disc' and `halo' in a synonymous way. Strictly speaking, the thick disc extends all the way to the midplane because it is a second component that we see superposed on to the thin disc. In contrast, the halo refers to heights above $\approx$1~kpc, excluding everything closer to the midplane.}
\begin{table*}
\caption{General properties of our sample galaxies.\label{tab:sample}}
\begin{tabular}{l ccc cc ccccc c}
\hline
Galaxy & $i^{\rm a}$ & $d_{25}^{\rm b}$ & $D^{\rm c}$ & $M_B^{\rm d}$ & $\rm Type^e$ & $\rm Nucleus^f$ & $\rm SFR^g$ & $r_{\star}^{\rm h}$ & $\log_{10}(\Sigma_{\rm SFR})^{\rm i}$ & $\log_{10}(M_{\rm tot})^{\rm j}$ & $V_{\rm rot}^{\rm k}$ \\
& ($\degr$) & ($\arcmin$) & (Mpc) & (mag) & & & ($\rm M_{\sun}\,yr^{-1}$) & (kpc) & ($\rm M_{\sun}\,yr^{-1}\,kpc^{-2}$) &  ($\rm M_{\sun}$)& ($\rm km\,s^{-1}$)\\
\hline
NGC 55   & $80.0$ & $34.1$ & $1.9 $ & $-18.89$ & SBm  &             & $0.16 $ & $6.2 $  & $-2.87$ & $10.26$ & $91 $ \\
NGC 253  & $78.5$ & $25.8$ & $3.9 $ & $-20.64$ & SABc & \ion{H}{ii} & $6.98 $ & $11.8$  & $-1.80$ & $11.16$ & $205$ \\
NGC 891  & $89.0$ & $12.2$ & $10.2$ & $-20.09$ & Sb   & \ion{H}{ii} & $4.72 $ & $11.7$  & $-1.96$ & $11.28$ & $212$ \\
NGC 3044 & $90.0$ & $4.4 $ & $18.7$ & $-19.53$ & SBc  &             & $2.03 $ & $9.8 $  & $-2.17$ & $10.81$ & $153$ \\
NGC 3079 & $88.0$ & $7.7 $ & $19.7$ & $-20.94$ & SBcd & Sy2         & $9.09 $ & $13.3$  & $-1.79$ & $11.35$ & $208$ \\
NGC 3628 & $87.0$ & $14.8$ & $7.9 $ & $-20.01$ & Sb   & T2          & $1.73 $ & $10.7$  & $-2.32$ & $11.26$ & $215$ \\
NGC 4565 & $90.0$ & $16.2$ & $24.7$ & $-22.37$ & Sbc  & Sy$1.9$     & $5.69 $ & $37.1$  & $-2.88$ & $11.91$ & $244$ \\
NGC 4631 & $85.0$ & $14.7$ & $6.9 $ & $-20.12$ & SBcd & \ion{H}{ii} & $2.89 $ & $12.1$  & $-2.20$ & $10.82$ & $138$ \\
NGC 4666 & $76.0$ & $4.2 $ & $26.6$ & $-21.14$ & SABc & LINER       & $16.19$ & $14.0$  & $-1.58$ & $11.15$ & $193$ \\
NGC 5775 & $84.0$ & $3.9 $ & $26.9$ & $-20.71$ & Sbc  & \ion{H}{ii} & $9.98 $ & $14.4 $ & $-1.81$ & $11.09$ & $187$ \\
NGC 7090 & $89.0$ & $6.8 $ & $10.6$ & $-19.70$ & Sc   &             & $0.62 $ & $7.2  $ & $-2.41$ & $10.57$ & $124$ \\
NGC 7462 & $90.0$ & $5.0 $ & $13.6$ & $-18.57$ & SBbc &             & $0.28 $ & $7.5  $ & $-2.80$ & $10.46$ & $112$ \\
\hline
\end{tabular}
\flushleft{{\bf Notes.} Data are from \citet{tully_88a} unless otherwise noted. Available on Vizier  (\href{http://vizier.cfa.harvard.edu}{http://vizier.cfa.harvard.edu}).\\
$^{\rm a}$ Inclination angles were measured from kinematic analysis of the \ion{H}{i} line in NGC~55 \citep{westmeier_13a} and in NGC~891 \citep{oosterloo_07a}, and in NGC~253 of the Balmer H\,$\alpha$ line \citep{pence_81a}. They have uncertainties of $\pm 1\degr$. In the remaining galaxies they have been measured by \citet{tully_88a} (NGC~3044--5775) and \citet{karachentsev_13a} (NGC~7090 and 7462) from the ratio of the minor to major axis assuming an intrinsic disc height. They have larger uncertainties, typically $\pm 3\degr$.\\ 
$^{\rm b}$ Optical diameter measured at the $25~\rm mag\,arcsec^{-2}$ isophote in $B$-band.\\
$^{\rm c}$ Distances are from the NASA/IPAC Extragalactic Database (NED) (Virgo infall
  only), assuming $H_0=73~\rm km\,s^{-1}$. For the most nearby galaxies NGC~55 and 253, we used the distances from \citet{pietrzynski_06a} (NGC~55) and \citet{karachentsev_03a} (NGC~253), who used the Cepheid period-luminosity relation and the tip of the red giant branch, respectively.\\
$^{\rm d}$ Absolute $B$-band magnitude from \citet{tully_88a} and corrected to our distances.\\ 
$^{\rm e}$ Galaxy morphological classification.\\
$^{\rm f}$ Optical classification of the nuclear spectrum, from \citet{ho_97a} where $\rm Sy = Seyfert$ and $\rm T = transition$ object. The classification of NGC~253 is based on the assumption that the nuclear activity is starburst dominated \citep{westmoquette_11a}. The classification of NGC~4666 is from \citet{veilleux_95a}.\\
$^{\rm g}$ Star-formation rate is ${\rm SFR}=0.39\times L_{\rm TIR}/10^{43}~\rm erg\,s^{-1}$, where $L_{\rm TIR}$ is the $3$--$1100~\umu \rm m$ total infrared luminosity. The conversion factor is from \citet{kennicutt_12a} and has an uncertainty of $0.3$~dex. See Appendix~\ref{physical} for details which infrared data were used.\\
$^{\rm h}$ Radius of the actively star forming disc (within the last $\approx$100~Myr), estimated from the radial extent of the radio continuum emission. A cross-check of six galaxies with GALEX FUV catalogue by \citet{gil_de_paz_07a} found agreement within 10~per cent.\\
$^{\rm i}$ Star-formation rate surface density is $\Sigma_{\rm SFR}={\rm SFR}/(\upi r_{\star}^2)$. The uncertainty is $0.3$~dex.\\
$^{\rm j}$ Total masses are calculated $M_{\rm tot}=0.233\cdot R_{\rm kpc}\cdot V_{\rm rot,100}^2\times 10^{10}~{\rm M_{\sun}}$, where $R_{\rm kpc}$ is the radius of the galaxy estimated from $d_{25}$ using the distance $D$, and $V_{\rm rot,100}$ is the rotation speed in units of $100~\rm km\, s^{-1}$. They have uncertainties of $0.1$~dex.\\
$^{\rm k}$ Maximum rotation speeds are from \citet{westmeier_13a} (NGC~55), \citet{pence_81a} (NGC~253), \citet{oosterloo_07a} (NGC~891), the HyperLeda database (\href{http://leda.univ-lyon1.fr}{http://leda.univ-lyon1.fr})  (NGC~3044--5775) \citep{makarov_14a} and \citet{dahlem_05a} (NGC~7090 and 7462). They have uncertainties of $\pm 3~\rm km\, s^{-1}$.}
\end{table*}

Most of our sample galaxies have been studied in the radio (continuum and
line) in detail before, so we will give here only a brief summary of their
most notable properties (general properties are summarized in Table~\ref{tab:sample}). NGC~55, a Magellanic-type (irregular, of
intermediate mass between dwarf and spiral galaxies) galaxy and member of the
Sculptor group  has a very
thick \ion{H}{i} disc extending up to a height of 10~kpc into the halo
\citep{westmeier_13a}. Thick \ion{H}{i} discs could be a sign of a galactic fountain or on-going gas accretion. NGC~253, another member of the Sculptor group, is a
prototypical nuclear starburst galaxy and is most notable for its `dumbbell
shaped' radio halo caused by strong synchrotron cooling in the centre
of the galaxy \citep{heesen_09a,heesen_09b,heesen_11a}. This galaxy
  is also notable for a very bright X-ray halo, which shows the presence of a
  `disc wind', which is unrelated to the nuclear outflow \citep{bauer_08a}. The distinction between nuclear outflows and disc winds is important for this study: nuclear outflows emanate from the central few 100~pc driven by a combination of concentrated star formation and jets from active galactic nuclei (AGNs). In our work,
we concentrate on the disc winds, which extend over a large fraction of
the star-forming disc over several kpc at least, although the scale can be smaller than 10-kpc \citep[see][for a discussion of nuclear outflows vs.\ disc winds]{westmoquette_11a}.

NGC~891 has a prominent thick \ion{H}{i} disc, the kinematics of which have been studied in detail by \citet{oosterloo_07a}. They showed that this galaxy still accretes intergalactic matter, so that the halo is only in part caused
by the galactic fountain. NGC~891 is the best case to study the influence of on-going accretion on to a radio halo. NGC~3044 hosts a large
\ion{H}{i} supershell that extends to a height of 3~kpc above the midplane
\citep{lee_97a}. These \ion{H}{i} supershells require dozens of supernovae (SNe) and are
essential for an active disc--halo interface. We expect these kind of features to be abundant in our galaxies, but the spatial resolution of our observations is not high enough to resolve these features.

NGC~3079 is a case in point for a galaxy with a nuclear outflow driven by a combination of starburst and low-luminosity AGN. Jets from the AGN entrain the surrounding material which is subsequently transported away from the nucleus \citep{middelberg_07a,shafi_15a}. Above and below the nucleus, there are well-defined kpc-sized lobes of
non-thermal emission, surrounded by filamentary H\,$\alpha$ emission that have counterparts in X-ray emission \citep{cecil_01a,cecil_02a}. While the filamentary emission is clearly connected to the nuclear outflow and AGN-driven jets, there are also indications for a disc wind in this galaxy. First, there are vertical H\,$\alpha$ and dust filaments extending over the full extent of the disc \citep{cecil_01a}, reminiscent of a `boiling disc' in NGC~253 \citep{sofue_94a}, which is indicative of an active disc--halo interface. Second, the X-ray emission in the halo extends, as in NGC~253, for several kpc away from the nuclear outflow \citep{strickland_04a}; it is questionable that the two are related. In order to study the disc wind only, we have masked the radio continuum emission stemming from the nuclear outflow in the following analysis.

NGC~3628, a nuclear starburst galaxy, is a member
of the Leo Triplet and possesses a 140-kpc long tidal tail of atomic hydrogen detected in the 21-cm \ion{H}{I} line
\citep{haynes_79a}. NGC~4565 has a `lagging' \ion{H}{I} halo, where the
rotational velocity of the neutral gas decreases with increasing distance from
the midplane. This makes it likely that this galaxy has an on-going galactic fountain, in spite of the galaxy having only little extra-planar \ion{H}{i} emission \citep{zschaechner_12a}. NGC~4631 is the galaxy with
arguably the most prominent radio halo. This is in part
caused by gravitational interaction with its neighbour NGC~4656, where a 'spur' of ordered magnetic fields leads to the neighbour galaxy \citep{hummel_88a,mora_13a}. This merger has probably caused a starburst in the past that led to an outflow the result of which we see now \citep{irwin_11a}. NGC~4666 is a starburst
galaxy with outflowing ionized H\,$\alpha$ emission \citep{dahlem_97a, voigtlaender_13a}, while NGC~5775 is the prototypical X-shaped halo magnetic field galaxy
\citep{tuellmann_00a,soida_11a}. These galaxies have magnetic field structures that look in projection like an `X'. The field shapes are thought to be caused by either a galactic dynamo in the halo or by the interaction of the magnetic field lines with a galactic wind \citep{heesen_09b}. Finally, NGC~7090 and NGC~7462 are rather unremarkable galaxies although both possess a layer of extra-planar H\,$\alpha$ and \ion{H}{i} emission \citep{rossa_03a,rossa_03b,dahlem_05a}. 

In this paper, we present observations at $L$, $C$ and $X$ band
($1.4$, 5 and $8.5$~GHz) in order to study vertical profiles of the non-thermal intensities and
the non-thermal radio spectral index. We use exponential and Gaussian functions to model the vertical profiles of the non-thermal radio emission. These profiles are then fitted with
our cosmic-ray transport models, in order to measure advection speeds,
diffusion coefficients and magnetic field scale heights. We estimate magnetic field
strengths from energy equipartition \citep{beck_05a} and use the  total infrared
luminosity to determine photon energy densities. Thus, we can take both the
synchrotron and inverse Compton (IC) losses of the CR electrons (CREs)
into account.

This paper is organized as follows: in Section~\ref{observations} we describe our 
observations and data reduction techniques. In Section~\ref{transport}, we motivate and
present our cosmic-ray transport models. Results are presented in
Section~\ref{results}, followed by the discussion in
Section~\ref{discussion}. We finish off with our conclusions in
Section~\ref{conclusions}. Figure~\ref{fig:maps} shows the radio continuum maps for NGC~4631,
followed by Figs~\ref{fig:fit} and \ref{fig:chi}, which contain a step-by-step guide on how
we obtained the best-fitting cosmic-ray transport model for this particular
galaxy. Figure~\ref{fig:par} summarizes the results for our sample with an
atlas of maps and models presented in Appendix~\ref{sec:image_atlas}. Throughout the paper, the radio spectral index
$\alpha$ is defined in the sense $S_{\nu}\propto\nu^{\alpha}$ and quantities
averaged across the sample are error   m and presented with an
uncertainty of $\pm 1\sigma$. 
\section{Observations and data reduction}
\label{observations}
\begin{table*}
\caption{Observation details for the galaxies presented in this paper.\label{tab:obs}}
\begin{tabular}{l ccccc c ccc}
\hline
Galaxy & Band$^{\rm a}$\ & $\nu^{\rm b}$ & $\rm Telescope^{\rm c}$ & $\rm Configuration^{\rm d}$\ & $\rm Project^{\rm e}$ & $\rm  Time^{\rm f}$ & $\rm Date^{\rm g}$ & $\rm Notes^{h}$ & $\rm Reference^{\rm i}$\ \ \\
& & (GHz) & & & & (h) & & & \\
\hline
NGC~55    & $L$          & $1.37$   & ATCA       & 750D                & C287    & $8.7$            & 1993 Aug 1           & Mosaic         & 17                 \\     
$\ldots$  & $\ldots$     & $\ldots$ & $\ldots$   & 375                 & C287    & $11.2$           & 1995 Jan 12          & $\ldots$       & $\ldots$           \\
$\ldots$  & $\ldots$     & $\ldots$ & $\ldots$   & 750A                & C287    & $11.1$           & 1995 Oct 25          & $\ldots$       & $\ldots$           \\
$\ldots$  & $\ldots$     & $\ldots$ & $\ldots$   & H75                 & C1341   & $5.0$            & 2005 Jul 17          & Mosaic         & This work          \\      
$\ldots$  & $\ldots$     & $\ldots$ & $\ldots$   & EW352               & C1341   & $9.4$            & 2005 Oct 7           & $\ldots$       & $\ldots$           \\
$\ldots$  & $C$          & $4.80$   & $\ldots$   & 375                 & C287    & $3.6$            & 1994 Mar 29          & Mosaic         & 17                 \\     
$\ldots$  & $\ldots$     & $\ldots$ & $\ldots$   & 375                 & C287    & $10.2$           & 1994 Mar 30          & $\ldots$       & $\ldots$           \\
$\ldots$  & $\ldots$     & $\ldots$ & $\ldots$   & 375                 & C287    & $7.8$            & 1994 Mar 31          & $\ldots$       & $\ldots$           \\
$\ldots$  & $\ldots$     & $\ldots$ & $\ldots$   & 375                 & C287    & $12.5$           & 1994 Nov 23          & $\ldots$       & $\ldots$           \\
$\ldots$  & $\ldots$     & $\ldots$ & $\ldots$   & 750A                & C287    & $5.1$            & 1995 Mar 1           & $\ldots$       & $\ldots$           \\
$\ldots$  & $\ldots$     & $\ldots$ & $\ldots$   & 375                 & C287    & $5.3$            & 1995 Aug 16          & $\ldots$       & $\ldots$           \\
$\ldots$  & $\ldots$     & $\ldots$ & $\ldots$   & 375                 & C287    & $10.2$           & 1995 Nov 24          & $\ldots$       & $\ldots$           \\
$\ldots$  & $\ldots$     & $4.67$   & $\ldots$   & EW352               & C1974   & $7.6$            & 2008 Nov 22          & $\ldots$       & This work          \\    
$\ldots$  & $\ldots$     & $\ldots$ & $\ldots$   & EW364               & C1974   & $9.9$            & 2009 Feb 13          & $\ldots$       & $\ldots$           \\
$\ldots$  & $C$          & $5.60$   & $\ldots$   & H168                & C1974   & $7.6$            & 2010 Mar 27          & $\ldots$       & $\ldots$           \\
$\ldots$  & $C$          & $4.80$   & Parkes     & single-dish         & P697    & $16.0$           & 2010 Oct 7           & Merged         & $\ldots$           \\
NGC~253   & $L$          & $1.46$   & VLA        & B+C+D               & AC278   & $4.1$            & 1990 Sep--1991 Mar   & Mosaic         & 2                  \\
$\ldots$  & $C$          & $4.86$   & $\ldots$   & D                   & AH844   & $35.8$           & 2004 Jul 4--24       & Mosaic         & 10                  \\
$\ldots$  & $\ldots$     & $4.85$   & Effelsberg & single-dish         & N/A     & N/A              & 1997                 & Merged         & $\ldots$           \\
NGC~891   & $L$          & $1.39$   & WSRT       & Multiple            & R02B    & 240              & 2002 Aug--Dec        &                & 13                 \\
$\ldots$  & $C$          & $4.86$   & VLA        & D                   & AA94    & $11.2$           & 1988 Aug 29          &                & 16                 \\
$\ldots$  & $\ldots$     & $4.85$   & Effelsberg & single-dish         & 44--95  & $9.1$            & 1996 Feb--Aug        &                & 6                  \\
NGC~3044  & $L$          & $1.49$   & VLA        & B                   & AI28    & $3.1$            & 1986 Aug 1           &                & This work          \\
$\ldots$  & $\ldots$     & $\ldots$ & $\ldots$   & C                   & AI23    & $0.8$            & 1985 Jul 25          &                & 11                 \\
$\ldots$  & $\ldots$     & $\ldots$ & $\ldots$   & D                   & AI31    & $1.1$            & 1987 Apr 28/30       &                & $\ldots$           \\
$\ldots$  & $C$          & $4.86$   & $\ldots$   & C                   & AB676   & $0.8$            & 1993 Jun 13          &                & 4                  \\
$\ldots$  & $\ldots$     & $\ldots$ & $\ldots$   & D                   & AM573   & $1.1$            & 1997 Nov 6           &                & This work          \\
$\ldots$  & $\ldots$     & $\ldots$ & $\ldots$   & D                   & AI31    & $1.0$            & 1987 Apr 28          &                & 11                 \\
NGC~3079  & $L$          & $1.66$   & VLA        & B                   & BS44    & $1.0$            & 1997 Mar 8           &                & This work          \\
$\ldots$  & $\ldots$     & $1.41$   & $\ldots$   & CD                  & BS44    & $2.4$            & 1997 Oct 2           &                & $\ldots$           \\
$\ldots$  & $\ldots$     & $1.43$   & $\ldots$   & C                   & AB740   & $1.3$            & 1996 Feb 17          &                & $\ldots$           \\
$\ldots$  & $C$          & $4.71$   & $\ldots$   & C                   & AC277   & $3.9$            & 1990 Dec 9           &                & 3                  \\
$\ldots$  & $\ldots$     & $4.86$   & $\ldots$   & D                   & AD177   & $2.5$            & 1986 Jan 16          &                & This work          \\
NGC~3628  & $L$          & $1.49$   & VLA        & CD                  & AS300   & $4.3$            & 1988 Mar 25          &                & 14                 \\
$\ldots$  & $\ldots$     & $\ldots$ & $\ldots$   & D                   & AS300   & $8.4$            & 1987 Apr 7           &                & $\ldots$           \\
$\ldots$  & $C$          & $4.86$   & $\ldots$   & D                   & AK243   & $7.7$            & 1991 Mar 28          &                & 7                  \\
NGC~4565  & $L$          & $1.49$   & VLA        & B                   & AS326   & $3.8$            & 1988 Jan 29          &                & 16                 \\
$\ldots$  & $\dots$      & $1.48$   & $\ldots$   & D                   & AS326   & $10.6$           & 1988 Aug 28          &                & $\ldots$           \\
$\ldots$  & $C$          & $4.86$   & $\ldots$   & D                   & AK424   & $3.4$            & 1996 Sep 28          &                & 6                  \\
NGC~4631  & $L$          & $1.37$   & WSRT       & maxi-short          & N/A     & $6.0$            & 2003 Apr 3           &                & 1                  \\
$\ldots$  & $C$          & $4.86$   & VLA        & D                   & AH369   & $12.1$           & 1989 Nov 22/26       & Mosaic         & 9                  \\    
$\ldots$  & $\ldots$     & $\ldots$ & $\ldots$   & D                   & AD896   & $4.3$            & 1999 Apr 14          & Mosaic         & 12                 \\    
$\ldots$  & $\ldots$     & $4.85$   & Effelsberg & single-dish         & 55--94  & $6.3$            & 1996 Feb--Aug        & Merged         & 6                  \\       
NGC~4666  & $L$          & $1.43$   & VLA        & CD                  & AD346   & $3.5$            & 1994 Nov 20          &                & 5                  \\   
$\ldots$  & $\ldots$     & $1.49$   & $\ldots$   & D                   & AS199   & $0.2$            & 1984 Aug 31          &                & This work          \\
$\ldots$  & $C$          & $4.86$   & $\ldots$   & D                   & AD326   & $12.5$           & 1993 Dec 20/24       &                & 5                  \\   
NGC~5775    & $L$          & $1.49$   & VLA      & B                   & AI0028  & $3.2$            & 1986 Aug 1           &                & 8                  \\     
$\ldots$  & $\ldots$     & $1.48$   & $\ldots$   & B                   & AB492   & $1.2$            & 1989 Aug 4           &                & $\ldots$           \\
$\ldots$  & $\ldots$     & $1.49$   & $\ldots$   & C                   & AH368   & $3.6$            & 1990 Nov 19/24       &                & $\ldots$           \\
$\ldots$  & $\ldots$     & $\ldots$ & $\ldots$   & D                   & AI31    & $1.9$            & 1987 Apr 27/30       &                & 11                 \\
$\ldots$  & $X$          & $8.45$   & $\ldots$   & D                   & AD455   & $13.4$           & 2001 Dec 14          &                & 15                 \\
\hline
\end{tabular}
\flushleft{{\bf Notes.}\\ $^{\rm a}$observing band; $^{\rm b}$observing centre frequency; $^{\rm c}$telescope; $^{\rm d}$configuration of the interferometer or single-dish telescope (the maxi-short configuration of the WSRT allows for optimum imaging performance for very extended sources within a single track observation); $^{\rm e}$project ID; $^{\rm f}$on-source time; $^{\rm g}$observation date; $^{\rm h}$comments (some interferometric observations were done in mosaic mode and the single-dish maps were merged with the interferometric maps (Section~\ref{sec:imaging})); $^{\rm i}$reference where the data were first published. For all entries, `$\ldots$' stands for an identical entry as above and `N/A' denotes an entry that is not available. Details for the observations of NGC~7090 and 7462 can be found in \citetalias{heesen_16a}. \\
{\bf References.} (1) \citealt{braun_07a}; (2) \citealt{carilli_92a}; (3) \citealt{cecil_01a}; (4) \citealt{colbert_96a}; (5) \citealt{dahlem_97a}; (6) \citealt{dumke_97a}; (7) \citealt{dumke_98a}; (8) \citealt{duric_98a}; (9) \citealt{golla_94a}; (10) \citealt{heesen_09a}; (11) \citealt{irwin_99a}; (12) \citealt{mora_13a}; (13) \citealt{oosterloo_07a}; (14) \citealt{reuter_91a}; (15) \citealt{soida_11a}; (16) \citealt{sukumar_91a}; (17) \citealt{wells_97a}.  }
\end{table*}

\subsection{Radio continuum maps}
We used a variety of radio continuum observations, most of them already published, which we
obtained from the respective telescope archives. We re-reduced these data, which we describe in the
following. In Table~\ref{tab:obs}, we present a summary of
the observations used, including references where the data were published
first.
\subsubsection{Australia Telescope Compact Array}
Observations of NGC~55 with the ATCA were calibrated
following standard procedures with {\small MIRIAD} \citep*{sault_95a},
where we set the flux
density of the primary calibrator J1938$-$634 with {\small MFCAL} (part of {\small MIRIAD}). At $L$ band we combined previously published data from \citet{wells_97a} with newly obtained data as part of
the Local Volume \ion{H}{i} Survey (LVHIS). At $C$ band, we used again the
previously published data of \citet{wells_97a} and added new observations; these data were observed in 2008 November
and 2009 February in the radio continuum mode with a bandwidth of
256~MHz, split into two IFs of 128~MHz bandwidth each. More data were
observed in 2010 with the upgraded Compact Array Broad-band Backend \citep[CABB;][]{wilson_11a} receiver ($4.7$--$6.6$~GHz). Because NGC~55 has an angular extension that exceeds the
primary beam both at $L$ and $C$ band, we used several pointings to mosaic the
galaxy. The calibrated $(u,v)$ data were precessed to J$2000.0$ if necessary and exported into the {\small FITS}
format. The observations and data reduction of NGC~7090 and
7462 were already presented in \citetalias{heesen_16a}.

\subsubsection{Very Large Array}
Observations with the VLA were 
calibrated following standard procedures with {\small AIPS}.\footnote{{\scriptsize AIPS}, the Astronomical Image Processing Software, is free software available from NRAO.} We set
the flux density of the primary calibrator (either 3C~48 or 286) according
to the model by \citet{baars_77a}. Since the data were observed only with the
`historical' VLA prior to the upgrade of the correlator, observations had
bandwidths of 100~MHz split into two IFs of 50~MHz each. We used data at $L$
band (NGC 253, 3044, 3079, 3628, 4565, 4666 and 5775), $C$ band (NGC 253, 891,
3044, 3079, 3628, 4565, 4631 and 4666) and $X$ band (NGC 5775). We used maps of NGC~253 published earlier
without either re-calibrating or re-imaging of the data. The observations at $L$ and $C$ band were already
described in \citet{carilli_92a} and \citet{heesen_09a}, respectively. The
calibrated data were precessed to J$2000.0$ where necessary and exported into
the {\small FITS} format.

\subsubsection{Westerbork Synthesis Radio Telescope}
We used a $1.4$-GHz map of NGC~891 observed with the WSRT, which was already presented by
\citet{oosterloo_07a}. Furthermore, we used the $1.4$-GHz map of NGC~4631
again observed with the WSRT. This galaxy was observed as part of the WSRT SINGS
survey \citep{braun_07a} and we re-imaged the map as described below.
\begin{table*}
\caption{Radio properties of our sample galaxies.\label{tab:radio}}
\begin{tabular}{l c cc cccc cc}
\hline
Galaxy & $\Theta_{\rm FWHM}^{\rm a}$ & $\nu_1^{\rm b}$ & $\nu_2^{\rm b}$ & $S_1^{\rm c}$ & $S_{\rm 1,nt}^{\rm d}$ & $S_2^{\rm c}$ & $S_{\rm 2,nt}^{\rm d}$ & $\alpha^{\rm e}$ & $\alpha_{\rm nt}^{\rm f}$ \\
& (arcsec) & \multicolumn{2}{c}{(GHz)} & (Jy) & (Jy) & (Jy) & (Jy) & & \\
\hline
NGC 55   & $41.10$ & $1.36$ & $5.56$ & $0.6082\pm 0.0172$ & $0.5219\pm 0.0152$ & $0.2966\pm 0.0107$ & $0.2216\pm 0.0084$ & $-0.51\pm 0.03$ & $-0.61\pm 0.03$ \\
NGC 253  & $30.00$ & $1.46$ & $4.85$ & $6.3000\pm 0.1409$ & $6.2133\pm 0.1390$ & $2.7100\pm 0.0678$ & $2.6331\pm 0.0659$ & $-0.70\pm 0.03$ & $-0.72\pm 0.03$ \\
NGC 891  & $17.30$ & $1.39$ & $4.86$ & $0.7720\pm 0.0155$ & $0.7663\pm 0.0154$ & $0.2720\pm 0.0056$ & $0.2670\pm 0.0056$ & $-0.83\pm 0.02$ & $-0.84\pm 0.02$ \\
NGC 3044 & $7.80 $ & $1.49$ & $4.86$ & $0.1040\pm 0.0021$ & $0.0990\pm 0.0020$ & $0.0450\pm 0.0009$ & $0.0406\pm 0.0009$ & $-0.71\pm 0.02$ & $-0.75\pm 0.03$ \\
NGC 3079 & $8.30 $ & $1.53$ & $4.71$ & $0.8360\pm 0.0169$ & $0.8262\pm 0.0167$ & $0.3430\pm 0.0350$ & $0.3342\pm 0.0341$ & $-0.79\pm 0.09$ & $-0.80\pm 0.09$ \\
NGC 3628 & $17.70$ & $1.49$ & $4.86$ & $0.5482\pm 0.0198$ & $0.5433\pm 0.0196$ & $0.2287\pm 0.0051$ & $0.2243\pm 0.0050$ & $-0.74\pm 0.04$ & $-0.75\pm 0.04$ \\
NGC 4565 & $13.10$ & $1.49$ & $4.86$ & $0.1599\pm 0.0116$ & $0.1585\pm 0.0115$ & $0.0557\pm 0.0012$ & $0.0544\pm 0.0013$ & $-0.89\pm 0.06$ & $-0.90\pm 0.06$ \\
NGC 4631 & $23.10$ & $1.37$ & $4.86$ & $1.2850\pm 0.0287$ & $1.2451\pm 0.0283$ & $0.4654\pm 0.0251$ & $0.4303\pm 0.0236$ & $-0.80\pm 0.05$ & $-0.84\pm 0.05$ \\
NGC 4666 & $15.70$ & $1.45$ & $4.86$ & $0.3870\pm 0.0087$ & $0.3712\pm 0.0083$ & $0.1370\pm 0.0039$ & $0.1230\pm 0.0035$ & $-0.86\pm 0.03$ & $-0.91\pm 0.03$ \\
NGC 5775 & $8.25 $ & $1.49$ & $8.46$ & $0.2680\pm 0.0097$ & $0.2630\pm 0.0095$ & $0.0598\pm 0.0022$ & $0.0556\pm 0.0020$ & $-0.86\pm 0.03$ & $-0.89\pm 0.03$ \\
NGC 7090 & $13.80$ & $1.38$ & $4.67$ & $0.0563\pm 0.0030$ & $0.0519\pm 0.0029$ & $0.0153\pm 0.0023$ & $0.0114\pm 0.0018$ & $-1.07\pm 0.13$ & $-1.24\pm 0.14$ \\
NGC 7462 & $13.90$ & $1.38$ & $4.67$ & $0.0256\pm 0.0009$ & $0.0238\pm 0.0012$ & $0.0066\pm 0.0002$ & $0.0050\pm 0.0006$ & $-1.10\pm 0.04$ & $-1.27\pm 0.11$ \\
\hline
\end{tabular}
\flushleft{{\bf Notes.} Uncertainties quoted in this table were calculated as explained in Section~\ref{sec:uncertainties}.\\
$^{\rm a}$ angular resolution,  referred to as the full width at half maximum (FWHM), of the circular synthesized beam.\\
$^{\rm b}$ observing frequencies $\nu_1$ and $\nu_2$.\\
$^{\rm c}$ integrated flux densities, $S_1$ and $S_2$, at observing frequencies $\nu_1$ and $\nu_2$.\\
$^{\rm d}$ non-thermal integrated flux densities, $S_{\rm 1,nt}$ and $S_{\rm 2,nt}$, after the subtraction of the thermal radio continuum emission (Sect.~\ref{sec:thermal_radio_continuum_maps}).\\
$^{\rm e}$ integrated radio spectral index between $\nu_1$ and $\nu_2$.\\
$^{\rm f}$ integrated non-thermal radio spectral index between $\nu_1$ and $\nu_2$.}
\end{table*}

\begin{table*}
\centering
\caption{Further radio properties of our sample galaxies.\label{tab:radio_maps}}
\begin{tabular}{l cccccccc}
\hline
Galaxy & $d_{\rm maj}^{\rm a}$ & $d_{\rm min}^{\rm a}$ & $\rm PA^{b}$ & $\sigma_1^{\rm c}$ & $\sigma_2^{\rm c}$ & $W_{\rm stripe}$$^{\rm d}$ & $W_{\rm stripe}$$^{\rm d}$ & Stripe centre$^{\rm e}$\\
& \multicolumn{2}{c}{(arcmin)} & ($\degr$) & \multicolumn{2}{c}{($\umu \rm Jy\,beam^{-1}$)} & (arcmin) & (kpc)\\\hline
NGC 55   & $24.0$ &  $9.3 $ & $108.0$ & $130$ & $40$ & $24.6$ & $14.0$ & RA~$\rm 00^{h}15^{m}07\fs 637$ Dec.~$-39\degr 12\arcmin 58\farcs 46$ \\
NGC 253  & $21.9$ &  $11.3$ & $52.0 $ & $250$ & $30$ & $20.3$ & $24.0$ & RA~$\rm 00^{h}47^{m}33\fs 190$ Dec.~$-25\degr 17\arcmin 16\farcs 14$\\
NGC 891  & $10.9$ &  $5.9 $ & $23.0 $ & $23 $ & $22$ & $4.9 $ & $15.0$ & RA~$\rm 02^{h}22^{m}33\fs 118$ Dec.~$+42\degr 20\arcmin 54\farcs 88$\\
NGC 3044 & $3.9 $ &  $2.2 $ & $114.0$ & $30 $ & $24$ & $2.7 $ & $15.0$ & RA~$\rm 09^{h}53^{m}40\fs 953$ Dec.~$+01\degr 34\arcmin 45\farcs 50$\\
NGC 3079 & $4.5 $ &  $2.1 $ & $167.0$ & $81 $ & $23$ & $4.4 $ & $26.0$ & RA~$\rm 10^{h}01^{m}57\fs 695$ Dec.~$+55\degr 40\arcmin 48\farcs 80$\\
NGC 3628 & $11.0$ &  $5.3 $ & $105.0$ & $80 $ & $22$ & $4.6 $ & $11.0$ & RA~$\rm 11^{h}20^{m}16\fs 642$ Dec.~$+13\degr 35\arcmin 21\farcs 76$\\
NGC 4565 & $13.7$ &  $2.8 $ & $135.5$ & $26 $ & $25$ & $7.6 $ & $56.0$ & RA~$\rm 12^{h}36^{m}20\fs 911$ Dec.~$+25\degr 59\arcmin 11\farcs 95$\\
NGC 4631 & $13.3$ &  $8.1 $ & $86.0 $ & $35 $ & $50$ & $5.8 $ & $12.0$ & RA~$\rm 12^{h}42^{m}07\fs 164$ Dec.~$+32\degr 32\arcmin 32\farcs 18$\\
NGC 4666 & $4.3 $ &  $2.5 $ & $40.0 $ & $50 $ & $20$ & $3.6 $ & $29.0$ & RA~$\rm 12^{h}45^{m}08\fs 707$ Dec.~$-00\degr 27\arcmin 44\farcs 23$\\
NGC 5775 & $4.1 $ &  $2.4 $ & $145.0$ & $35 $ & $15$ & $3.6 $ & $29.0$ & RA~$\rm 14^{h}53^{m}57\fs 526$ Dec.~$+03\degr 32\arcmin 42\farcs 35$\\
NGC 7090 & $5.4 $ &  $3.5 $ & $128.0$ & $28 $ & $13$ & $2.7 $ & $8.5$ & RA~$\rm 21^{h}36^{m}28\fs 774$ Dec.~$+54\degr 33\arcmin 26\farcs 70$\\
NGC 7462 & $3.8 $ &  $1.9 $ & $73.0 $ & $25 $ & $9 $ & $2.9 $ & $12.0$ & RA~$\rm 23^{h}02^{m}46\fs 751$ Dec.~$+40\degr 50\arcmin 08\farcs 09$\\
\hline
\end{tabular}
\flushleft{{\bf Notes.}\\ 
$^{\rm a}$ extent of the radio continuum emission along the major and minor axes.\\ 
$^{\rm b}$ position angle of the major axis.\\ 
$^{\rm c}$ rms map noises at observing frequencies $\nu_1$ and $\nu_2$.\\
$^{\rm d}$ stripe width to measure the vertical intensity profile in arcmin and kpc.\\
$^{\rm e}$ central position of the stripe in J$2000.0$ coordinates.}
\end{table*}

\begin{table}
\centering
\caption{Literature flux densities.\label{tab:flux}}
\begin{tabular}{l ccccc}
\hline
Galaxy & $\nu_1$ & $\nu_2$ & $S_1$ & $S_2$ & Reference\\
& \multicolumn{2}{c}{(GHz)} & \multicolumn{2}{c}{(mJy)} & ($\nu_1$ / $\nu_2$) \\
\hline
NGC 55   & $1.49$ & $4.85$ & $381 $ & $197 $ & 3 / 10\\
NGC 253  & $1.46$ & $4.85$ & $6300$ & $2710$ & 7 / 7\\
NGC 891  & $1.40$  & $4.85$ & $701 $ & $286 $ & 4 / 8\\
NGC 3044 & $1.40$  & $4.85$ & $114 $ & $43  $ & 9 / 5\\
NGC 3079 & $1.40$  & $4.85$ & $865 $ & $321 $ & 4 / 1\\
NGC 3628 & $1.40$  & $4.85$ & $525 $ & $247 $ & 4 / 8\\
NGC 4565 & $1.40$  & $4.80 $ & $134 $ & $54  $ & 4 / 8\\
NGC 4631 & $1.37$ & $4.80 $ & $1290$ & $476 $ & 2 / 8\\
NGC 4666 & $1.40$  & $4.85$ & $434 $ & $161 $ & 4 / 6\\
NGC 5775 & $1.40$  & $4.85$ & $284 $ & $94  $ & 4 / 8\\
\hline
\end{tabular}
\flushleft{{\bf References.} (1) \citealt*{becker_91a};  (2) \citealt{braun_07a}; (3) \citealt{condon_96a}; (4) \citealt{condon_02a}; (5) \citealt{gregory_91a}; (6) \citealt{griffith_95a}; (7) \citealt{heesen_09a}; (8) \citealt{stil_09a}; (9) \citealt{white_92a}; (10) \citealt{wright_94a}.}
\end{table}

\subsubsection{Single-dish telescopes}
In order to correct for the missing `zero-spacings flux' we use additional $C$ band maps
from the Effelsberg 100-m telescope (NGC~253, 891, 4631) and from the
Parkes 64-m telescope (NGC~55).\footnote{Radio interferometers are only sensitive to emission not exceeding a certain angular scale, which is related to the shortest baseline for which visibilities are measured. The (u,v)-plane has no measurements at a baseline of zero length, which can be only measured with a single-dish telescope. This is needed to measure the true integrated flux density. The difference between the interferometric and single-dish flux density is hence usually referred to as the `missing zero-spacings' flux.} The Effelsberg maps have been published before (see Table~\ref{tab:obs} for references), but the Parkes map is so far unpublished, so we
describe these observations here in more detail. Observations with the
$C$ band AT Multi-band
receiver ($4.5$--$5.1$~GHz) were carried out in 2010 September. Maps were taken
in standard fashion using the `basket weaving' procedure with scans alternating in the right ascension and declination directions. The data were calibrated with scans in azimuth
and elevation of the primary calibrator J1938$-$634, before they were combined using the single-dish mapping algorithm described in \citet{carretti_10a}. The resulting Parkes map of NGC~55 is presented in
Appendix~\ref{additional_maps} along with the Effelsberg maps of NGC~253, 891 and
4631.

\subsubsection{Imaging}
\label{sec:imaging}
We imported the calibrated $(u,v)$ data into the Common Astronomy Software Applications \citep[{\small CASA;}][]{mcmullin_07a}
and formed and deconvolved images of them using the multi-scale multi-frequency (MS--MFS) algorithm in the {\small CLEAN} task \citep{rau_11a}. We
did not use the frequency dependence of the sky model (nterms=1) because this is not necessary for our small fractional bandwidths (mostly $<$10 per 
cent). The obvious exception are the ATCA data, where the fractional bandwidth is highest with 34~per cent. But because we created a `joint deconvolution' mosaic \citep*{sault_96a} in {\scriptsize CASA} (imagermode=mosaic), we could not use the frequency dependence; this option can thus far not be combined with a frequency dependent skymodel ($\rm nterms\geq 2$). Since we do not detect any obvious artefacts left over after the  deconvolution, we think that our approach is sufficient for our data.\footnote{We do not advocate this procedure as generally the best deconvolution technique for mosaicked observations. Depending on the data, a linear mosaic (e.g.\ using {\scriptsize LTESS} in {\scriptsize AIPS}) of the individual pointings deconvolved with nterms=2 may give a better result.} We used the multi-scale option with angular scales ranging typically
between one synthesized beam size and the size of the galaxy. We cleaned the maps down to
2$\sigma$, where $\sigma$ is the rms noise level. Finally, we applied the primary beam correction. 

We deconvolved 
one map with {\small CLEAN} using Briggs weighting (robust=0) and adjusted the
weighting of the other map to robust values between $0$ and $0.5$, in order to
have a similar resolution. For NGC~55 and 4631, we had several pointings of a
mosaic. They were processed within {\small CASA} using the joint deconvolution
option of the {\small CLEAN} algorithm. Maps were imported back into {\small AIPS} to convolve them to a common
circular beam and re-grid them to the same coordinate system, for which we used
{\small PARSELTONGUE} \citep{kettenis_06a} to batch process them. The $C$-band maps of NGC~55 and 4631 were combined with the single-dish maps from the
Parkes and Effelsberg telescopes, respectively, using the {\small IMERG} task in {\small AIPS}. The $C$
band map of NGC~891 had no flux missing since the single-dish and interferometric integrated fluxes were in good agreement, so that we did not merge it with the
Effelsberg map. The properties of the radio maps are summarized in Tables~\ref{tab:radio} and \ref{tab:radio_maps}.

\begin{table*}
\caption{Physical properties of our sample galaxies.\label{tab:phys}}
\begin{tabular}{lcccc cccc ccc}
\hline
Galaxy & $L_{\rm TIR}^{\rm a}$ & $B_0^{\rm b}$ & $U_{\rm TIR}^{\rm c}$ & $U_{\rm IRF}^{\rm d}$ & $U_{\rm B}^{\rm e}$ & $U_{\rm IRF}/U_{\rm B}^{\rm f}$ & $U_{\rm rad}/U_{\rm B}^{\rm g}$ & $F_{\rm H\,\alpha}^{\rm h}$ & $E(B-V)^{\rm i}$ & Ref.\ H\,$\alpha$\\
& $(10^{43}~\rm erg\, s^{-1})$ & ($\umu$G) & \multicolumn{3}{c}{($10^{-13}~\rm erg\, cm^{-3})$} & & & ($10^{12}~\rm erg\,s^{-1}\,cm^{-2}$) & (mag) & Map / Flux\\
\hline
NGC 55   & $0.41 $ & $7.9 $ & $0.6 $ & $1.6 $ & $24.5 $ & $0.07$ & $0.24$ & $78.1\pm 3.6$ & $0.012$ & 2 / 5     \\
NGC 253  & $17.90$ & $14.0$ & $7.1 $ & $19.5$ & $78.3 $ & $0.25$ & $0.30$ & $79.0\pm 3.6$ & $0.017$ & 5 / 5    \\
NGC 891  & $12.10$ & $14.7$ & $4.9 $ & $13.4$ & $86.0 $ & $0.16$ & $0.20$ & $5.1 \pm 1.1$ & $0.058$ & 7 / 5     \\
NGC 3044 & $5.20 $ & $13.1$ & $3.0 $ & $8.2 $ & $68.1 $ & $0.12$ & $0.18$ & $4.5 \pm 0.5$ & $0.025$ & 7 / 7     \\
NGC 3079 & $23.30$ & $19.9$ & $7.3 $ & $20.0$ & $157.5$ & $0.13$ & $0.15$ & $9.0 \pm 0.4$ & $0.361$ & 11 / 6     \\
NGC 3628 & $4.43 $ & $12.6$ & $2.2 $ & $5.9 $ & $63.1 $ & $0.09$ & $0.16$ & $4.5 \pm 0.4$ & $0.020$ & $24~\umu\rm m$ / 7     \\
NGC 4565 & $14.60$ & $8.7 $ & $0.6 $ & $1.6 $ & $29.8 $ & $0.05$ & $0.19$ & $1.3 \pm 0.4$ & $0.015$ & $24~\umu\rm m$ / 8    \\
NGC 4631 & $7.40 $ & $13.5$ & $2.8 $ & $7.7 $ & $72.3 $ & $0.11$ & $0.16$ & $36.1\pm 5.0$ & $0.017$ & 5 / 5     \\
NGC 4666 & $41.50$ & $18.2$ & $11.9$ & $32.4$ & $132.3$ & $0.24$ & $0.28$ & $14.4\pm 0.6$ & $0.028$ & $24~\umu\rm m$ / 10  \\
NGC 5775 & $25.60$ & $16.3$ & $6.9 $ & $18.8$ & $105.9$ & $0.18$ & $0.22$ & $4.5 \pm 0.5$ & $0.037$ & 1 / 4  \\
NGC 7090 & $1.60 $ & $9.8 $ & $1.7 $ & $4.8 $ & $38.2 $ & $0.12$ & $0.23$ & $4.0 \pm 0.6$ & $0.020$ & 5 / 5   \\
NGC 7462 & $0.72 $ & $9.7 $ & $0.7 $ & $2.0 $ & $37.1 $ & $0.05$ & $0.16$ & $1.6 \pm 0.8$ & $0.009$ & 9 / 3   \\
\hline
\end{tabular}
\flushleft{{\bf Notes.} See Appendix~\ref{physical} for details how we calculated the presented values.\\
$^{\rm a}$ total infrared luminosity $3$--$1100~\umu\rm m$ from \emph{Spitzer} and \emph{IRAS} data.\\ 
$^{\rm b}$ magnetic field strength in the disc plane, calculated from energy equipartition using {\scriptsize BFIELD} \citep{beck_05a}, available on \href{http://www3.mpifr-bonn.mpg.de/staff/mkrause/}{http://www3.mpifr-bonn.mpg.de/staff/mkrause/}.\\ 
$^{\rm c}$ total infrared photon energy density, calculated from the total infrared luminosity.\\ 
$^{\rm d}$ radiation energy density of the interstellar radiation field, excluding the contribution from the cosmic microwave background.\\ 
$^{\rm e}$ magnetic energy density $U_{\rm B}=B_0^2/(8\upi)$.\\ 
$^{\rm f}$ ratio of the energy densities of the interstellar radiation field (excluding the contribution from the cosmic microwave background) to that of the magnetic field.\\ 
$^{\rm g}$ ratio of the energy densities of the total interstellar radiation field (including the cosmic microwave background) to that of the magnetic field.\\ 
$^{\rm h}$ Balmer H\,$\alpha$ line flux, corrected for foreground absorption.\\ 
$^{\rm i}$ the $E(B-V)$ values are from NED to calculate the foreground absorption as $A_{V}=2.59\times E(B-V)~\rm mag$.\\
{\bf References.} (1) \citealt{collins_00a}; (2) (D.~J.~Bomans et al.\ 2016, priv.\ comm.); (3) \citetalias{heesen_16a}; (4) \citealt{james_04a}; (5) \citealt{kennicutt_08a}; (6) \citealt{moustakas_06a}; (7) \citealt{rand_11a}; (8) \citealt{robitaille_07a}, (9) \citealt{rossa_03a}; (10) \citealt{voigtlaender_13a}; (11) \citealt{young_96a}. Where no H\,$\alpha$} map was available, we used a \emph{Spitzer} 24-$\umu$m map instead (marked as $24~\umu\rm m$) from \citet{dale_09a} (NGC~3628) and \citet{bendo_12a} (NGC~4565 and 4666). 
\end{table*}

\subsubsection{Integrated flux densities}
\label{sec:integrated_flux_densities}
We integrated the flux densities of our galaxies using rectangular boxes with {\small IMSTAT} in {\small AIPS}. Highly inclined galaxies have box-shaped radio haloes (rather then elliptical haloes), so that this method is the most suitable one (rather than integrating in ellipses as for moderately inclined galaxies). The angular extent of the radio emission along the major and minor axis, which is the region that encompasses the 3$\sigma$ contour line of the most sensitive map can be found in Table~\ref{tab:radio_maps}. We chose the box size to integrate the emission to be slightly larger than this region, checking that the choice has only little influence ($\leq$2~per cent).

We checked the integrated flux
densities (Table~\ref{tab:radio}) of our
maps and found most of them to agree within 5 per cent with published values
in the literature (Table~\ref{tab:flux}). These values
are interferometric measurements at $L$ band and single-dish measurements at
$C$ band (so that missing fluxes play no role). We detect significantly higher (50 per cent) flux densities in NGC~55 at both a
$L$ and $C$ band. At $L$ band, our observations have a better $(u,v)$-coverage
than the comparison snapshot observation with the VLA by \citet{condon_96a}, so that we are less
affected by the missing zero-spacings flux. At $C$ band, the flux density of \citet{wright_94a} is integrated in an area with a major axis of only 7~arcmin, whereas the source size detected by us is 24~arcmin; faint, diffuse emission was missed so that the flux density is lower. Other notable exceptions
are NGC~4565, where we detect a 20 per cent higher flux density at $L$ band, and NGC~4666,
where our $C$ band flux density is 15 per cent lower than the literature value. In case
of NGC~4565 it is likely that we have detected more flux since we have used a
12~h long observation in D-configuration, whereas the literature value of
\citet{condon_02a} is based on a NVSS snapshot observation. NGC~4666 has an optical size of
$4.6$~arcmin, which is comparable to the largest angular scale (LAS) detectable at
$C$ band with the VLA (5~arcmin) in D-configuration, so that it is possible that we have missed
some flux.
%

%
Notably, our other $C$ band flux densities agree within 5 per cent with
single-dish data, so that missing zero-spacings are not an issue for us. At
$L$ band, we see indications of missing flux in NGC~55 and NGC~253. In these two galaxies the
spectral index is affected by missing flux in $L$ band, which in NGC~55 causes
a flat spectral index in the northern halo and in NGC~253 a flat
spectral index for distances exceeding 4~kpc from the midplane. We restrict our
spectral index analysis to the unaffected areas. In the remaining galaxies at $L$ band,
we do not expect it to be an issue. The largest galaxy of those is NGC~4565, which has an optical angular size of $16.2$~arcmin. This galaxy has been measured with a 12~h long integration with the VLA in D-array, where the LAS is 16~arcmin, so that missing flux is likely not an issue here.

\subsection{Thermal radio continuum maps}
\label{sec:thermal_radio_continuum_maps}
To correct for the contribution of thermal radio continuum emission, we used
Balmer H\,$\alpha$ line observations following the procedure described in
\citet{heesen_14a}. Where no H\,$\alpha$ map was available, we used a
\emph{Spitzer} 24-$\umu$m map instead and scaled it to a published H\,$\alpha$
flux density (NGC~3628, 4565 and 4666). The \emph{Spitzer} mid-infrared map
traces the dust-obscured star-formation, so that we cannot expect a good
H\,$\alpha$--mid-infrared correlation on a spatially resolved basis. But since we
are averaging (radially and along the line of sight) over several kpc in order to measure the vertical profiles of the
non-thermal radio continuum, the exact distribution does not matter
for our study. Also, we find that the thermal radio continuum contribution at $C$ band is typically
less than 10~per cent in the halo, where our
study is focused (at $L$ band it is even lower). The non-thermal radio spectral index is hence only by a
value of $0.06$ systemically steeper between $1.4$ and $4.85$~GHz (steepening from
$\alpha=-1.00$ to $-1.06$, for instance) than the total radio spectral index. This is similar to the size of our error bars on the radio
spectral index, so that the contribution of the thermal radio continuum
emission is actually not that important in the halo. Consequently, we do not expect the
difference between the Balmer H\,$\alpha$ and \emph{Spitzer} mid-infrared map to
make a large difference for our analysis.

This can also be seen when considering that the thermal contribution to the radio continuum intensity at $1.4$~GHz
(it varies slowly with frequency as $\nu^{-0.1}$) is:
\begin{equation}
  I_{\rm th, 1.4} = 2.5 \times 10^{-9} \cdot {\rm EM}
  \cdot \Theta_{\rm FWHM}^2~{\rm Jy\,beam^{-1}},
\end{equation}
where $\rm EM$ is the H\,$\alpha$ emission measure in units of $\rm
cm\,pc^{-6}$ and $\Theta_{\rm FWHM}$ is the angular resolution of the radio map, referred to as the full width at half maximum, in units of arcsec. The above relation is obtained when combining equation~(2) from \citet{voigtlaender_13a} with equation~(4) from
\citet{heesen_14a}, assuming an
electron temperature of $T_{\rm e}=10^4$~K and an integration area of $\Omega =1.133\cdot \Theta_{\rm FWHM}^2~\rm arcsec^2$. Typically, the emission measure
drops below 100~$\rm cm\,pc^{-6}$ at a height of 1~kpc, so that at 10~arcsec resolution the
thermal radio continuum intensity is smaller than $25~\rm \umu Jy\,beam^{-1}$ \citep{collins_00a}, similar to
the rms noise of our maps. The emission
measure drops even further at larger heights (the scale heights are only $\approx$$0.5$~kpc) and the thermal contribution becomes entirely negligible.

We corrected our maps for foreground absorption using $E(B-V)$ values, but
these corrections are only a few per cent (it is largest for NGC~891 with 15
per cent). We do not correct for internal absorption by dust (internal to the observed galaxy) except in NGC~3079 and
4666, where the Balmer decrement has been used to estimate it \citep{voigtlaender_13a}. The correction is a factor of $2.4$ in NGC~3079 and $2.9$ in
NGC~4666, hence it totally dominates the estimate of the H\,$\alpha$ flux. But
since the absorption by
dust is negligible in the halo \citep{collins_00a}, we 
do not have to correct for internal absorption. Hence, we have possibly
overcorrected the thermal flux in NGC~3079 and 4666, but the fractions are low
($<$10 per cent). We also do not correct for the
contribution of [\ion{N}{II}], which in our sample is between 25 (NGC~55)
and 45 per cent (NGC~4666). Using the absolute $B$-band
magnitude as proxy \citep{kennicutt_08a}, we expect an average of 35 per cent
of [\ion{N}{II}] contribution; this means, we slightly overestimate the
thermal radio continuum emission.\footnote{The [\ion{N}{II}] emission line falls into the bandpass of the H\,$\alpha$ filter; the contribution from this emission cannot be separated from the H\,$\alpha$ flux which is hence an overestimate.} We conclude that the
thermal contribution is negligible in the halo and our measurements of the non-thermal intensities are conservative \emph{lower} limits.

\subsection{Masking}
\label{sec:masking}
We masked unrelated background sources, which we identified as point-like sources in the halo that have no counterpart in the H\,$\alpha$ map. Furthermore, we masked nuclear starbursts and AGNs in those galaxies that have a prominent, dominating nucleus in the radio continuum maps (NGC~253, 3079 and 3628). In NGC~3079, we have masked the nuclear outflow as well, as far as we could distinguish it from the remainder of the disc and halo emission. In NGC~891, we have masked emission from SN1986J \citep{rupen_87a}. For all galaxies, the same mask was then applied to all maps (radio continuum at both frequencies and thermal radio continuum) before the construction of the spectral index maps.

\subsection{Uncertainties}
\label{sec:uncertainties}
The error of the integrated flux densities were calculated with two contributions. First, we assumed a 2 per cent relative calibration error for the WSRT and VLA observations \citep{braun_07a,perley_13a}. For the single-dish observations, the calibration error is also 2 per cent for the Parkes observations at $C$ band \citep{griffith_93a} and for the Effelsberg observations at $C$ and $X$ band (R.~Beck 2017, priv.\ comm.). Second, we added the uncertainty arising from the baselevel error, which is $\sigma_{\rm N}=\sigma_{\rm b}N=\sigma\sqrt{N}$, where $N$ is the number of beams within the integration region and the baselevel uncertainty for the intensities is $\sigma_{\rm b}=\sigma/\sqrt{N}$.  This is a crude estimate since if an image contains large-scale artefacts due to insufficient calibration (interferometric data) or due to scanning effects (single-dish data), the rms variation in the image, $\sigma$, is larger than just thermal noise and does not have Gaussian characteristics. Hence, we checked this error by measuring the average flux intensity in areas surrounding the galaxies, where the box size was selected to be similar to the galaxy size and found approximate agreement. The calibration and background errors were quadratically added with $(\delta S_{\nu})^2 = (\epsilon_{S} S_{\nu})^2 + \sigma_{\rm N}^2$, where $\epsilon_{S}=0.02$. For the subtraction of the thermal emission we used the error of the H\,$\alpha$ flux density, which was then propagated into the non-thermal flux densities and spectral indices.

For our main analysis, we created vertical profiles of the radio continuum intensity averaged in stripes. The stripes were spaced by $0.5\cdot \theta_{\rm FWHM} \cdot D$ in the vertical direction and had stripe widths between $0.5$ and $1.0$ of the length of the major axis (see Table~\ref{tab:radio_maps} for the stripe widths). For the vertical intensity profiles, we used a 5 per cent calibration error, owing to the calibration uncertainty and the deconvolution process. The error contributions are again added in quadrature $(\delta I_{\nu})^2=(\epsilon_{I} I_{\nu})^2 + \sigma_{\rm b}^2$ with $\epsilon_{I}=0.05$. This uncertainty neglects the variation of the intensities within stripe width; the reason is that we assume in the following a 1D approach of cosmic-ray transport and any variation as function of galactocentric radius is not included in our model. The error of the radio spectral index is then calculated with following equation:
\begin{equation}
	\Delta\alpha = \frac{1}{\ln\left (\frac{\nu_1}{\nu_2}\right )}\cdot \sqrt{\left ( \frac{\delta I_1}{I_1}\right )^2 +  \left (\frac{\delta I_2}{I_2}\right )^2},
\end{equation}
where $\delta I_1$ and $\delta I_2$ are the intensity errors and $I_1$ and $I_2$ the intensities at the observing frequencies $\nu_1$ and $\nu_2$, respectively.

\section{Cosmic-ray transport models}
\label{transport}
\subsection{Motivation}
\subsubsection{Key assumptions}
\label{sec:assumptions}
Our goal is to study the vertical cosmic-ray transport using the 1D models for
pure advection and diffusion from \citetalias{heesen_16a}. These models assume
that the CREs are injected at a galactic height of $z=0$ and then transported
away from the disc by either pure advection in a galactic wind or by
diffusion along vertical magnetic field lines. We use these models with the following two assumptions:
\begin{itemize}
\item[] (i) we fit the radio spectral index in the halo only. This is motivated by the fact that the
  vertical profiles of the radio spectral index are not affected by the limited
  angular resolution for heights $|z|\gtrapprox 1$~kpc and the correction for thermal radio continuum emission becomes negligible.
\item[] (ii) we model the magnetic field as a two-component exponential function
  with a thin and a thick disc. This is motivated by the fact that in most galaxies the
  vertical non-thermal intensity profiles can be best fitted by a
  two-component exponential function \citep[e.g.][]{dahlem_94a,oosterloo_07a,heesen_09a,soida_11a,mora_13a}. Exceptions can be
  explained by low angular resolution and inclination angles, or by a diffusion-dominated
  cosmic-ray transport, which causes Gaussian intensity profiles (\citetalias{heesen_16a}).
\end{itemize}
We discuss these assumptions in more detail in Sects \ref{sec:disc_halo_interface} and
\ref{sec:magnetic_field_structure}. The 1D treatment is motivated by the many studies of cosmic-ray
driven winds that have
used the `flux tube' approximation \citep[e.g.][]{breitschwerdt_91a, breitschwerdt_93a,everett_08a,dorfi_12a,
  recchia_16a}. They assume straight, open magnetic field lines rising
above the disc. In this geometry, the wind flows along a tube of approximately
constant cylindrical cross-section up to a height $z\approx z_{\rm break}$,
after which the area increases (usually assumed to scale as $\propto z^{2}$) and hence the
magnetic field strength decreases as well. \citet{everett_08a} found $z_{\rm
  break}=4.5$~kpc in the Milky Way, which is of the order of the scale height of the magnetic field
\citep{haverkorn_12a}. These models have been able to reproduce the X-ray data
in the Milky Way and nearby galaxies, where it was shown that the wind speed
in the halo is of the order of the escape velocity and
its value does not change by more than a factor of a few 
\citep{everett_08a,breitschwerdt_12a}. This is important, because we assume a
constant advection speed that is able to describe the data to first order. We note that choosing a constant advection speed will give the minimum magnetic field strength in the halo. An accelerating flow dilutes the CRE number density by longitudinal expansion and adiabatic losses \citep{heesen_18a}, so that the magnetic field estimate would have to increase in order to fit the observed intensities.

\begin{figure}
  \includegraphics[width=1.0\hsize]{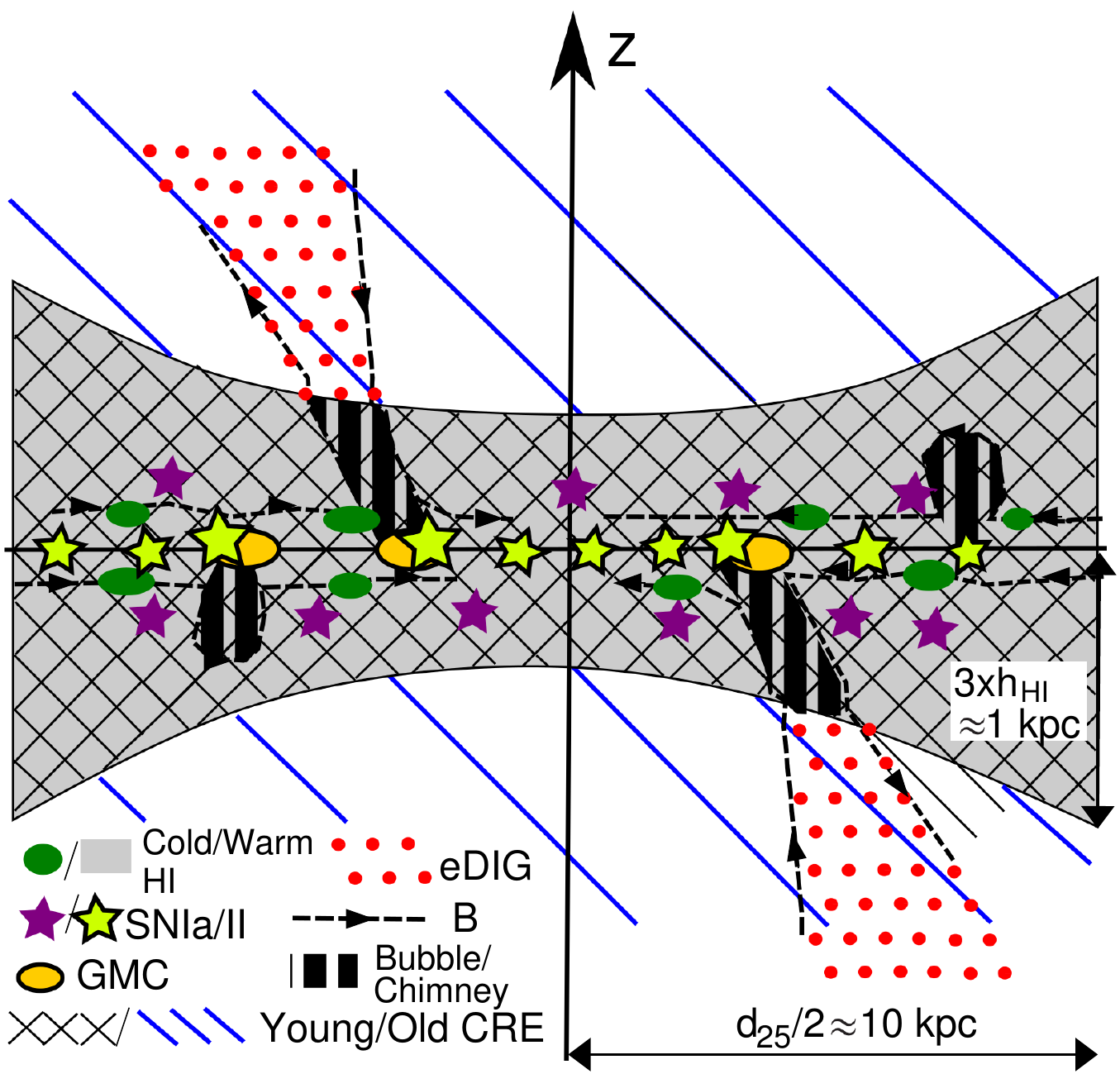}
  \caption{A diagram of the disc--halo interface in galaxies with $\Sigma_{\rm SFR}\ll 10^{-1}~\rm M_{\sun}\,yr^{-1}\,kpc^{-2}$. CREs are injected
    in the disc plane by Type II SNe (and Type Ib/Ic) and
    in the halo by Type Ia SNe. They accumulate in superbubbles
    which can break out from the flaring disc of warm \ion{H}{i} and
    form a `chimney', which are cospatial with plumes of extra-planar diffuse
    ionized gas (eDIG). The magnetic field (dashed lines with arrows) is anchored to the disc in the cold
    \ion{H}{i} clouds. The vertical scale is stretched by a factor of 5.}
\label{fig:phase}
\end{figure}

\subsubsection{Disc--halo interface}
\label{sec:disc_halo_interface}
This paper is about haloes, but in order to investigate them, we need to take into account the full vertical extent of a galaxy, so we give here with a brief overview of the vertical structure of the interstellar medium (ISM). In the context of radio haloes, the transition from the thin to the thick disc is important, so we start with what is commonly referred to as the disc--halo interface. In Fig.~\ref{fig:phase}, we show a conceptual diagram of the disc--halo interface and
we discuss in the following how our model can fit into this picture. This diagram is
most realistic (although still very simplifying) for galaxies with low
$\Sigma_{\rm SFR}\ll 10^{-1}~\rm M_{\sun}\,yr^{-1}\,kpc^{-2}$, where cosmic
rays are most likely to play a role in launching a wind. Galaxies with higher $\Sigma_{\rm SFR}$ have thermally driven so-called superwinds \citep{heckman_00a}, which are not included in our study \citep[see][for a radio continuum study of the superwind galaxy M82]{adebahr_13a}. 

Cosmic rays
are accelerated and injected into the ISM via SNe, where the
ejected shells have speeds of the order of 10,000~$\rm km\, s^{-1}$, which
create a strong shock while sweeping up material and slowing down. The turbulent magnetic field
lines, which surround the shock, deflect the cosmic rays, resulting in an efficient Fermi-type I
acceleration with the cosmic rays crossing the shock many times
\citep{bell_78a}. On average, the kinetic energy per SN is $10^{51}$~erg, a few
per cent of which is used for the acceleration of cosmic rays \citep[e.g.][]{rieger_13a}. Of the
energy stored in the cosmic rays, between 1 and 2 per cent goes into the CREs with the rest into
protons and heavier nuclei \citep[e.g.][]{beck_05a}. The SNe can be either Type Ia (thermonuclear
detonation) or Type II (core-collapse, also Type Ib and Ic). This is of significance since the SNe
of Type Ia can occur at galactic heights of a few 100~pc, whereas SNe of other types
are restricted to within 100~pc from the midplane \citep[effectively, the scale height
of the molecular gas;][]{ferriere_01a}.

In the thin disc, the CRE population is very diverse. Approximately 10 per cent
of the radio continuum emission stems from individual SN remnants, with the remaining 90 per cent made up by the
diffuse emission \citep{lisenfeld_00a}. The radio spectral index varies strongly between the spiral arms
and the inter-arm regions, which is caused by spectral ageing of the CREs
with increasing distance from star formation sites \citep{tabatabaei_13a,heesen_14a}. Furthermore, CREs can be found
within superbubbles, which can be observed during later stages as \ion{H}{i} holes with sizes ranging from 100~pc to 2~kpc \citep{bagetakos_11a}. They are the result of dozens of SNe and some of
them can
contain the CREs long enough that they display a curved
non-thermal radio continuum spectrum \citep{heesen_15a}. They may also be the site of `chimneys',
where star formation activity in the disc drives advective hot gas flows
upward into the halo, carrying magnetic fields along with the hot gas motion
and forming cavities in the disk that are observable as \ion{H}{i} holes
\citep{heald_12a}. We would expect at any given time a few dozens of superbubbles per
galaxy with a diameter of a few 100 pc, so that the volume porosity is a few
per cent but can be up to 20 per cent \citep{bagetakos_11a}. It hence becomes
clear that the CREs in the thin disc are in fact a 
superposition of many spectral ages (young CREs in the vicinity of star forming regions, old CREs in
the inter-arm regions, young and old CREs in non-thermal superbubbles). Consequently, the radio spectral index of the thin
disc is not equivalent to the injection energy spectral index of the CREs as expected from theory, which is $\gamma_{\rm inj}\approx 2$, where the CRE number density is a power-law as function of energy $N\propto E^{-\gamma_{\rm inj}}$, but is steeper. Indeed, the integrated non-thermal
radio continuum spectrum of galaxies between 1 and 10~GHz has a spectral index of $\alpha_{\rm
  nt}\approx -1$, which corresponds to $\gamma_{\rm inj}=3$ \citep{niklas_97b,heesen_14a,basu_15a,tabatabaei_17a}.
  
In the thick disc, the CRE population becomes more uniform since the
distance to the star forming sites becomes more similar and the contrast
between spiral arm and
inter-arm regions diminishes. The height at which the transition from thin to thick disc
happens is probably of the order of a few 100~pc; this is
related to the
scale height $h_{\rm \ion{H}{I}}$ of the warm \ion{H}{i} disc and thus to the largest height the
superbubbles can expand to before they blow-out \citep[at $z=3\times h_{\rm \ion{H}{I}}$;][]{maclow_99a}. The
\ion{H}{i} discs show a flaring, where the scale height increases with
galactocentric distance \citep[$h_{\rm \ion{H}{I}}=100$--500~pc;][]{bagetakos_11a}, so that the superbubble blow-out would 
occur at heights between $0.3$ and $1.5$~kpc. 

As we will show below, we see the
transition from thin to thick disc at a height of 1--2~kpc as traced by a
change of the slope in the radio
spectral index. Since we cannot really resolve the
thin disc, we can only probe the upper limit of the transition height, but our measured values are at least consistent with the described picture. \emph{Our first
  assumption is hence that we fit the radio 
spectral index in the halo only. In the midplane, the radio continuum emission can be described by a power law with an
`injection' spectral index, where this is effectively the
 spectral index of the diffuse non-thermal radio continuum emission.}

\subsubsection{Magnetic field structure}
\label{sec:magnetic_field_structure}
The magnetic field in galaxy discs is
in approximate energy equipartition with the turbulent gas motions of the
(almost) neutral gas and its
associated kinetic energy density \citep{beck_16a}. If the magnetic field is associated with the warm
\ion{H}{i} disc, as suggested by the superbubble picture, the resulting scale
height would be  $0.2$--$1.0$~kpc in the thin disc as the magnetic field scale
height is a factor of 2 higher than the magnetic pressure scale height.\footnote{For an exponential magnetic field distribution $B\propto \exp(-z/h_{\rm B1})$, the magnetic energy density is $U_{\rm
  B}= B^2/(8\upi)\propto \exp(-2z/h_{\rm B1})$.} The magnetic field scale height in the thick disc is
larger, from equipartition estimates the scale heights are of the order of 4--8~kpc. It
is unclear whether the field is associated with either the hot X-ray emitting
gas, which has similar scale heights \citep{strickland_04a,hodges-kluck_13a},
or the thick \ion{H}{i} disc that is seen in some galaxies with scale
heights of 2--4~kpc \citep{zschaechner_15a,vollmer_16a}. We caution though that in
both cases accretion from the intergalactic medium
probably plays a role as well, so that part of the gas in the halo has a different
origin from the magnetic field which is generated by processes 
\emph{within} the galaxy.

A discussion of
the magnetic field structure would be incomplete without mentioning the
results from polarization. The halo field consists of a turbulent and ordered
component, with the ordered component having the larger scale height. The
turbulent component possibly stems from Parker-type loops, which form when buoyant superbubbles inflated with cosmic rays rise from the disc and opposing magnetic field lines reconnect at its base. This allows a closed magnetic field line to detach from the disc magnetic field \citep{parker_92a}.  In that case the trajectory of the loop would be
ballistic, such as for clouds of \ion{H}{i} gas in a `Galactic fountain'. The
  ordered magnetic field could be from blown-out superbubbles, where the
  magnetic field opens up in the halo, with overlapping bubbles in the thick
  disc creating the halo field \citep{heald_12a,mao_15a,mulcahy_17a}. An alternative scenario is that the turbulent magnetic field is generated in the thin disc by the small-scale dynamo, driven by star-formation related turbulence, and advected into the halo by the wind. A lateral pressure gradient or a galactic wind
  would then be able to shape the field lines into the often observed
  X-shaped pattern
  \citep[e.g.][]{dahlem_97a,tuellmann_00a,krause_06a,heesen_09b,krause_09a,soida_11a,mora_13a,chyzy_16a}. Such
  a magnetic field structure has been also seen in models of lagging magnetized haloes \citep{henriksen_16a}. Of course, our simple model cannot take all of this
  into account, and we attempt in this work only to fit for the total magnetic
  field strength in the halo, neglecting this sub-division.

\emph{Our second assumption is hence, that the magnetic field strength can be described by a two-component exponential function.} In the thin disc, the magnetic field strength is regulated by the pressure of the warm \ion{H}{i} gas, which has an exponential distribution. This is expected for a constant velocity dispersion in a constant gravity field as found near the midplane. In the thick disc, the dominating turbulent magnetic field is in pressure equilibrium with either the hot X-ray emitting gas or the thick \ion{H}{i} disc. If the halo gas is in a hydrostatic equilibrium, the density distribution is close to an exponential function. This is corroborated by observations of the soft X-ray emission in nearby galaxies, which showed that the vertical distribution of the hot ionized gas can be best described by an exponential function, rather than a Gaussian or a power-law function \citep{strickland_04a}. This suggests that X-ray haloes are `hot atmospheres' surrounding galaxies. If the magnetic field is frozen into the ionized plasma, the magnetic field strength will relate to the ionized gas density, so that exponential magnetic fields can be expected.

\begin{figure*}
  \includegraphics[width=1.0\hsize]{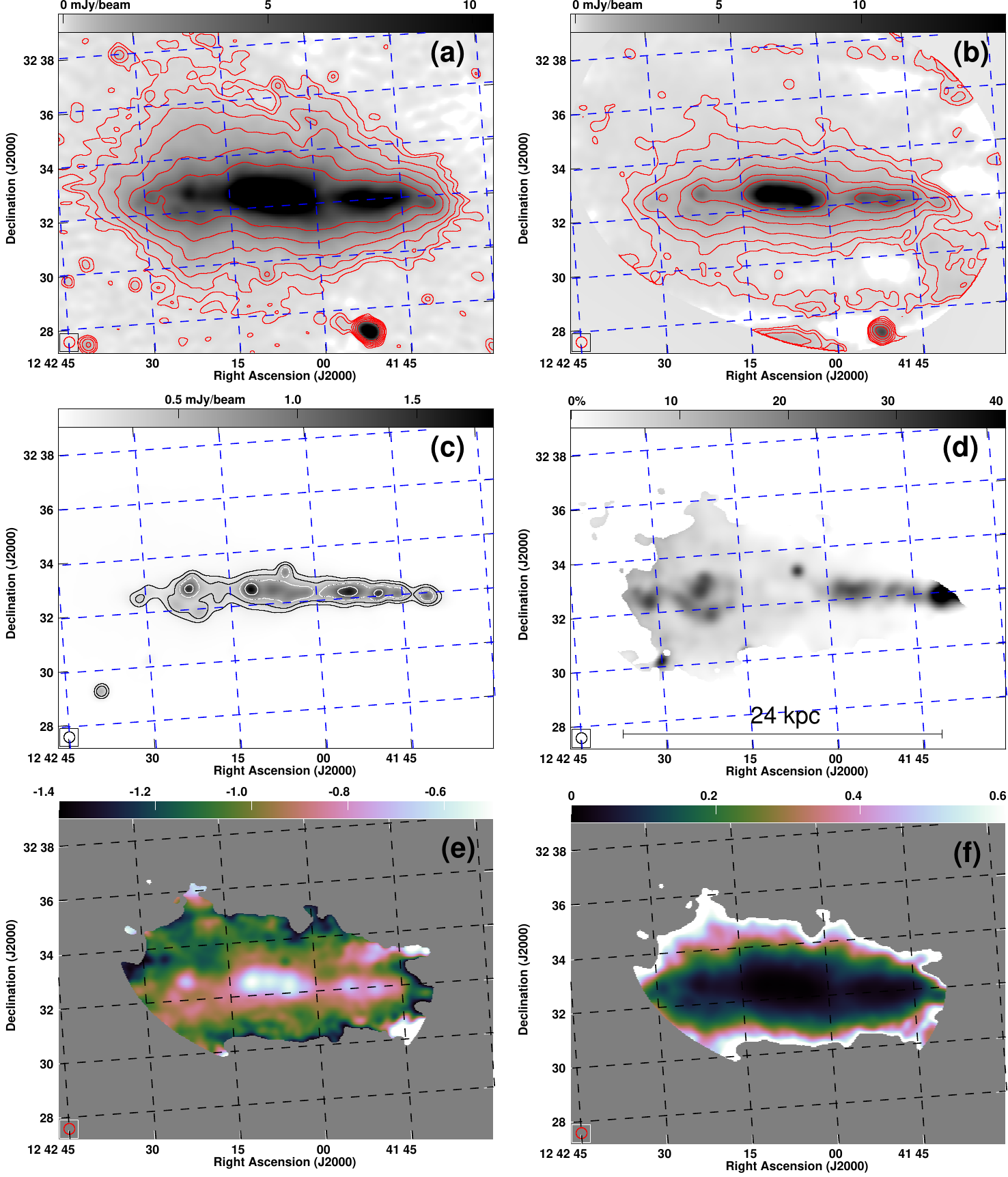}
  \caption{NGC~4631. (a) and (b) radio continuum emission at
    $1.37$ and $4.86$~GHz, respectively. The high noise level near the edge of the $4.86$-GHz map stems from the correction for the primary beam attenuation of the VLA. (c) the thermal radio continuum emission at
    $4.86$~GHz. Contours in panels (a)--(c) are at (3, 6, 12, 24, 48 and 96)
    $\times~\sigma$, where $\sigma$ is the rms map noise. In panel (c), the same contour levels as in (b) are used to ease the comparison. (d) thermal
  fraction at $4.86$~GHz, where they grey-scale ranges from 0 to 40 per cent. (e) non-thermal radio spectral index between $1.37$
  and $4.86$~GHz, where the colour-scale ranges from $-1.4$ to $-0.5$. (f) error of the non-thermal radio spectral index, where the colour-scale ranges from 0 to $0.6$. In all panels, the synthesized beam is shown in the bottom
left corner and the maps are rotated so that the major axis
is horizontal.}
\label{fig:maps}
\end{figure*}

\subsection{Cosmic-ray transport equations}
\label{sec:cosmic_ray_transport_equations}
We are using 1D transport equations for the CRE number density $N(E,z)$, where $E$ is the CRE energy and $z$ is the distance to the midplane. We assume a fixed inner boundary condition with $N(E,0)=N_0E^{-\gamma_{\rm inj}}$, where $\gamma_{\rm inj}$ is the injection CRE energy spectral index and $N_0$ is a normalization constant. The transport equation for advection is:
\begin{equation}
  \frac{\upartial N(E,z)} {\upartial z} = \frac{1}{V}\left \lbrace
    \frac{\upartial}{\upartial E}\left [ b(E)
    N(E,z)\right ]\right \rbrace\qquad ({\rm Advection}),
\label{eq:adv}
\end{equation}
where $V$ is the advection speed, assumed here to be constant. Similarly, for
diffusion we have:
\begin{equation}
  \frac{\upartial^2N(E,z)}{\upartial z^2} = \frac{1}{D}\left \lbrace\frac{\upartial}{\upartial
    E}\left [ b(E) N(E,z)\right ]\right\rbrace\qquad ({\rm Diffusion}),
\label{eq:diff}
\end{equation}
where we parametrize the diffusion coefficient as function of the CRE energy
as $D=D_0E_{\rm GeV}^{\umu}$, where $E_{\rm GeV}$ is the CRE energy in units
of GeV and $D_0$ is the diffusion coefficient at 1~GeV. We assume $\umu=0.5$ in agreement with
what is used for modelling the Milky Way \citep*{strong_07a}. There is some
debate as to whether this energy dependence applies to CREs with energies of a few
GeV \citep[e.g.][]{recchia_16b,mulcahy_16a}, but using $\umu=0$ instead changes
the results only slightly and we cannot distinguish with our data one way or
the other (see also \citetalias{heesen_16a}). The combined synchrotron and IC loss rate for CREs is given by \citep[e.g.][]{longair_11a}:
\begin{equation}
-\left (\frac{{\rm d}E}{{\rm d}t}\right )=b(E)=\frac{4}{3} \sigma_{\rm T} c \left (\frac{E}{m_{\rm
      e}c^2} \right )^2 (U_{\rm rad}+U_{\rm B}),
\label{eq:be}
\end{equation}
where $U_{\rm rad}$ is the radiation energy density, $U_{\rm B}$ is the magnetic
field energy density, $\sigma_{\rm T}=6.65\times 10^{-25}~\rm cm^2$ is the
Thomson cross-section and $m_{\rm e}=511~\rm keV\,c^{-2}$ is the electron
rest mass. 

The radiation energy density is the sum of the cosmic microwave background (CMB) radiation energy density and the interstellar radiation field (IRF) energy density; the latter includes contributions from starlight radiation energy density and the total infrared radiation energy density from emission by dust. The magnetic field strength $B_0$ in the midplane was calculated from energy equipartition \citep{beck_05a}. The energy density of the magnetic field is then $U_{\rm B}=B^2/(8\upi)$. Furthermore, we assumed that the ratio $U_{\rm IRF}/U_{\rm B}$ is constant everywhere. The resulting values of these calculations can be found in Table~\ref{tab:phys} and the details of the calculations are explained in Appendix~\ref{physical}.

We assume a two-component exponential magnetic field
distribution:
\begin{equation}
  B(z) = B_1\cdot \exp ( -|z|/ h_{\rm B1}) + (B_0 - B_1) \cdot \exp(-|z|/h_{\rm B2}).
\label{eq:b_distribution}
\end{equation}
Here, $h_{\rm B1}$ and $h_{\rm B2}$ are the magnetic field scale heights in
the thin and thick disc, respectively, $B_0$ is the magnetic field strength in
the midplane and $B_1$ is the magnetic field strength of the thin disc
component. Equations (\ref{eq:adv}) and (\ref{eq:diff}) are integrated
numerically from the inner boundary, so that no outer boundary condition is
required. In order to calculate the non-thermal radio continuum intensities the CRE number
density is convolved with the synchrotron emission profile of an individual
CRE (see \citetalias{heesen_16a} for details).

\begin{figure*}
  \includegraphics[width=1.0\hsize]{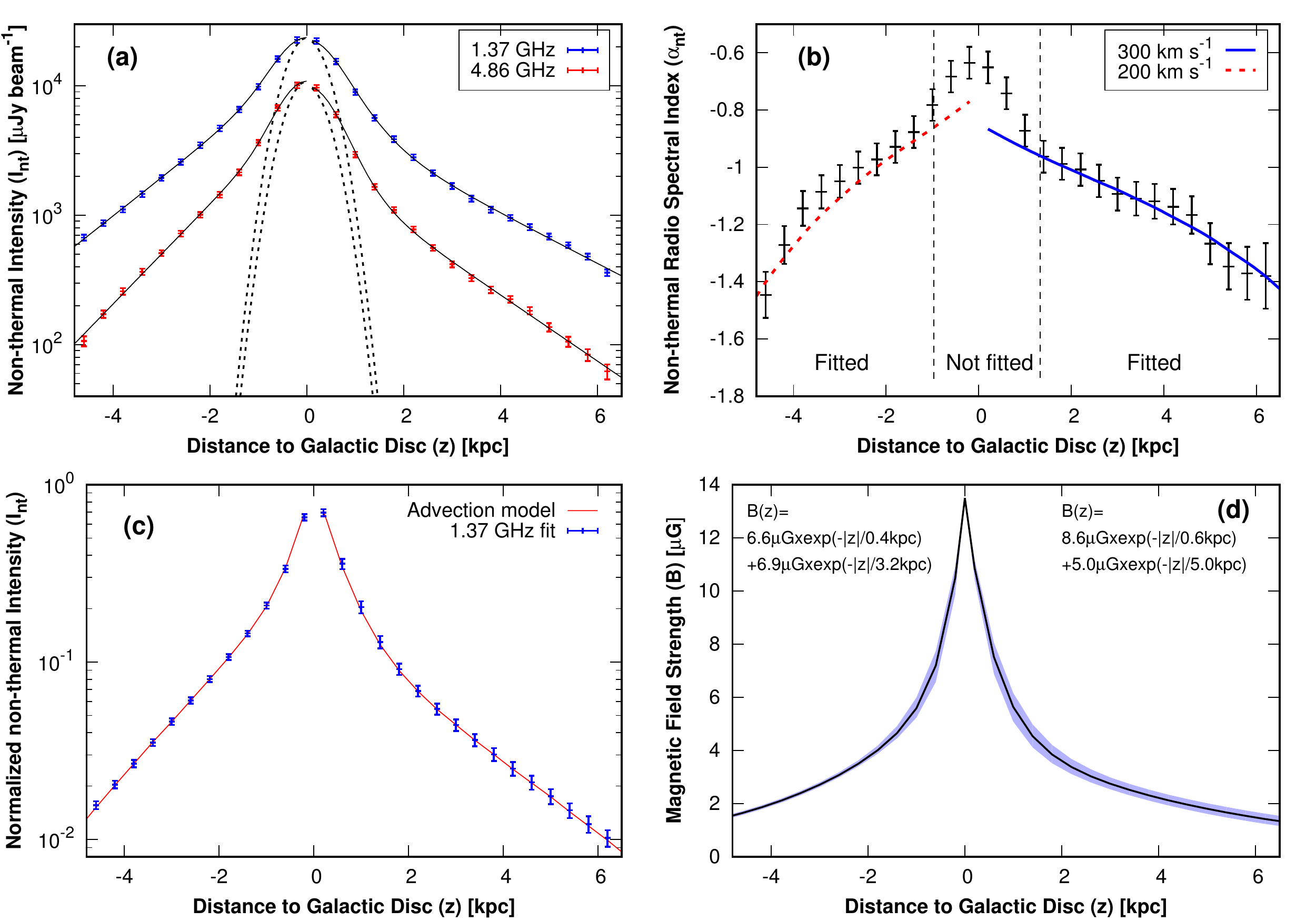}
  \caption{Vertical profiles in NGC~4631. (a) non-thermal radio
    continuum intensities at $1.37$ and $4.86$~GHz. We fitted a two-component exponential
    function (solid line) to the profiles. Black dashed
  lines show the effective beam, which is $I_{\rm beam}(z)= \exp(-z^2/(2b^2))$ (see Section~\ref{sec:vertical_intensity_profiles}) (b) non-thermal radio spectral
    index between $1.37$ and $4.86$~GHz. Lines show the best-fitting advection
    models, with an advection speed of $300~\rm km\,s^{-1}$ in the northern
    halo (blue solid line) and $200~\rm km\,s^{-1}$ in the southern halo (red dashed line). (c) normalized non-thermal intensity model profile at
    $1.37$~GHz, which is deconvolved from the synthesized beam.  The red line shows the intensities of the best-fitting
    advection model. (d) model of the magnetic field strength with
    uncertainties (blue-shaded area). Distances $z>0$ refer
    to the northern halo and distances $z<0$ to the southern one. In panels (a)--(c), the size of the symbols is equivalent to the size or the error bars. }
\label{fig:fit}
\end{figure*}

\subsection{Fitting procedure}
\subsubsection{Vertical intensity profiles}
\label{sec:vertical_intensity_profiles}
In this section, we outline the fitting procedure of the cosmic-ray transport models, where present a step-by-step analysis of NGC~4631. In Fig.~\ref{fig:maps}, we show the radio continuum maps of NGC~4631 -- a galaxy that has one of the most impressive radio haloes. We have created
vertical profiles of the non-thermal radio continuum intensity, which we
present in Fig.~\ref{fig:fit}(a). 

In our work, the scale height of the thin disc is comparable to the spatial resolution of the maps. This is why we use the analytical functions of \citet{dumke_95a} that convolve the Gaussian point-spread function (PSF) with either exponential or Gaussian intensity distributions. This Gaussian PSF is exact for interferometric maps and it is a good approximation for single-dish maps. Furthermore, it has become `standard practice' to include a contribution of the inclined disc (if $i<90\degr$), measuring a Gaussian radio intensity distribution along the major axis and projecting that on to the minor axis. This larger PSF is then referred to as the `effective beam' \citep[e.g.][]{heesen_09a,adebahr_13a}.

We fit
two-component exponential and Gaussian functions to vertical non-thermal intensity profiles, which we present as
solid and dashed lines, respectively (see Appendix~\ref{sec:intensity_models} for details). For this galaxy, a two-component
exponential function fits better than a two-component Gaussian function with a
reduced $\chi^2_{\rm exp}=0.4$ as opposed to $\chi^2_{\rm Gauss}=3.0$ (average of the northern and
southern halo at both frequencies). Using the fits to the data, we
  create vertical non-thermal intensity model profiles using either of the two frequencies, depending which
  data has the better quality. We calculate the error bars
  using the uncertainties of the maxima and scale heights of the thin and
  thick discs (see Appendix~\ref{sec:intensity_models} for details).

In Fig.~\ref{fig:fit}(b), we show the vertical profile of the non-thermal
spectral index in NGC~4631. The spectral index is fairly flat in
the midplane with $\alpha\approx -0.7$ and steepens in the halo to values of
$\alpha\approx -1.4$. The profile shows a conspicuous `shoulder', where the
spectral index slope flattens at heights $|z|\gtrapprox 1$~kpc. This distance
corresponds to the transition from the thin to the thick disc seen in the intensity
profiles. We note that the slope of the spectral index profile cannot be
resolved in the thin disc because the effective beam is too large. Furthermore, we are overestimating the non-thermal radio spectral index in the thin disc since we did not correct the Balmer H\,$\alpha$ line emission for internal absorption due to dust (Section~\ref{sec:thermal_radio_continuum_maps}).
In the halo, however, these influences can be
neglected, so that we can study the spectral index profile there. In order to illustrate the influence of the effective beam, we
  have shown its contribution to the vertical intensity profile in Fig.~\ref{fig:fit}(a); in case of
  NGC~4631 we restrict
  the fitting of the spectral index to $|z|\geq 1.4$~kpc. We show the fitting
  areas for the rest of the sample in Appendix~\ref{sec:image_atlas}. In
  NGC~55 and 253, the fitting area has also an outer limit, because the $L$
  band maps are affected by missing fluxes (Section~\ref{sec:integrated_flux_densities}). In NGC~253, we do not exclude this
  thin disc in the fit since we do not resolve it in the spectral index
  anyway and there would be otherwise not enough data points for a
  meaningful fit.

\begin{figure*}
  \includegraphics[width=0.9\hsize]{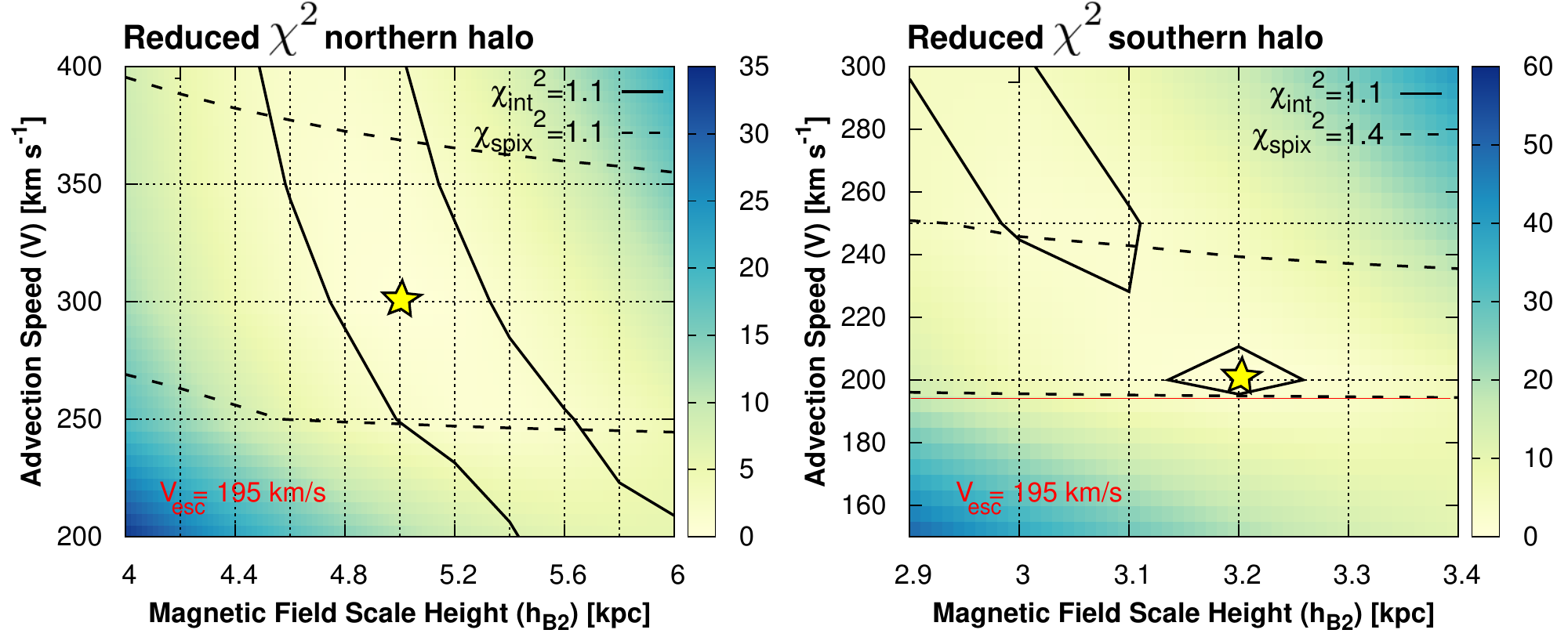}
  \caption{NGC~4631. Reduced $\chi^2$ of the advection model fitting in
   the thick disc as function of
    magnetic field scale height and the advection speed. The colour scale shows
    $\chi^2=\chi^2_{\rm int}+\chi^2_{\rm spix}$, where $\chi^2_{\rm int}$
    refers to the model intensity profile (Fig.~\ref{fig:fit}c) and $\chi^2_{\rm spix}$ refers to
    the spectral index profile (Fig.~\ref{fig:fit}b). Solid lines show $\chi^2_{\rm
      int,min}+1$ and dashed lines show $\chi^2_{\rm spix,min}+1$. The
    best-fitting models (with uncertainties) lie in the intersection of the two
  contours and are marked by yellow stars. The left panel is for the northern
  halo ($z>0$) and the right one for the
  southern halo ($z<0$). The red line shows the escape velocity near the
  midplane ($V_{\rm esc}=\sqrt{2}\times V_{\rm rot}$).}
\label{fig:chi}
\end{figure*}

\subsubsection{Finding the best-fitting model}
We simultaneously fit for the advection speed (or diffusion coefficient), the
magnetic field scale height and the injection spectral index. In order to find an `initial guess',  we first
vary the magnetic field profile to fit the intensity model at one frequency,
varying the scale height and choosing an advection speed (or
diffusion coefficient) that approximately fits the spectral index profile (together with the chosen injection spectral index). For NGC~4631, we
show the best-fitting advection intensity profile in Fig.~\ref{fig:fit}(c) and the best-fitting advection
spectral index profile in Fig.~\ref{fig:fit}(b). The corresponding vertical profile of the magnetic field strength is shown in
Fig.~\ref{fig:fit}(d). The error interval of the latter
stems from the uncertainties of the magnetic field scale heights in the thin and thick disc, it does not include any uncertainty from the total magnetic field strength
estimate from the equipartition assumption (we discuss the influence of the equipartition estimate in Section~\ref{sec:magnetic_field_uncertainties}). The magnetic field strength decreases
rapidly and reaches a value of approximately half the maximum field strength
at $|z|=1$~kpc from where on it decreases more slowly.


From the best-fitting model we can work out the uncertainties of the advection
speed and magnetic field strength. This is shown in Fig.~\ref{fig:chi}, where we
plot the reduced $\chi^2$ for both the intensity and spectral index fitting in
NGC~4631. The fitting of the intensities depends almost exclusively on the
magnetic field profile, although a higher advection speed can offset a lower
magnetic field scale height. In contrast, the fitting of the spectral index
profile depends mostly on the advection speed and hardly on the magnetic
field scale height. This means that best-fitting areas are almost
perpendicularly aligned, so that the advection speed and magnetic field scale
height can be constrained well by a simultaneous fit.

\section{Results}
\label{results}
In Table~\ref{tab:results}, we present non-thermal intensity scale heights in the thick disc at $L$ and $C$ band, magnetic field scale heights in the thin and thick discs, advection speeds and diffusion coefficients for the entire sample. In some cases, we could find a best-fitting value for the magnetic field scale height and/or advection speed, but not an upper and/or lower limit. For these values, the given error interval is a lower limit; they are plotted in Fig.~\ref{fig:par} for the parameter studies with arrows denoting the error bars, but are not taken into account for the statistical analysis. Maps and plots are presented in Appendix~\ref{sec:image_atlas}.

\subsection{The distinction between advection and diffusion}
At first we present our findings whether the cosmic-ray transport in haloes is advection- or diffusion-dominated. We note that this distinction is a common one since advection will take
over diffusion in the halo quickly if there is a galactic wind
\citep{ptuskin_97a,recchia_16a}. The effective boundary $s_{\star}$ at which ($|z|>s_{\star}$) advection dominates over diffusion is:
\begin{equation}
  \frac{s_{\star}^2}{D} \approx \frac{s_{\star}}{V} \Rightarrow
    s_{\star}\approx 0.3 \times \frac{D_{28}}{V_{100}}~\rm kpc,
\label{eq:scrit}
\end{equation}
where $D_{28}$ is the diffusion coefficient in units of $10^{28}~\rm cm^2\,
s^{-1}$ and $V_{\rm 100}$ is the advection speed in units of $100~\rm
km\,s^{-1}$. Diffusion coefficients in galaxies have values of a few $10^{28}~\rm cm^2\,s^{-1}$ \citep[e.g.][]{berkhuijsen_13a,tabatabaei_13b,mulcahy_14a,mulcahy_16a} in agreement with the Milky Way value of $3\times 10^{28}~\rm cm^2\,s^{-1}$ \citep{strong_07a}. For a diffusion coefficient of $3\times
10^{28}~\rm cm^2\,s^{-1}$ and an advection speed of $200~\rm km\,s^{-1}$, the
transition happens at a height of $0.45~\rm kpc$. Hence, it is justified to assume
an advection only model in the halo if there is a wind.

If there is no
wind, the radio halo may be diffusion dominated. As \citetalias{heesen_16a}
showed, the vertical profile of the radio spectral index can be used in order to distinguish between advection and diffusion. Advection has spectral index profiles, which gradually steepen as function of height. In contrast, diffusion leads to spectral index profiles that steepen only very little at small heights, but have steep cut-offs at large heights. This is caused by a steep cut-off of the CRE number density at large heights in case of diffusion, which reflects the fact that the CREs cannot escape the galaxy. Thus, the vertical diffusion spectral index profiles have `parabolic' shapes, which should distinguish them from the `linear' advection profiles. In practice it can, however, be difficult to allow for a
distinction since the uncertainties of the radio spectral index are too
large. Therefore, we fitted all our galaxies with both an advection and
diffusion model. 
\begin{table*}
\centering
\caption{Non-thermal intensity scale height $z_0$ at $L$ and $C$ band, advection speed $V$, diffusion coefficient $D_0$ ($D=D_0E_{\rm GeV}^{0.5}$) and magnetic field scale height of the thin ($h_{\rm B1}$) and thick disc ($h_{\rm B2}$) in the northern (N) and southern (S) halo.\label{tab:results}}
\begin{tabular}{l cccc cccc cc}
\hline
Galaxy & $z_0$ ($L$/N)$^{\rm a}$ & $z_0$ ($C$/N)$^{\rm b}$ & $z_0$ ($L$/S)$^{\rm c}$ & $z_0$ ($C$/S)$^{\rm d}$ & $h_{\rm B1}$ (N)$^{\rm e}$ & $h_{\rm B1}$ (S)$^{\rm f}$ & $h_{\rm B2}$ (N)$^{\rm g}$ & $h_{\rm B2}$ (S)$^{\rm h}$ & $V$ (N)$^{\rm i}$ & $V$ (S)$^{\rm j}$ \\
& (kpc) & (kpc) & (kpc) & (kpc) & (kpc) & (kpc) & (kpc) & (kpc) & ($\rm km\,s^{-1}$) & ($\rm km\,s^{-1}$) \\
\hline
NGC 55   & $2.03\pm 1.46$ & $1.72\pm 0.22$ & $1.06\pm 0.14$ & $2.01\pm 0.71$ & $0.3\pm 0.1$ & $0.2\pm 0.1$ & $4.0^{++2.0}_{-1.2}$  & $2.5^{+0.8}_{-0.7}$ & $150^{++200}_{--100}$  & $100^{+80}_{-10}$          \B\\ 
NGC 253  & $1.26\pm 0.07$ & $1.14\pm 0.04$ & $1.33\pm 0.04$ & $1.28\pm 0.02$ & N/A          & N/A          & $2.8^{+0.2}_{-0.7}$   & $3.0^{+0.1}_{-0.5}$ & $400^{+200}_{-20}$    & $500^{++400}_{-120}$       \T\B\\ 
NGC 891  & $1.33\pm 0.02$ & $1.30\pm 0.02$ & $1.25\pm 0.01$ & $1.24\pm 0.03$ & $0.1\pm 0.1$ & $0.2\pm 0.1$ & $3.0^{+0.2}_{-0.2}$   & $2.5^{+0.2}_{-0.2}$ & $600^{++600}_{-200}$   & $600^{++600}_{-200}$        \T\B\\ 
NGC 3044 & $3.64\pm 0.95$ & $1.47\pm 0.26$ & $1.30\pm 0.08$ & $1.15\pm 0.14$ & $0.5\pm 0.1$ & $0.5\pm 0.1$ & $4.5^{++4.5}_{-1.7}$  & $3.0^{+0.7}_{-0.6}$ & $200^{++200}_{--100}$  & $200^{+200}_{-50}$         \T\B\\ 
NGC 3079 & $5.14\pm 8.86$ & $1.00\pm 0.19$ & $3.41\pm 0.63$ & $1.54\pm 0.10$ & $1.2\pm 0.1$ & $1.0\pm 0.1$ & $4.5^{+7.5}_{-2.0}$   & $8.0^{+7.0}_{-3.0}$ & $300^{+80}_{-40}$     & $400^{+80}_{-80}$          \T\B\\ 
NGC 3628 & $1.23\pm 0.05$ & $1.16\pm 0.04$ & N/A            & $0.87\pm 0.02$ & $0.1\pm 0.1$ & $0.1\pm 0.1$ & $2.6^{+0.3}_{-0.3}$   & $2.0^{+0.1}_{-0.3}$ & $250^{+250}_{-50}$    & $250^{++400}_{-50}$      \T\B\\ 
NGC 4565 & $1.90\pm 0.10$ & $1.63\pm 0.07$ & $3.11\pm 0.66$ & $1.75\pm 0.10$ & N/A          & $0.8\pm 0.1$ & $3.5^{+1.1}_{-0.5}$   & $4.0^{+0.2}_{-0.7}$ & $500^{++300}_{-220}$  & $250^{+150}_{-20}$        \T\B\\ 
NGC 4631 & $2.24\pm 0.07$ & $1.70\pm 0.05$ & $1.47\pm 0.02$ & $1.13\pm 0.02$ & $0.6\pm 0.1$ & $0.4\pm 0.1$ & $5.0^{+0.6}_{-0.5}$   & $3.2^{+0.1}_{-0.2}$ & $300^{+80}_{-50}$    & $200^{+50}_{-20}$           \T\B\\ 
NGC 4666 & $1.51\pm 0.05$ & $1.54\pm 0.03$ & $2.01\pm 0.08$ & $1.70\pm 0.11$ & $0.5\pm 0.1$ & $0.5\pm 0.1$ & $3.0^{+0.3}_{-0.3}$   & $4.0^{+0.4}_{-0.5}$ & $700^{++200}_{-100}$  & $500^{+80}_{-70}$          \T\B\\ 
NGC 5775 & $2.72\pm 0.12$ & $1.27\pm 0.06$ & $1.91\pm 0.12$ & $2.03\pm 0.80$ & $1.0\pm 0.1$ & $0.4\pm 0.1$ & $7.0^{+1.5}_{-1.0}$   & $4.0^{++2.0}_{-0.8}$ & $400^{+100}_{-50}$   & $450^{++100}_{--150}$    \T\B \\
NGC 7090 & $2.07\pm 0.18$ & $1.96\pm 0.50$ & $1.53\pm 0.39$ & $1.14\pm 0.44$ & $0.6\pm 0.1$ & $0.4\pm 0.1$ & $5.5^{+1.0}_{-1.2}$   & $4.5^{+3.5}_{-2.7}$ & $200^{+310}_{-10}$   & $300^{++300}_{-120}$       \T\B\\
NGC 7462 & $1.62\pm 0.15$ & $1.38\pm 0.03$ & $1.50\pm 0.03$ & $1.46\pm 0.01$ & N/A          & N/A          & $3.0^{+1.0}_{-0.2}$   & $3.5^{+0.7}_{-0.5}$ & $<80$             & $<100$                 \T\B\\
\hline
         &              &              &              &              &              &              & $h_{\rm B2}$ (N)$^{\rm g}$ & $h_{\rm B2}$ (S)$^{\rm h}$ & $D_0$ (N)$^{\rm i}$ & $D_0$ (S)$^{\rm j}$  \\
         &              &              &              &              &              &              & (kpc)     & (kpc)     & \multicolumn{2}{c}{($\rm 10^{28}cm^2\,s^{-1}$)} \\
\hline
NGC 7462 &              &              &              &              &              &              & $3.0^{+1.0}_{-0.2}$ & $3.5^{+0.7}_{-0.5}$ & $2.2^{+0.4}_{-0.4}$ & $2.6^{+0.1}_{-0.6}$  \T\B\\
\hline
\end{tabular}
\flushleft{{\bf Notes.} Values with `$^{++}$' or `$_{--}$' have no upper or lower limits. Where no upper limit for $V$ was measured, we use $I_{\rm nt}\propto B^{1-\alpha_{\rm nt}}$ for freely escaping CREs to estimate the lower limit of $h_{\rm B2}$.\\
  $^{\rm a}$ non-thermal exponential (Gaussian for NGC~7462) intensity scale height at $L$ band in the thick disc in the northern halo.\\
  $^{\rm b}$ non-thermal exponential (Gaussian for NGC~7462) intensity scale height at $C$ band in the thick disc in the northern halo.\\
  $^{\rm c}$ non-thermal exponential (Gaussian for NGC~7462) intensity scale height at $L$ band in the thick disc in the southern halo.\\
  $^{\rm d}$ non-thermal exponential (Gaussian for NGC~7462) intensity scale height at $C$ band in the thick disc in the southern halo.\\
  $^{\rm e}$ magnetic field scale height in the thin disc in the northern halo. NGC~253, 4565 and 7462 have only a thick disc in the northern halo.\\
  $^{\rm f}$ magnetic field scale height in the thin disc in the southern halo. NGC~253 and 7462 have only a thick disc in the southern halo.\\
  $^{\rm g}$ magnetic field scale height in the thick disc in the northern halo.\\
  $^{\rm h}$ magnetic field scale height in the thick disc in the southern halo.\\
  $^{\rm i}$ advection speed (diffusion coefficient for NGC~7462) in the northern halo.\\
  $^{\rm j}$ advection speed (diffusion coefficient for NGC~7462) in the southern halo.}
\end{table*}

We find that only NGC~7462 can be fitted with a diffusion
coefficient roughly in agreement of what we would expect, namely $D_0\approx 3\times
10^{28}~\rm cm^2\,s^{-1}$. All other galaxies required values in excess of
this, with values between $1.2\times
10^{29}~\rm cm^2\,s^{-1}$ (NGC~55) and $4\times
10^{30}~\rm cm^2\,s^{-1}$ (NGC~4666). There is no physical reason why the
diffusion coefficient should deviate so much from the Milky Way value, because the magnetic field structure is
not too dissimilar.\footnote{The ratio of ordered (vertical to the line of sight) magnetic field strength to total magnetic field strength in late-type spiral galaxies is $B_{\rm ord,\perp}/B_{\rm tot}=0.3$ with some fluctuations within the galaxies \citep[maxima in the inter-arm regions, mini ma in the spiral arms;][]{fletcher_10a}. This is similar to what has been found in radio haloes (see Section~\ref{sec:magnetic_field_structure} for references). Hence, we expect similar diffusion coefficients to what is found in galactic discs.} Such high diffusion coefficients would also lead to the
cosmic rays leaving the galaxy without interaction, so that they do not transfer energy and momentum to the ionized gas. This is at odds with cosmic-ray driven wind
models, which now have become very popular
\citep[e.g.][]{breitschwerdt_91a,everett_08a,samui_10a,salem_14a}. Their advantage over previous models is that they can explain the existence of winds in galaxies with $\Sigma_{\rm SFR}\ll 10^{-1}~\rm M_{\sun}\,yr^{-1}\,kpc^{-2}$, where thermally driven winds suffer from too strong radiative cooling. Thus, we rule out models with $D_0>10^{29}~\rm cm^2\,s^{-1}$
and assume that haloes are in this case advection dominated.



\subsection{Non-thermal intensity scale heights}
\label{scale_heights}
We find that the vertical non-thermal intensity profiles in 7 out of the
12 sample galaxies are better fitted by exponential rather than
Gaussian functions, where $\chi^2_{\rm exp}-\chi^2_{\rm Gauss}\leq -1$ (NGC 55,
891, 3044, 3079, 3628, 4631 and 5775; the reduced $\chi^2$ values are the average of the northern and
southern halo at both frequencies). In
3 galaxies the fits are of equivalent quality, where $-1<\chi^2_{\rm exp}-\chi^2_{\rm
  Gauss}<1$ (NGC~253, 4565 and 7090). Only in 1 galaxy the Gaussian
fits significantly better than the exponential function, where $\chi^2_{\rm
  exp}-\chi^2_{\rm Gauss}\geq 1$ (NGC~7462). In 10 out of the 11 exponential haloes
(including the 7 better and the 3 equivalent fits) we
also detect a thin disc component. Only in NGC~4565 this is not the
case, which we believe is due to a combination of low spatial resolution
and disc intensity (NGC~4565 has the lowest average surface intensity). We
detect a thick disc in the Magellanic-type galaxy NGC~55
(Appendix~\ref{sec:image_atlas}), the first report of a radio halo in this galaxy.

Averaged across the sample, the
exponential scale height of the thin disc is $0.41\pm 0.21$~kpc at
$1.4$~GHz and $0.25\pm 0.13$~kpc at 5~GHz. The scale height of the thick disc
is $1.33\pm 1.25$~kpc at $1.4$~GHz and $1.22\pm 0.43$~kpc at 5~GHz; these values are in excellent agreement with those of \mbox{\citet{krause_17a}}. We found a
large variety of scale heights at $1.4$~GHz ranging between $1.2$~kpc and
$2.7$~kpc (excluding values with fractional errors larger than 10 per cent). At
5~GHz the scale heights of the thick disc vary between $0.9$ and
$2.0$~kpc. In NGC~7462, we find only a thick Gaussian disc with a scale height
of $\approx$$1.5$~kpc at both $1.4$ and 5~GHz.

\begin{figure*}
  \includegraphics[width=1.0\hsize]{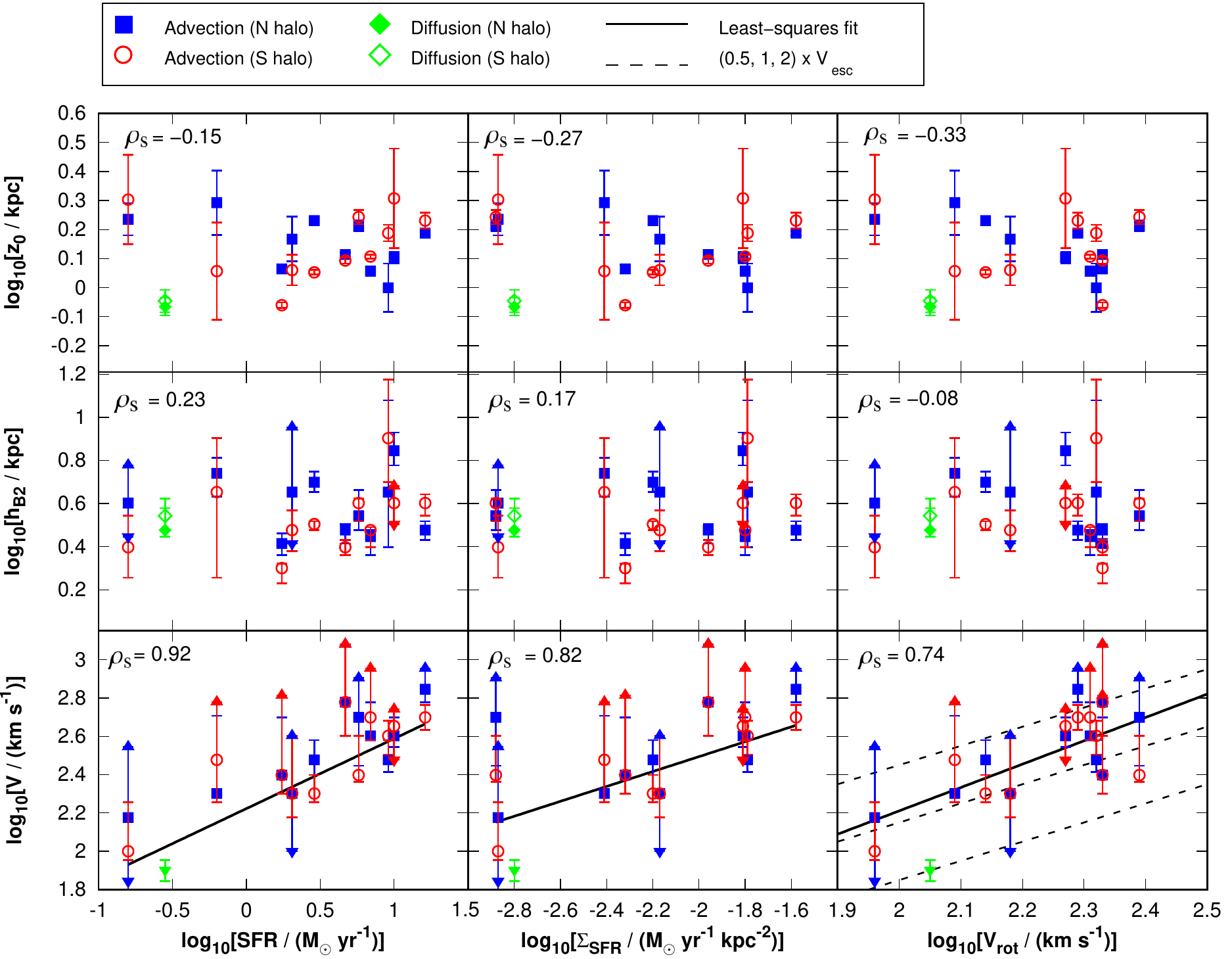}
  \caption{Parameter studies in log--log diagrams as function of SFR, SFR
    surface density ($\Sigma_{\rm SFR}$) and rotation speed $V_{\rm rot}$. \emph{Top panels:} non-thermal
  intensity scale height ($z_0$) at 5~GHz ($8.5$~GHz for NGC~5775) of the thick radio
  disc. \emph{Middle panels:} magnetic field scale height ($h_{\rm B2}$) of the thick radio
  disc. \emph{Bottom panels:} Advection speed ($V$), where solid lines show least-squares fits. In the bottom right panel the dashed lines show ($0.5$, 1,
    2)$\times V_{\rm  esc}$. In each panel, we also present Spearman's rank correlation coefficient, $\rho_{\rm s}$, which we derived from values that have both an upper and lower limits.}
\label{fig:par}
\end{figure*}

In Fig.~\ref{fig:par}, we show the dependence of the 5-GHz
intensity scale heights as function of the SFR, $\Sigma_{\rm SFR}$ and
$V_{\rm rot}$. These are log--log diagrams so that we can fit linear functions
to them, which are representative of power-laws. We find no correlation between the intensity scale height
and either parameter (Spearman's rank correlation coefficient $|\rho_{\rm
  s}|\leq 0.33$). Full results of the scale height fitting are presented in
Appendix~\ref{sec:intensity_models}.

\subsection{Non-thermal radio spectral indices}
We find that in those 7 galaxies that have both a thin and a thick disc, the
vertical non-thermal radio spectral index
profile shows the thin and the thick disc clearly separated as well (NGC~55, 891, 3044,
3628, 4631, 5775 and 7090). The exceptions are NGC 253, 4565 and 4666, where the thin and the
thick disc are difficult to separate in the intensity
profiles. This is caused by either low inclination angles ($<$$79\degr$
in NGC 253 and 4666) or low spatial resolution ($2.1$~kpc in
NGC~4565). In NGC~3079, we can separate the
thin and the thick disc in the intensity profile, but the slope of the
spectral index profile in the thick disc is
so high that there is no visible transition from the thin to the thick
disc. Even though we have masked the nuclear outflow, it is possible that additional diffuse emission associated with it can contaminate our radio continuum emission. Since this is more pronounced at $L$ band, where the scale heights in the thick radio discs are the largest in our sample, old CREs, possibly stemming from an earlier episode of AGN activity, could be responsible for this emission. Such emission can also be explained as the result of jet interactions with clumpy media \citep{middelberg_07a}, which can contaminate the radio emission around the thin/thick disc transition.  In all galaxies, the non-thermal radio spectral index
in the thick disc varies between $-1.4$ and $-1.0$, showing that
spectral ageing of the CREs plays an important role in the halo.

The integrated non-thermal radio spectral indices range from
$-1.3$ (NGC~7462) to $-0.6$ (NGC~55). The significance of the integrated non-thermal spectral index is that we can infer which energy losses of the CREs are dominating. If synchrotron and IC radiation losses are important, the CRE spectrum is $N(E,z)\propto E^{-\gamma_{\rm inj}-1}$. If, on the other hand, adiabatic losses dominate or the CREs escape freely, the CRE spectrum is unchanged from the injection spectrum with $N(E,z)\propto E^{-\gamma_{\rm inj}}$ \citep{lisenfeld_00a,longair_11a}. In case radiation losses are important, a galaxy is referred to as an \emph{electron calorimeter}, where the CREs are losing their energy before they can escape through radiation and ionization losses \citep{yoast_hull_13a}.

This shows that NGC~55 is not an electron calorimeter, whereas NGC~7462 is one. NGC~55 is a Magellanic-type dwarf
irregular galaxy and possibly has a strong wind, similar to the starburst dwarf irregular galaxy IC~10 \citep{chyzy_16a}. On the other hand, NGC~7462 is diffusion-dominated
(no wind) and all the CREs are confined in the halo, leading to high CRE
radiation losses. Our finding that galaxies have galactic winds means that galaxies are in general not electron calorimeters. This has some consequences for the radio continuum--SFR relation and its close corollary the radio continuum--far-infrared relation, which we discuss in Section~\ref{sec:radio_continuum_sfr_relation}.

\subsection{Magnetic field scale heights}
\label{sec:magnetic_field_scale_heights}
In the thin disc, magnetic field scale heights range from $0.1$~kpc in
NGC~3628 to $1.2$~kpc in NGC~3079. Averaged across the sample, the
magnetic field scale height in the thin disc is $0.50\pm 0.32$~kpc (assuming
an error of $0.1$~kpc for each measurement). In the thick disc, we find scale
heights between 2~kpc in NGC~3628
and 8~kpc in NGC~3079. In Fig.~\ref{fig:par} we show that the magnetic field scale height of the thick disc
does not depend on either SFR, $\Sigma_{\rm SFR}$ or $V_{\rm rot}$. The magnetic field in
the 
halo must be regulated by something else than the wind driven by the
star formation in the disc. Maybe the
geometry plays a 
role, as for instance predicted by the flux tube model used in the
aforementioned 1D cosmic-ray driven wind 
models. The large scale heights in NGC~3079 can possibly be explained by the strong interaction of this galaxy with
nearby group members \citep{shafi_15a}. Tidal interaction can `puff' up the thin and the thick discs by inflating the warm neutral medium, which in our picture results in a larger magnetic field scale height. Alternatively, the prominent starburst/AGN-driven nuclear outflow advects cosmic rays and magnetic fields, so that the magnetic field extends further into the halo \citep{cecil_01a,middelberg_07a}.

Averaged across our sample, the
magnetic field scale height in the thick disc is $3.0\pm
1.7$~kpc. Compared with the magnetic field scale height predicted by energy
equipartition, $h_{\rm B}\approx 4\cdot z_0$ (where $z_0$ is the 5-GHz
intensity scale height) for $\alpha_{\rm nt}\approx -1$, the magnetic field scale height
in the thick disc is 35~per cent lower. This is because the
vertical decrease of the CRE number density is smaller than that of the
magnetic field energy density (fig.~10 in
\citetalias{heesen_16a}).

\subsection{Advection speeds}

The advection
speeds range from $100~\rm km\,s^{-1}$ in NGC~55 to $700~\rm km\,s^{-1}$
in NGC~4666. The advection speeds rise as $V\propto \rm SFR^{0.36\pm 0.06}$ and $V\propto \rm
\Sigma_{\rm SFR}^{0.39\pm 0.09}$ as shown in Fig.~\ref{fig:par}. We found Spearman's rank correlation coefficients of $\rho_{\rm s} = 0.92$ and $\rho_{\rm s} = 0.82$, respectively; for this we used a least-squares fitting routine in the log--log diagram, taking only data
points with both upper and lower limits into account (that have no arrows in Fig.~\ref{fig:par}). These correlations
are robust with respect to performing the `jackknife' test, where we leave out NGC~55, the galaxy with the lowest SFR. We find another possible
correlation with $V_{\rm rot}$, which is, however, not robust to the jackknife test. We need more observations of dwarf irregular galaxies to
measure winds in them. A correlation between advection speed and SFR has been suggested earlier based on the observation that the 5-GHz radio continuum scale height is almost constant and does not depend on the magnetic field strength in the midplane \citep{krause_15a}. If one assumes that the scale height is a function of the CRE lifetime, neglecting the influence of the magnetic field, then scale height scales as $\propto B_0^{-1.5}$, where it is also assumed that synchrotron losses are dominating over IC losses (Appendix~\ref{physical}). If one now uses that the magnetic field strength scales as $B_0\propto \rm SFR^{0.3}$ \citep[e.g.][]{heesen_14a}, one can conclude that the scale height scales as $\propto \rm SFR^{-0.5}$. Hence, in order to have a constant scale height the advection speed has to scale as $V\propto \rm SFR^{0.5}$ \citep{krause_15a}. In reality, the magnetic field scale height plays also a role, so that the scaling is somewhat less with the SFR, but the general idea is a valid one.

The advection
speeds lie within the range $0.5\leq V/V_{\rm esc} \leq 2$
(including the uncertainties), where
$V_{\rm esc}=\sqrt{2}\times V_{\rm rot}$ is the escape velocity near the midplane (further away from the disc, in the dark matter halo, the escape velocity
is higher). In NGC~7462,
where diffusion dominates, we can calculate an upper
limit for the advection speed, by comparing how much advection can contribute
at most. This can be done either analytically using
equation~(\ref{eq:scrit}), inserting $s_{\star}=2.5$~kpc, the CRE scale height in
NGC~7462 (fig.~10 in \citetalias{heesen_16a}), or numerically by comparing
profiles of the CRE number density. Both ways give a similar result, namely that the
upper limit for the advection speed in NGC~7462 is between 80 and $100~\rm
km\,s^{-1}$.

Are we tracing really disc winds, or are the radio haloes only extensions to the nuclear outflows seen in galaxies with nuclear star bursts or active galactic nuclei (AGN)? A case in point is NGC~3079, a galaxy with a low-luminosity AGN. On the northern side of the galaxy a nuclear outflow is detected with an outflow speed of the warm ionized gas of up to $\approx$$1000~\rm km\,s^{-1}$ \citep{cecil_01a}. There is also a nuclear outflow on the southern side, which can be seen only in the radio continuum due to absorption by dust of the optical emission. A similar outflow, albeit less prominent in the radio continuum, can be seen in the nuclear region of NGC~253 \citep{heesen_11a,westmoquette_11a}. Here, the nuclear outflow speed is of the order of $\approx$300$~\rm km\, s^{-1}$, traced by the Doppler shift of the warm ionized gas. The other galaxies in our sample have not been studied in detail with regards to their nuclear outflows, although the study by \citet{ho_97a} of the [\ion{N}{ii}] emission line offers some clues. Our samples have in common NGC~891, 3079, 3628, 4565, 4631 and 5775. Their line widths are all smaller than our advection speeds, with the exception of NGC~3079. In NGC~891, 3628 and 5775 the line-width equivalent velocity is less than 50 per cent of our advection speed. Studying nuclear outflows and determining their kinematics is very challenging and is usually based on the presence of unusually broad or shifted lines (or lines with excess wing emission).  Even when such features are identified, it is extremely difficult to distinguish between rotation, inflow, or outflow (often outflows are inferred from the presence of velocities exceeding the escape velocity of the host galaxy). Hence, it is not possible to compare the properties of nuclear outflows and disc winds in detail for our sample.

\section{Discussion}
\label{discussion}

\subsection{Cosmic-ray driven winds}
\label{winds}
Our result of increasing advection speeds as function of the SFR is also
observed in studies of other
low-redshift galaxies, where the wind speed of either the warm ionized
gas \citep[e.g.][]{arribas_14a,heckman_15a,heckman_16a} or the neutral gas
\citep[e.g.][]{martin_05a,rupke_05a} is measured. This can be explained by a
star-formation driven wind due to a combination 
of a hot wind fluid driven by the thermalized
ejecta of massive
 stars and radiation pressure. Another explanation is
cosmic-ray driven winds, where the cosmic-ray pressure (via
Alfv\'en waves) in conjunction with the pressure of the thermal gas pushes the
material outwards. Hybrid forms of winds are also possible. Cosmic-ray driven winds are predominant for galaxies
with low $\Sigma_{\rm SFR}$, where the pressure of the thermal gas alone is
insufficient to launch a wind \citep{everett_08a}. The influence of the cosmic
ray pressure 
can be tested with our observations: the magnetic field strength in galaxies
scales as $B\propto \rm SFR^{0.30\pm 0.02}$ \citep{heesen_14a}, similar to the
relation of the advection speeds, which is $V\propto \rm SFR^{0.36\pm 0.06}$. Hence, the magnetic field strength and the advection velocity are nearly proportional to each other. This also means that $V^2 \propto B^2$, so that the kinetic
energy density
 of the wind $U_{\rm kin} = (1/2)\rho V^2$ is proportional to the magnetic energy density $U_{\rm B}=B^2/(8\upi)$; here, $\rho$ is the density of the hot ionized gas. This may tell
us that the wind is driven by cosmic rays, which are roughly in equipartition with the magnetic field.

Another important result is the remarkable agreement between advection speeds and escape
velocities. Again, this can be explained by cosmic-ray driven winds. Initially, the
wind speed is below the sound speed of the combined thermal and cosmic ray gas
(the so-called `compound sound speed'), but the flow accelerates in the halo where it goes through the
critical point (Mach number $M = 1$) at a distance of a few kpc away from the disc. Eventually the wind
accelerates further to reach an asymptotic velocity of a few times the escape
velocity \citep{breitschwerdt_91a,everett_08a}. Wind speeds of
the order of the escape velocity are also predicted for pressure-driven galactic winds without
the contribution from cosmic rays \citep{murray_05a,heckman_15a,heckman_16a}. The accelerating force in momentum-driven winds is a combination of radiation pressure, ram pressure from galactic winds and non-thermal pressure due to magnetic fields and cosmic rays.\footnote{Pressure-driven winds are also sometimes referred to as momentum-driven winds in the literature. In our case, the cosmic-ray driven wind implicitly means a momentum-driven wind since the cooling of the thermal gas is so strong that the influence of the cosmic-ray pressure becomes important.} It is an intriguing possibility
that galactic winds are driven by the combined pressure supplied by star formation (via SNe
ejecta and stellar winds) and cosmic rays.

Within the critical point, the compound sound speed can be approximated by \citep{breitschwerdt_91a}:
\begin{equation}
	c_{\rm comp}^2 \approx \frac{1}{2} g_{\rm eff} \frac{Z_0^2}{z_{\rm c}},
\end{equation}
where $g_{\rm eff}$ is the gravitational acceleration in the halo, $Z_0$ is the height where the flow becomes spherical and $z_{\rm c}$ is the height of the critical point. The gravitational acceleration can be approximated by $g_{\rm eff}\approx V_{\rm rot}^2/r_{\star}$ because the flow emanates as a disc wind from radii with active star formation ($r\leq r_{\star}$). We do not know where the critical point lies, but it must be of the order of the magnetic field scale height since our outflow speeds are already exceeding the escape velocity. Hence, setting $z_{\rm c}\approx h_{\rm B}$, we find for the compound sound speed:
\begin{equation}
	c_{\rm comp} = V_{\rm rot} \cdot \frac{Z_0}{\sqrt{2r_{\star}\cdot h_{\rm B}}}.
    \label{eq:v_cr}
\end{equation}
With $r_{\star}\approx 10~\rm kpc$, $Z_0\approx 15~\rm kpc$ and $h_{\rm B}\approx 3~\rm kpc$, we find $c_{\rm comp}\approx 2\cdot V_{\rm rot}$, in good agreement with our advection speeds. While the exact numerical value of equation~(\ref{eq:v_cr}) is uncertain, the prediction is that the advection speeds of cosmic-ray driven winds scale with the rotation speeds and are similar to the escape velocities.

The particular value of our radio continuum observations lies in the fact that we can measure
wind speeds for galaxies with $\Sigma_{\rm SFR}< 0.3\times 10^{-1}~\rm
M_{\sun}\,yr^{-1}\,kpc^{-2}$, whereas conventional absorption and emission line studies have
so far focused on galaxies exceeding this limit. All of our sample galaxies
fulfil $\Sigma_{\rm SFR}>10^{-3}~\rm
M_{\sun}\, yr^{-1}\, kpc^{-2}$, for which \citet{rossa_03a} detect extra-planar
  diffuse ionized gas (eDIG). Only in NGC~7462, which has $\Sigma_{\rm SFR}=1.6\times 10^{-3}~\rm
M_{\sun}\, yr^{-1}\, kpc^{-2}$, we find a diffusion-dominated halo.
This galaxy is only marginally above the $\Sigma_{\rm SFR}$-threshold. This
leaves the possibility that galaxies with no eDIG are
diffusion-dominated and have no outflows in them, whereas galaxies with eDIG are
advection-dominated and have outflows. We need to extend this
kind of study to galaxies with lower $\Sigma_{\rm SFR}$ to confirm this
theory.


\subsection{Radio--SFR relation}
\label{sec:radio_continuum_sfr_relation}

Radio continuum emission in galaxies emerges from two distinct processes: 
thermal free--free (bremsstrahlung) and non-thermal (synchrotron) radiation.
Both are related to the formation of massive stars. UV-radiation from massive stars ionizes the ISM, which gives rise to the free--free emission. The explanation of the non-thermal radio continuum emission (which dominates at frequencies below $\approx$30~GHz) is more involved: massive stars end their lives as SNe. When the blast wave of the explosion reaches the ISM, strong
shocks are formed, which accelerate protons, nuclei, and electrons. The CREs spiral
around the interstellar magnetic field lines, thereby emitting highly linearly polarized synchrotron emission. The relation between the radio continuum luminosity of a galaxy and its SFR (in the following, the radio--SFR relation) is due to the interplay of star formation, CREs and magnetic fields.

The radio--SFR relation is very
tight, as \citet{heesen_14a} have shown: using the relation of
\citet{condon_92a}, and converting $1.4$-GHz radio luminosities into radio derived
SFRs, they found agreement within 50 per cent with state-of-the-art star
formation tracers, such as far-UV, H\,$\alpha$ and mid- or far-infrared emission.
An even better agreement can be achieved if the radio spectrum is integrated over a wide frequency range \citep[`bolometric radio luminosity';][]{tabatabaei_17a}.
Moreover, these authors found that the radio luminosity is a non-linear function of the SFR,
as predicted by the model of \citet{niklas_97a} as detailed below. In radio haloes, the spatially resolved radio--SFR relation is super-linear as well if we take the $\lambda$850-$\umu\rm m$ far-infrared emission as a proxy for the SFR \citep{irwin_13a}.

It has been realised early on \citep{condon_92a}
for the non-thermal radio continuum emission to be
related to the SFR, (i) the CREs have either to emit all their energy within a galaxy, so that the galaxy constitutes an electron calorimeter; (ii) or the cosmic rays have to be in energy equipartition with the magnetic field \emph{and} there is a magnetic field--SFR or magnetic field--gas relation \citep{niklas_97a}. Model (i) predicts a linear non-thermal radio--SFR relation, model (ii) a non-linear one. Present-day observations of spiral galaxies favour model (ii), which is expected since galaxies in general are not electron calorimeters. Electron calorimetry might hold at best in starburst galaxies, but almost certainly not in low-mass dwarf irregular galaxies, which lose a large fraction of their CREs in galactic winds and
outflows \citep[e.g.][]{chyzy_16a}. 

We can now confirm that even normal massive late-type spiral galaxies do have such outflows, which can reasonably explain why energy equipartition is found in them. In this picture, the SFR determines the magnetic field strength. The model by \citet{niklas_97a} uses the elegant way of assuming a magnetic field--gas relation (the turbulent energy density of the magnetic field is equivalent to the magnetic field energy density), from which the magnetic field--SFR relation is the result of the Kennicutt--Schmidt relation. Then, a galaxy can only store as many cosmic rays as the magnetic field can contain. If the cosmic-ray pressure becomes too high, the buoyant cosmic-ray gas will escape together with the magnetic field from the galaxy. Cosmic-ray driven winds serve thus as a `pressure valve' that allow overabundant cosmic rays to escape and thus preserve the energy equipartition with the magnetic field.

\subsection{Magnetic field uncertainties}
\label{sec:magnetic_field_uncertainties}

The largest source of uncertainty in our study stems from the estimate of the magnetic field
strength assuming energy equipartition. This assumption is supported by the observed relation between radio continuum
luminosity and SFR (Section~\ref{sec:radio_continuum_sfr_relation}). Non-calorimetric models require energy equipartition, while
calorimetric models are not in good agreement with the observations
\citep{niklas_97a,heesen_14a,li_16a}. Nevertheless, it is interesting to note
what happens when we increase or decrease the
magnetic field strength by a factor of 2. Because the typical halo magnetic
field strength is $B_{\rm halo}\approx 3\mu$G, the IC losses in the radiation field of
the cosmic microwave background (CMB) become comparable to the synchrotron
losses of the CREs (the CMB equivalent
magnetic field strength is $B_{\rm CMB}=3.2~\umu$G). The CRE lifetime, as determined by synchrotron and IC radiation losses, can be expressed by (\citetalias{heesen_16a}):
\begin{equation}
t_{\rm rad} = 34.2 \left (\frac{\nu}{\rm 1~GHz}\right )^{-0.5}
\left (\frac{B}{\rm 10~\umu G}\right )^{-1.5} \left
  (1+\frac{U_{\rm rad}}{U_{\rm B}}\right )^{-1}~{\rm Myr}.
\label{eq:t_rad}
\end{equation}
If we now decrease and increase the magnetic field strength by a factor of 2, we have $B_{\rm halo,-}=1.5~\umu\rm G$, $B_{\rm halo}=3.0~\umu\rm G$ and $B_{\rm halo,+}=6.9~\umu\rm G$. Inserting this into equation~(\ref{eq:t_rad}) and neglecting the radiation energy density from the star-forming disc, so that $U_{\rm rad}=U_{\rm CMB}=(3.2~\umu\rm G)^2/(8\upi)$, we find $t_{\rm rad,-}=90$, $t_{\rm rad}=75$ and $t_{\rm rad,+}=48$~Myr. Hence, for the lower magnetic field strength the advection speed increases by 20 per cent, whereas for the higher magnetic field strength the advection speed increases by 60~per cent. In summary, the uncertainty of the magnetic field strength provides \emph{lower} limits for the advection speeds. Even if the magnetic field strength is in reality a factor of 2 lower, the advection speed decreases only little since the IC losses are compensating for the smaller synchrotron losses.


\section{Conclusions}
\label{conclusions}

In this paper, we present radio continuum observations of 12 nearby
($D=2$--27~Mpc) edge-on galaxies at two different frequencies, namely at $1.4$ and $5$~GHz (one galaxy at $8.5$~GHz instead of 5~GHz). Our sample includes 11 late-type spiral (Sb or Sc) galaxies and one Magellanic-type barred galaxy (SBm), which are all highly inclined ($i\geq 76\degr$). The angular resolution of our maps varies between $8.25$ and $41.1$~arcsec, which corresponds to spatial resolutions at the distances of the galaxies between $0.4$ and $2.1$~kpc. We subtracted the thermal radio continuum emission using Balmer H\,$\alpha$ maps (or \emph{Spitzer} 24-$\umu$m maps, scaled to H\,$\alpha$) to study
the non-thermal radio continuum emission in the halo (where internal absorption by dust can
be neglected). We fitted the vertical
intensity profiles with exponential and Gaussian functions, correcting for the
effective beam (the combined effect of angular resolution and projection). The intensity model profiles and the spectral index
profiles in the halo ($|z|\gtrapprox1$~kpc) were then fitted with 1D cosmic-ray transport
models from the software {\small SPINNAKER}. In this way, we simultaneously measured CRE advection speeds (or diffusion
coefficients) and magnetic field scale heights. These are our main conclusions:

\begin{enumerate}
\item we discover a previously unknown radio halo in the Magellanic-type
  galaxy NGC~55 (see Appendix~\ref{sec:image_atlas}). This galaxy is hence another example of a dwarf irregular galaxy with a non-thermal outflow.
\item in 7 out of 12 galaxies, we find exponential vertical intensity
  profiles and only in one galaxy we find a Gaussian profile. The profiles in the
  remaining galaxies can be equally well described by either an exponential or
  Gaussian distribution. We conclude that vertical intensity profiles in the thick radio discs are predominantly exponential.
\item In 11 out of 12 galaxies, we find a thin disc in addition to a thick
  disc. Averaged across the sample, the thin disc has an
  exponential scale height of $0.41\pm 0.21$~kpc at $1.4$ GHz and $0.25\pm
  0.13$~kpc at $5$~GHz. The thick disc has an exponential scale height of
  $1.33\pm 1.25$~kpc at $1.4$~GHz and $1.22\pm 0.43$~kpc at 5~GHz.
\item The magnetic field in the thin disc has a scale height of $0.50\pm 0.32$~kpc. This is in approximate agreement with the expected value if the magnetic field is in pressure equilibrium with the warm neutral medium (\ion{H}{i} disc).
\item The vertical spectral index profiles show a clear separation between the
  thin and the thick disc in those cases where we see the separation also in
  the intensities ($i>85\degr$, resolution $<$1~kpc). The spectral index
  profiles at $|z|\gtrapprox 1$~kpc are linear as expected for cosmic-ray advection. Only
  NGC~7462 shows the parabolic shape expected for diffusion. The spectral index separation shows two different CRE populations in the disc and halo. If superbubble break out is responsible for the separation, we would expect the transition to be happening at $0.5$--$1.5$~kpc (three times the \ion{H}{i} scale height).

\item The magnetic field scale height in the thick disc across our sample is $3.0\pm
  1.7$~kpc. In individual galaxies, it is not correlated with either SFR, $\Sigma_{\rm SFR}$ or
  $V_{\rm rot}$.
\item In 11 out of 12 galaxies, we find advection dominated haloes. The advection
  speeds correlate as $V\propto \rm SFR^{0.36\pm 0.06}$ and $V\propto \Sigma_{\rm SFR}^{0.39\pm 0.09}$ and agree remarkably well with the
  escape velocity ($0.5\leq V/V_{\rm esc}\leq 2$). The scaling relations and
  the good agreement with the escape velocities can
  be explained by cosmic-ray driven winds, with the pressure supplied by star formation
  (via SNe ejecta and stellar winds) and/or by cosmic rays (via Alfv\'en waves).
\item Radio haloes show the existence of disc winds that extend over the star-forming disc with an extension of several kpc at least. Cosmic rays are important to launch these winds, because averaged over the full extent of the disc, the $\Sigma_{\rm SFR}$ is low. In our sample,
  $10^{-3} \leq \Sigma_{\rm SFR} / ({\rm M_{\sun}\,yr^{-1}\,kpc^{-2}}) \leq 10^{-1}$, where we detect
  advective haloes in 11 out of 12 galaxies. The transition from
  diffusion- to advection-dominated haloes possibly happens at  $\Sigma_{\rm SFR}= 10^{-3}~\rm
 M_{\sun}\,yr^{-1}\,kpc^{-2}$, above which galaxies have winds, in
 agreement with the detection of eDIG in their haloes.
 \end{enumerate}

Our study demonstrates the usefulness of radio continuum observations to study galactic
winds, complementary to spectroscopic studies of optical and far-UV
emission and absorption lines. As a next step it would be
desirable to repeat this kind of analysis for a larger sample of late-type
spiral galaxies such as from the
CHANG-ES survey \citep{irwin_12a,wiegert_15a} and to include
dwarf irregular galaxies such as from the LITTLE THINGS survey
\citep{heesen_11a,hunter_12a}. Furthermore, low-frequency radio continuum
observations with LOFAR \citep{vanhaarlem_13a} will allow us to trace the oldest CREs far away from
the galactic disc, of which early results
show a prominent radio halo in NGC~5775 (Heald et al.\ 2018, in prep.).

\section*{Acknowledgements}
We would like to thank the anonymous referee for their detailed, helpful comments that have greatly improved the paper. We are grateful to Philip Schmidt for carefully reading the manuscript and
providing us with many useful suggestions that helped to improve the paper. We thank
Robert Braun for giving us his WSRT $(u,v)$ data of NGC~4631 and Tom
Oosterloo for letting us use his WSRT map of NGC~891. Mat Smith is thanked for
a discussion about the nature of supernovae. VH acknowledges support from the Science and Technology Facilities Council (STFC)
under grant ST/J001600/1. DJB, RB and RJD
are supported by the Deutsche Forschungsgemeinschaft (DFG) through Research
Unit FOR~1254. DDM acknowledges support from ERCStG 307215 (LODESTONE).The Australia Telescope is funded by the Commonwealth of Australia for operation as a National Facility managed by CSIRO. The   National Radio Astronomy Observatory is a facility of the National Science Foundation operated under cooperative agreement by Associated Universities,
  Inc. The Westerbork Synthesis Radio Telescope is operated by ASTRON (Netherlands Foundation for Research in Astronomy) with support from the Netherlands Foundation for Scientific Research (NWO). The Effelsberg 100-m
telescope is operated by the Max-Planck Institut f\"ur Radioastronomie
(MPIfR). The Parkes 64-m Radio Telescope is part of the Australia Telescope, funded by the Commonwealth of Australia for operation as a National Facility managed by CSIRO. This research has made use of the NASA/IPAC
Extragalactic Database (NED) which is operated by the Jet Propulsion
Laboratory, California Institute of Technology, under contract with the
National Aeronautics and Space Administration.

\bibliographystyle{mnras}

{\small  \bibliography{spix}
}

\appendix

\section{Physical parameters}
\label{physical}
In order to calculate the synchrotron and IC losses of the CREs, we need to
make estimates of both the magnetic field strength and the radiation energy
density. The magnetic field strength is calculated assuming energy
equipartition between the cosmic rays and the magnetic field using the revised
equipartition formula of \citet{beck_05a}. We calculate the radio continuum
surface intensity $I_{1}=S_{1}/(\upi r_{\star}^2)$ (using the $1.4$-GHz flux densities) and use this as
an input for the program {\small BFIELD}, assuming a
pathlength of 1~kpc, an inclination angle of $i=0\degr$, a polarization
degree of 10~per cent, a proton-to-electron ratio of $K_0=100$ and a
non-thermal radio spectral index of $\alpha_{\rm nt}=-1$.\footnote{{The program {\scriptsize BFIELD} is available on \href{http://www3.mpifr-bonn.mpg.de/staff/mkrause/}{http://www3.mpifr-bonn.mpg.de/staff/mkrause/}}} Resulting magnetic
field strengths in the midplane are tabulated in Table~\ref{tab:phys}.

The IRF radiation energy density is the sum of the starlight radiation energy density and the total infrared radiation energy density from dust. Following \citet{draine_11a}, the starlight radiation energy density is scaled to the total infrared radiation energy density as $U_{\rm star}=1.73\times U_{\rm TIR}$, so that we find the interstellar radiation field (IRF) energy density as $U_{\rm IRF}=U_{\rm TIR}+U_{\rm star} = 2.73\times U_{\rm TIR}$.\footnote{In \citet{draine_11a}, the ratio of $1.73$ of stellar radiation energy density to far-infrared radiation energy density is presented. We have used the total infrared luminosity instead in order to include also the emission from the hot dust.} The total infrared radiation energy density is then
calculated as $U_{\rm TIR}=L_{\rm TIR}/(2\upi r_{\star}^2c)$, where $L_{\rm TIR}$ is the total infrared luminosity and $c$ is the speed of light:
\begin{equation}
  U_{\rm TIR}({10^{-13}~\rm erg\,cm^{-3}}) = 175 \times \frac{L_{\rm TIR}(10^{43}~\rm erg\,s^{-1})}{\upi
    r_{\star}^2({\rm kpc})^2}.
\end{equation}
The total infrared luminosities are derived either from \emph{Spitzer}
(if available) or \emph{IRAS} data using the prescriptions by \citet{dale_02a}. The
\emph{Spitzer} flux densities are from \citet{engelbracht_08a} (NGC~3079) and
\citet{dale_09a} (NGC~4631 and 7090). The \emph{IRAS} flux densities are from \citet{rice_88a}
(NGC~55), \citet{soifer_89a}
(NGC~3044), \citet{moshir_90a} (NGC~7462) and \citet{sanders_03a} (NGC~253,
891, 3628, 4565, 4666 and 5775). 

The ratio of IRF to magnetic field
energy density $U_{\rm IRF}/U_{\rm B}$ with $U_{\rm B}=B_0^2/(8\upi)$ is assumed
to be constant everywhere and lies between $0.02$ (NGC~7462) and $0.25$
(NGC~253). The total radiation energy density is then $U_{\rm
  rad}=U_{\rm IRF}+U_{\rm CMB}$, where $U_{\rm CMB}=4.1\times 10^{-13}~\rm erg\,
cm^{-3}$ is the radiation energy density of the CMB. The ratio of radiation energy density to magnetic energy
density is then also the ratio of IC to synchrotron losses. In our sample
galaxies the ratio lies between $0.15$ (NGC~3079) and $0.30$ (NGC~253), hence
synchrotron losses dominate. We tabulate the radiation energy densities in
Table~\ref{tab:phys}. 

\section{Intensity models}
\label{sec:intensity_models}
\begin{table}
\centering
\caption{Effective beam sizes.\label{tab:sigma}}
\begin{tabular}{l cccc}
\hline
Galaxy & $i$ & $\rm FWHM_{\rm disc}$ & $\rm FWHM_{\rm comb}$ & $b$\\
& ($\degr$) & (kpc) & (kpc) & (kpc)\\
\hline
N 55   & $85.0$ & $2.2 $ & $0.44$ & $0.18$\\
N 253  & $78.5$ & $6.5 $ & $1.43$ & $0.60$\\
N 891  & $89.0$ & $11.6$ & $0.91$ & $0.38$\\
N 3044 & $90.0$ & $3.8 $ & $0.73$ & $0.31$\\
N 3079 & $88.0$ & $9.0 $ & $0.88$ & $0.37$\\
N 3628 & $87.0$ & $12.0$ & $0.95$ & $0.40$\\
N 4565 & $90.0$ & $40.0$ & $1.62$ & $0.68$\\
N 4631 & $85.0$ & $6.3 $ & $0.97$ & $0.41$\\
N 4666 & $76.0$ & $11.0$ & $3.39$ & $1.42$\\
N 5775 & $84.0$ & $15.0$ & $1.93$ & $0.81$\\
N 7090 & $89.0$ & $5.5 $ & $0.74$ & $0.31$\\
N 7462 & $90.0$ & $5.5 $ & $0.95$ & $0.40$\\
\hline
\end{tabular}
\end{table}

\begin{table*}
\caption{Exponential (top half of the table) and Gaussian (bottom half) fits to the vertical profiles of the non-thermal radio continuum intensities. In the top half of the table, $w20n1$, $w20n2$, $w20s1$ and $w20s2$ are the intensities at $\nu_1$ ($1.4$ or $1.5$~GHz) in the northern and southern halo for the thin and the thick disc respectively. Similarly, $w6n1$, $w6n2$, $w6s1$ and $w6s2$ are the values at $\nu_2$ (mostly $\approx$5~GHz, except N~5775 where $\nu_2=8.46$~GHz). The corresponding exponential scale heights are $h20n1$, $h20n2$, $h20s1$ and $h20s2$. In the bottom half of the table, the analogue values with an additional `$g$' are presented for Gaussian fits to the intensities. Reduced $\chi^2$ are also presented for each fit.\label{tab:fits}}
\begin{tabular}{lccc cccc ccc}
\hline
Galaxy & $w20n1$ & $w20n2$ & $h20n1$ & $h20n2$ & $\chi^2$ & $w6n1$ & $w6n2$ & $h6n1$ & $h6n2$ & $\chi^2$\\
& \multicolumn{2}{c}{($\umu\rm Jy\, beam^{-1}$)} & \multicolumn{2}{c}{(kpc)} & & \multicolumn{2}{c}{($\umu\rm Jy\, beam^{-1}$)} & \multicolumn{2}{c}{(kpc)} & \\
\hline
N 55   & $6048 \pm 233$  & $292\pm 239$    & $0.36\pm 0.03$ & $2.03\pm 1.46$ & $0.63$ & $3807\pm 51$    &  $260\pm 37$   & $0.22\pm 0.01$ & $1.72\pm 0.22$ & $0.41$\\
N 253  & $-$             & $16991\pm 1242$ & $0.44\pm 0.06$ & $1.26\pm 0.07$ & $5.86$ & $-$             &  $6023\pm 389$ & $0.38\pm 0.06$ & $1.14\pm 0.04$ & $5.85$\\
N 891  & $33576\pm 5493$ & $11463\pm 493$  & $0.10\pm 0.05$ & $1.33\pm 0.02$ & $1.59$ & $17675\pm 7801$ &  $3258\pm 137$ & $0.11\pm 0.05$ & $1.30\pm 0.02$ & $0.82$\\
N 3044 & $2481 \pm 58$   & $113\pm 25$     & $0.60\pm 0.30$ & $3.64\pm 0.95$ & $0.39$ & $1351\pm 90$    &  $202\pm 63$   & $0.32\pm 0.04$ & $1.47\pm 0.26$ & $0.75$\\
N 3079 & $3598 \pm 106$  & $111\pm 149$    & $0.77\pm 0.05$ & $5.14\pm 8.86$ & $0.13$ & $1051\pm 254$   &  $519\pm 311$  & $0.43\pm 0.11$ & $1.00\pm 0.19$ & $0.22$\\
N 3628 & $7950 \pm 1503$ & $2949\pm 208$   & $0.10\pm 0.10$ & $1.23\pm 0.05$ & $1.63$ & $4160\pm 506$   &  $932\pm 63$   & $0.10\pm 0.10$ & $1.16\pm 0.04$ & $1.40$\\
N 4565 & $-$             & $1799\pm 164$   & $-$            & $1.90\pm 0.10$ & $5.00$ & $-$             &  $689\pm 44$   & $-$            & $1.63\pm 0.06$ & $2.08$\\
N 4631 & $31222\pm 1567$ & $6085\pm 387$   & $0.47\pm 0.03$ & $2.24\pm 0.07$ & $0.48$ & $16976\pm 1112$ &  $2469\pm 158$ & $0.33\pm 0.02$ & $1.70\pm 0.05$ & $0.60$\\
N 4666 & $13439\pm 4252$ & $16167\pm 1509$ & $0.41^{\rm a}$         & $1.51\pm 0.05$ & $0.69$ & $18535\pm 918$  &  $3714\pm 171$ & $0.23^{\rm a}$         & $1.54\pm 0.03$ & $0.13$\\
N 5775 & $5364 \pm 164$  & $968\pm 97$     & $0.70\pm 0.03$ & $2.72\pm 0.12$ & $0.21$ & $5827\pm 538$   &  $560\pm 56$   & $0.10\pm 0.30$ & $1.27\pm 0.06$ & $0.96$\\
N 7090 & $1585 \pm 201$  & $616\pm 76$     & $0.31\pm 0.07$ & $2.07\pm 0.18$ & $0.93$ & $576\pm 85$     &  $124\pm 44$   & $0.32\pm 0.09$ & $1.96\pm 0.50$ & $2.23$\\
N 7462 & $-$             & $864 \pm 50$    & $-$            & $0.99\pm 0.04$ & $1.72$ & $-$             &  $223\pm 17$   & $-$            & $0.86\pm 0.06$ & $2.31$\\
\hline
& $w20s1$ & $w20s2$ & $h20s1$ & $h20s2$ & $\chi^2$ & $w6s1$ & $w6s2$ & $h6s1$ & $h6s2$ & $\chi^2$\\
\hline
N 55   & $5067\pm 809$   & $1859\pm 401$  & $0.20\pm 0.05$ & $1.06\pm 0.14$ & $2.15$ & $4136\pm 602$   &  $231\pm 71$   & $0.17\pm 0.02$ & $2.01\pm 0.71$ & $3.92$\\
N 253  & $-$             & $15587\pm 625$ & $0.44\pm 0.06$ & $1.33\pm 0.04$ & $1.70$ & $-$             &  $5484\pm 172$ & $0.38\pm 0.06$ & $1.28\pm 0.02$ & $1.47$\\
N 891  & $43174\pm 2381$ & $9433\pm 195$  & $0.10\pm 0.05$ & $1.25\pm 0.01$ & $0.33$ & $13587\pm 3081$ &  $2797\pm 169$ & $0.18\pm 0.04$ & $1.24\pm 0.03$ & $1.11$\\
N 3044 & $1786\pm  106$  & $849\pm 125$   & $0.38\pm 0.05$ & $1.30\pm 0.08$ & $0.30$ & $1223\pm 76$    &  $318\pm  90$  & $0.33\pm 0.05$ & $1.15\pm 0.14$ & $0.44$\\
N 3079 & $3511\pm  182$  & $513\pm 143$   & $0.72\pm 0.07$ & $3.41\pm 0.63$ & $0.98$ & $1100\pm 61$    &  $447\pm  74$  & $0.49\pm 0.05$ & $1.54\pm 0.10$ & $0.24$\\
N 3628 & $5082\pm  371$  & $56\pm 238$    & $0.73\pm 0.12$ & $594 \pm 4.e5$ & $3.13$ & $3296\pm 374$   &  $1229\pm 64$  & $0.10\pm 0.10$ & $0.87\pm 0.02$ & $0.57$\\
N 4565 & $1531\pm  312$  & $505\pm 159$   & $0.80\pm 0.80$ & $3.11\pm 0.66$ & $3.71$ & $696\pm 177$    &  $500\pm  47$  & $0.25$         & $1.75\pm 0.10$ & $0.91$\\
N 4631 & $26395\pm 1537$ & $14554\pm 476$ & $0.32\pm 0.02$ & $1.47\pm 0.02$ & $0.11$ & $17953\pm 3730$ &  $6752\pm 362$ & $0.19\pm 0.05$ & $1.13\pm 0.02$ & $0.53$\\
N 4666 & $25119\pm 3712$ & $9393\pm 967$  & $0.41^{\rm a}$         & $2.01\pm 0.08$ & $1.06$ & $17737\pm 3418$ &  $3487\pm 569$ & $0.23^{\rm a}$         & $1.70\pm 0.11$ & $2.20$\\
N 5775 & $6593\pm 2164$  & $1857\pm 293$  & $0.35\pm 0.15$ & $1.91\pm 0.12$ & $2.21$ & $7993\pm 318$   &  $269\pm  25$  & $0.10\pm 0.30$ & $2.03\pm 0.80$ & $0.62$\\
N 7090 & $1500\pm 129$   & $443\pm 27$    & $0.40\pm 0.06$ & $1.53\pm 0.39$ & $1.96$ & $439\pm 53$     &  $148\pm  13$  & $0.29\pm 0.10$ & $1.14\pm 0.44$ & $2.17$\\
N 7462 &  $-$            & $1164\pm 103$  & $-$            & $0.88\pm 0.05$ & $4.59$ &   $-$           &  $257\pm 27$   & $-$            & $0.90\pm 0.08$ & $5.52$\\
\hline
& $w20gn1$ & $w20gn2$ & $h20gn1$ & $h20gn2$ & $\chi^2$ & $w6gn1$ & $w6gn2$ & $h6gn1$ & $h6gn2$ & $\chi^2$\\
\hline
N 55   & $3877\pm 178$   & $568 \pm 69 $  & $0.51\pm 0.02$ & $1.71\pm 0.11$ & $0.91$ & $2364\pm 157 $ &  $211  \pm 19$  & $0.39\pm 0.02$ & $2.09\pm 0.14$ & $1.51$\\
N 253  & $8944\pm 5027$  & $3323\pm 5474$ & $1.57\pm 0.50$ & $2.92\pm 1.25$ & $3.64$ & $3666\pm 439 $ &  $994  \pm 439$ & $1.27\pm 0.21$ & $3.16\pm 0.47$ & $6.03$\\
N 891  & $13568\pm 2341$ & $3444\pm 275 $ & $0.71\pm 0.10$ & $3.10\pm 0.07$ & $5.23$ & $6408\pm 910 $ &  $1007 \pm 70 $ & $0.60\pm 0.07$ & $2.99\pm 0.07$ & $3.54$\\
N 3044 & $1561\pm 110$   & $255 \pm 46  $ & $0.84\pm 0.06$ & $2.67\pm 0.23$ & $2.20$ & $920 \pm 90  $ &  $125  \pm 20 $ & $0.55\pm 0.05$ & $2.38\pm 0.21$ & $2.10$\\
N 3079 & $2208\pm 164$   & $646 \pm 99  $ & $0.86\pm 0.07$ & $2.44\pm 0.15$ & $1.34$ & $928 \pm 29  $ &  $222  \pm 17 $ & $0.67\pm 0.03$ & $2.04\pm 0.06$ & $0.18$\\
N 3628 & $4650\pm 1512$  & $1329\pm 135 $ & $0.37\pm 0.14$ & $2.35\pm 0.11$ & $3.57$ & $2959\pm 1327$ &  $423  \pm 33 $ & $0.24\pm 0.11$ & $2.24\pm 0.08$ & $2.47$\\
N 4565 & $-$             & $1243\pm 92  $ & $-$            & $2.15\pm 0.14$ & $2.74$ & $-$            &  $494  \pm 39 $ & $-$            & $1.65\pm 0.13$ & $1.72$\\
N 4631 & $19405\pm 1910$ & $2617\pm 200 $ & $0.99\pm 0.06$ & $4.33\pm 0.14$ & $3.80$ & $9677\pm 1201$ &  $891  \pm 83 $ & $0.76\pm 0.06$ & $3.60\pm 0.14$ & $4.80$\\
N 4666 & $19698\pm 1657$ & $3949\pm 198 $ & $0.96\pm 0.10$ & $3.81\pm 0.05$ & $0.11$ & $7709\pm 877 $ &  $720  \pm 116$ & $1.07\pm 0.14$ & $4.11\pm 0.19$ & $0.61$\\
N 5775 & $3609\pm 246$   & $446 \pm 40  $ & $1.30\pm 0.08$ & $5.14\pm 0.21$ & $1.71$ & $3742\pm 1744$ &  $204  \pm 11 $ & $0.24\pm 0.12$ & $2.77\pm 0.06$ & $0.24$\\
N 7090 & $1172\pm 47$    & $353 \pm 11  $ & $0.62\pm 0.03$ & $3.27\pm 0.07$ & $0.28$ & $398 \pm 28  $ &  $74   \pm 7  $ & $0.60\pm 0.04$ & $3.10\pm 0.19$ & $0.97$\\
N 7462 & $-$             & $536\pm 77$    & $-$            & $1.62\pm 0.15$ & $13.9$ & $-$            &  $149 \pm 4$    & $-$            & $1.38\pm 0.03$ & $0.47$\\
\hline
& $w20gs1$ & $w20gs2$ & $h20gs1$ & $h20gs2$ & $\chi^2$ & $w6gs1$ & $w6gs2$ & $h6gs1$ & $h6gs2$ & $\chi^2$\\
\hline
N 55   & $3728 \pm 423 $ & $947\pm 96 $   & $0.40\pm 0.04$ & $1.84\pm 0.11$ & $2.84$ & $2630  \pm 453 $ &  $188 \pm 33$  & $0.30\pm 0.04$ & $2.26\pm 0.35$  & $7.23$\\
N 253  & $6561 \pm 1213$ & $5426\pm 1456$ & $1.22\pm 0.24$ & $2.74\pm 0.23$ & $0.82$ & $3296  \pm 146 $ &  $1231\pm 99 $ & $1.19\pm 0.06$ & $3.22\pm 0.09$  & $0.74$\\
N 891  & $13171\pm 2308$ & $2421\pm 236 $ & $0.75\pm 0.10$ & $3.06\pm 0.08$ & $6.90$ & $6241  \pm 832 $ &  $783 \pm 65 $ & $0.65\pm 0.07$ & $2.93\pm 0.08$  & $3.95$\\
N 3044 & $1476 \pm 73  $ & $372 \pm 23  $ & $0.69\pm 0.03$ & $2.50\pm 0.07$ & $0.68$ & $905   \pm 60  $ &  $155 \pm 18 $ & $0.57\pm 0.04$ & $2.15\pm 0.13$  & $1.05$\\
N 3079 & $2284 \pm 184 $ & $400 \pm 41  $ & $1.17\pm 0.07$ & $4.62\pm 0.27$ & $2.91$ & $885   \pm 71  $ &  $198 \pm 24 $ & $0.85\pm 0.07$ & $3.00\pm 0.17$  & $1.69$\\
N 3628 & $3792 \pm 1026$ & $1389\pm 252 $ & $0.47\pm 0.16$ & $1.92\pm 0.14$ & $4.61$ & $1890  \pm 252 $ &  $427 \pm 32 $ & $0.40\pm 0.06$ & $1.89\pm 0.06$  & $0.98$\\
N 4565 & $-$             & $1201\pm 80  $ & $-$            & $1.77\pm 0.09$ & $1.68$ & $-$              &  $522 \pm 62 $ & $-$            & $1.21\pm 0.17$  & $2.22$\\
N 4631 & $20322\pm 2099$ & $5066\pm 430 $ & $0.83\pm 0.08$ & $3.12\pm 0.10$ & $2.50$ & $10913 \pm 837 $ &  $1949\pm 107$ & $0.65\pm 0.04$ & $2.62\pm 0.05$  & $1.14$\\
N 4666 & $25617\pm 6272$ & $3132\pm 215 $ & $0.74\pm 0.20$ & $4.52\pm 0.10$ & $0.45$ & $(1.1\pm 2.1)e5$ &  $1125\pm 56 $ & $0.06\pm 1.20$ & $3.92\pm 0.06$  & $0.18$\\
N 5775 & $3573 \pm 510 $ & $505 \pm 83  $ & $1.16\pm 0.15$ & $4.49\pm 0.29$ & $5.32$ & $2343  \pm 621 $ &  $106 \pm 11 $ & $0.45\pm 0.12$ & $4.08\pm 0.22$  & $1.78$\\
N 7090 & $1206 \pm 105 $ & $190 \pm 26  $ & $0.69\pm 0.06$ & $190 \pm 26  $ & $2.17$ & $389   \pm 28  $ &  $54  \pm 6  $ & $0.50\pm 0.04$ & $2.84\pm 0.24$  & $0.80$\\
N 7462 & $-$             & $731 \pm 29 $  & $-$            & $1.50\pm 0.03$ & $1.22$ & $-$              &  $172 \pm 2$   & $-$            & $1.46\pm 0.01$  & $0.10$\\
\hline
\end{tabular}
\flushleft{{\bf Notes.} Exponential scale heights in the southern halo of NGC~7090 are from \citetalias{heesen_16a}.\\
$^{\rm a}$ Average scale height of the sample, since the exponential scale height in the thin disc in NGC~4666 is unconstrained.}
\end{table*}

Following \citet{dumke_95a}, we used analytical
functions that allow the fitting of vertical radio continuum emission profiles with one or two
component Gaussian or exponential functions, where the instrumental PSF is
assumed to be of Gaussian shape, which is exact for interferometric maps (synthesized beam). In addition to the PSF, a correction for
projected disc emission can be added, for instance by fitting a Gaussian and
adding the FWHM in quadrature. We have done this, so that the combined
resolution is ${\rm FWHM_{comb}}=\sqrt{{\rm FWHM^2}+{(\cos(i)\cdot \rm
    FWHM_{disc})^2}}$. The effective beam size is then $b = 0.5 \cdot
{\rm FWHM_{\rm comb}} / \sqrt
{2\ln(2)}\approx 0.425\cdot {\rm FWHM_{\rm comb}}$. For an exponential
profile we fit the following function:
\begin{eqnarray} 
  W_{\rm exp}(z) & = & \frac{w_0}{2}\exp\left (-\frac{z^2}{2b^2}\right
  )\nonumber\\
& & \times\left [ \exp\left ( \frac{b^2-zz_0}{\sqrt{2}b z_0} \right
  )^2
{\rm erfc} \left (\frac{b^2-zz_0}{\sqrt{2}b z_0} \right )\right .\nonumber\\
& & \left . + \exp\left ( \frac{b^2+zz_0}{\sqrt{2}b z_0} \right  
  )^2
{\rm erfc} \left (\frac{b^2+zz_0}{\sqrt{2}b z_0} \right )\right ],
\end{eqnarray}
where $b$ is the effective beam size, $w_0$ is the maximum of the
distribution and $z_0$ the (exponential) scale height. We have introduced the
complementary error function as:
\begin{equation}
{\rm erfc}(x) = 1 - {\rm erf}(x) = \frac{2}{\sqrt{\upi}}\int_0^\infty
\exp(-r^2){\rm d}r.
\end{equation}
And for a Gaussian intensity
profile, we fit the following function:
\begin{equation}
  W_{\rm Gauss}(z) = \frac{w_o z_0}{\sqrt{2b^2+z_0^2}} \exp\left (-\frac{z^2}{2b^2+z_0^2} \right ).
\end{equation}
Again, $b$ is the effective beam size, $w_0$ the maximum of the
distribution and $z_0$ the (Gaussian) scale height. We have fitted for a
thin and a thick disc in all those cases where this provided a better fit than
a single thick disc. The resulting fit
parameters are presented in Table~\ref{tab:fits}. Here, the maxima of the exponential thin and thick disc at $L$ band in the northern halo are $w20n1$ and $w20n2$, with the corresponding scale heights of $h20n1$ and $h20n2$. Similarly, at $C$ band the maxima are $w6n1$ and $w6n2$ with scale heights of $h6n1$ and $h6n2$. For Gaussian fits, the same parameters at $L$ band are $w20gn1$, $w20gn2$, $h20gn1$ and $h20gn2$. The parameters for $C$ band and the southern haloes are defined in a similar way.

Using these fits, one can now construct the intensity models. For instance, the
resulting (normalized) intensity model for $C$ band ($\nu=\nu_2$) and an
exponential intensity profile in the northern halo is:
\begin{equation}
 I_{\nu}(z) = \frac{w6n1\cdot \exp(-z/h6n1) + w6n2 \cdot \exp(-z/h6n2)}{w6n1+w6n2}.
\end{equation}
The other cases can be calculated accordingly.

\section{Additional radio continuum maps}
\label{additional_maps}
\begin{table*}
\caption{Additional maps produced in our study.\label{tab:additional}}
\begin{tabular}{l ccccc}
\hline
Galaxy & Telescope & $\nu$ & $\theta_{\rm FWHM}$ & $\sigma_{\rm rms}$& $S_{\nu}$\\
& & (GHz) & (arcsec) & ($\rm mJy\,beam^{-1}$) & (mJy)\\\hline
N 55    & Parkes    & $4.80$ & $255.0$                               & $1.50$  & 333 \\
N 891   & Effelsberg    & $4.85$ & $147.0$                               & $0.50$  & 281 \\
N 4157  & VLA    & $1.49$ & $25.1\times 13.7$ & $0.10$  & 185 \\
N 4217  & VLA    & $4.86$ & $15.8\times 10.9$  & $0.04$ & 44  \\
N 4631  & Effelsberg    & $4.85$ & $147.0$                               & $0.50$  & 472 \\
N 4631  & VLA    & $1.35$ & $52.0$                                & $0.04$ & 1201\\
N 4631  & VLA    & $1.65$ & $52.0$                                & $0.05$ & 975 \\
N 4634  & VLA    & $4.86$ & $15.4\times 13.6$   & $0.06$ & 14  \\
\hline
\end{tabular}
\flushleft{{\bf Notes.} $S_{\nu}$ denotes the integrated flux density of each galaxy.}
\end{table*}

\begin{figure*}
  \includegraphics[width=1.0\hsize]{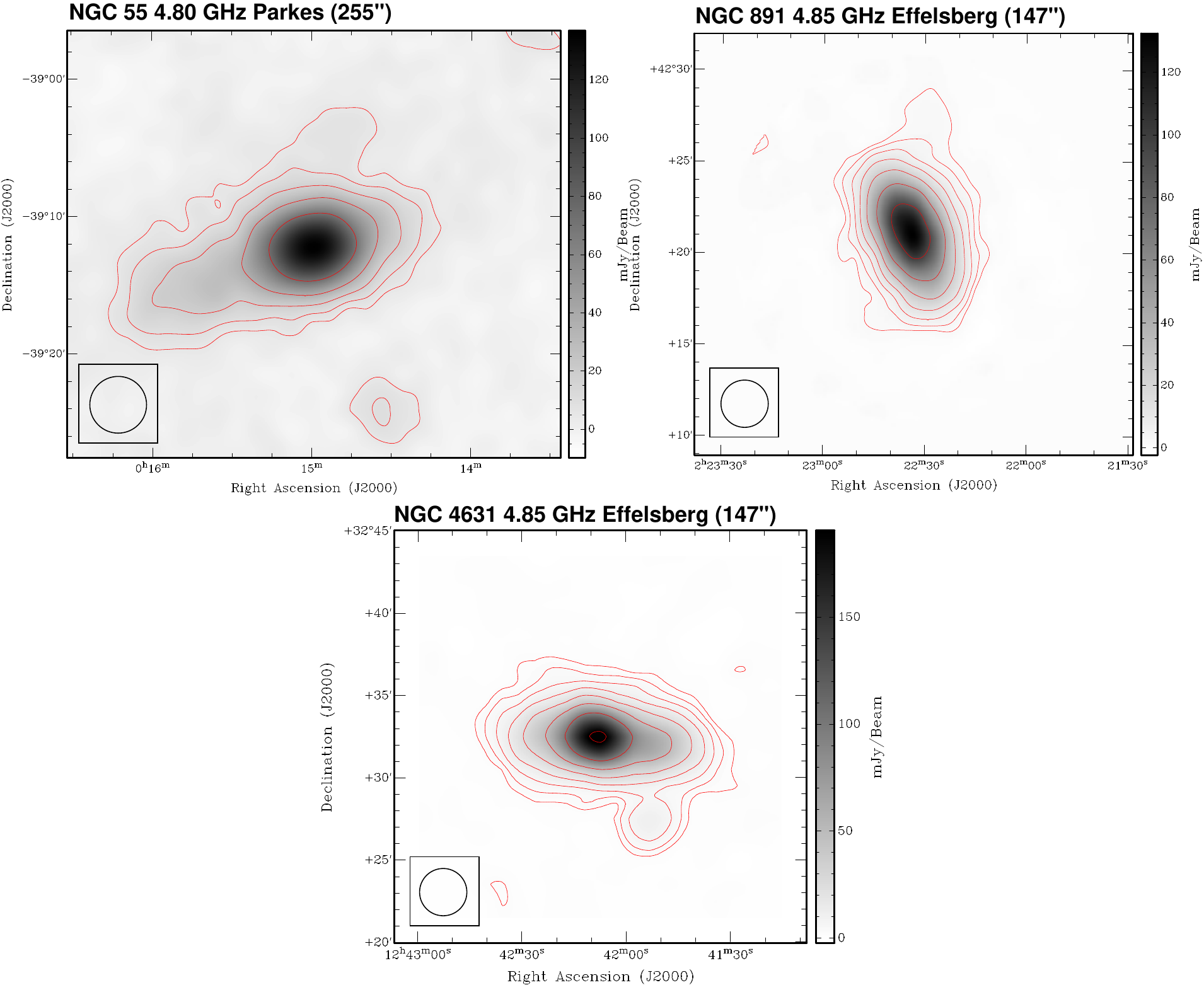}
  \caption{Single-dish maps of NGC~55 (\emph{top left}), NGC~891 (\emph{top right}) and NGC~4631 (\emph{bottom}). Contours show the radio continuum emission at levels of
  $3\sigma\cdot2^n$ ($n=0, 1, 2, \ldots$). The size of the primary beam is shown in the bottom left corner.}
\label{fig:sd1}
\end{figure*}
As part of our study we have obtained several additional radio continuum maps, which we present
in this Appendix. We will make these maps publicly available (as well as all
the other radio continuum maps in this paper); see Table~\ref{tab:additional} for the map properties.\footnote{The maps will be made available on the website of the Centre de Donn\'ees astronomiques de Strasbourg (CDS) (\href{http://cds.u-strasbg.fr}{http://cds.u-strasbg.fr}).} For 4 galaxies (NGC~55, 253, 891 and 4631) we have used
single-dish maps, to correct for the missing zero-spacing flux where
necessary. The Effelsberg maps of NGC~253 and 4631 were already presented in
\citet{heesen_09a} and \citet{mora_13a}, respectively, and the Effelsberg map of
NGC 891 was already presented
in \citet{dumke_97a}. We present these maps for completeness here again in
Fig.~\ref{fig:sd1}. The $4.80$-GHz map of NGC~55 
obtained with the 64-m Parkes telescope is so far unpublished and shown here also
in Fig.~\ref{fig:sd1}.

\begin{figure*}
  \includegraphics[width=1.0\hsize]{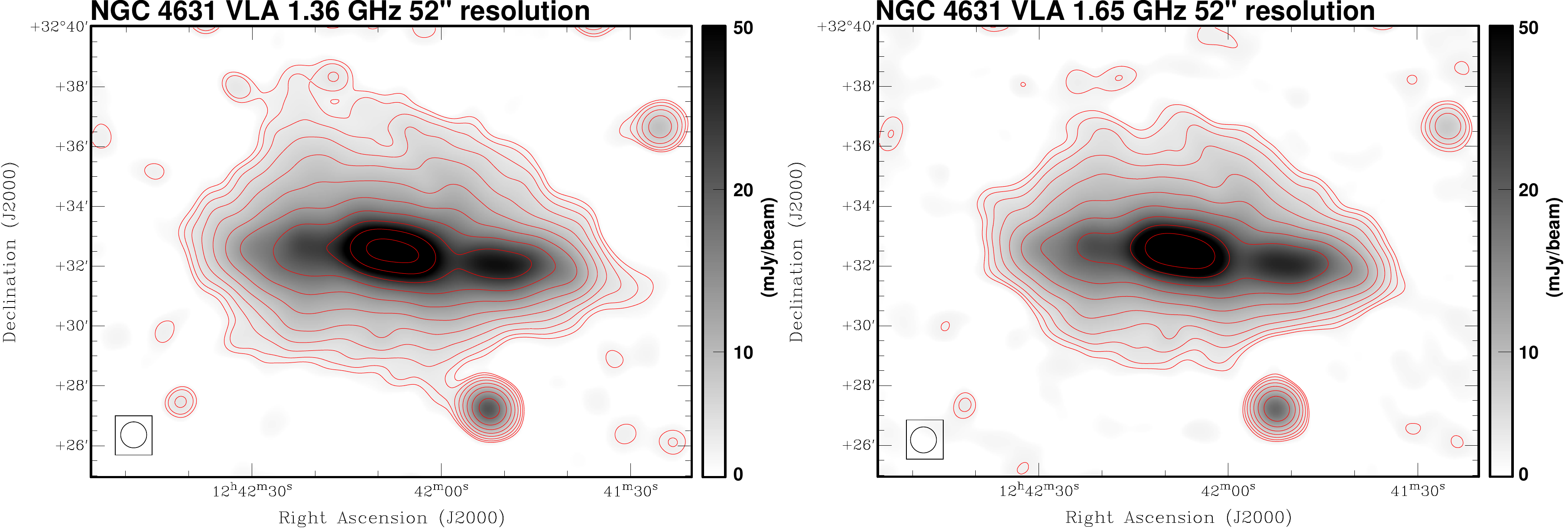}
  \caption{VLA maps of NGC~4631 at $1.37$~GHz (\emph{left}) and $1.65$~GHz (\emph{right}). Contours show the radio continuum emission at levels of
  $3\sigma\cdot2^n$ ($n=0, 1, 2, \ldots$). The size of the synthesized
  beam is shown in the bottom left corner.}
\label{fig:n4631_rainer}
\end{figure*}

\begin{figure*}
  \includegraphics[width=0.9\hsize]{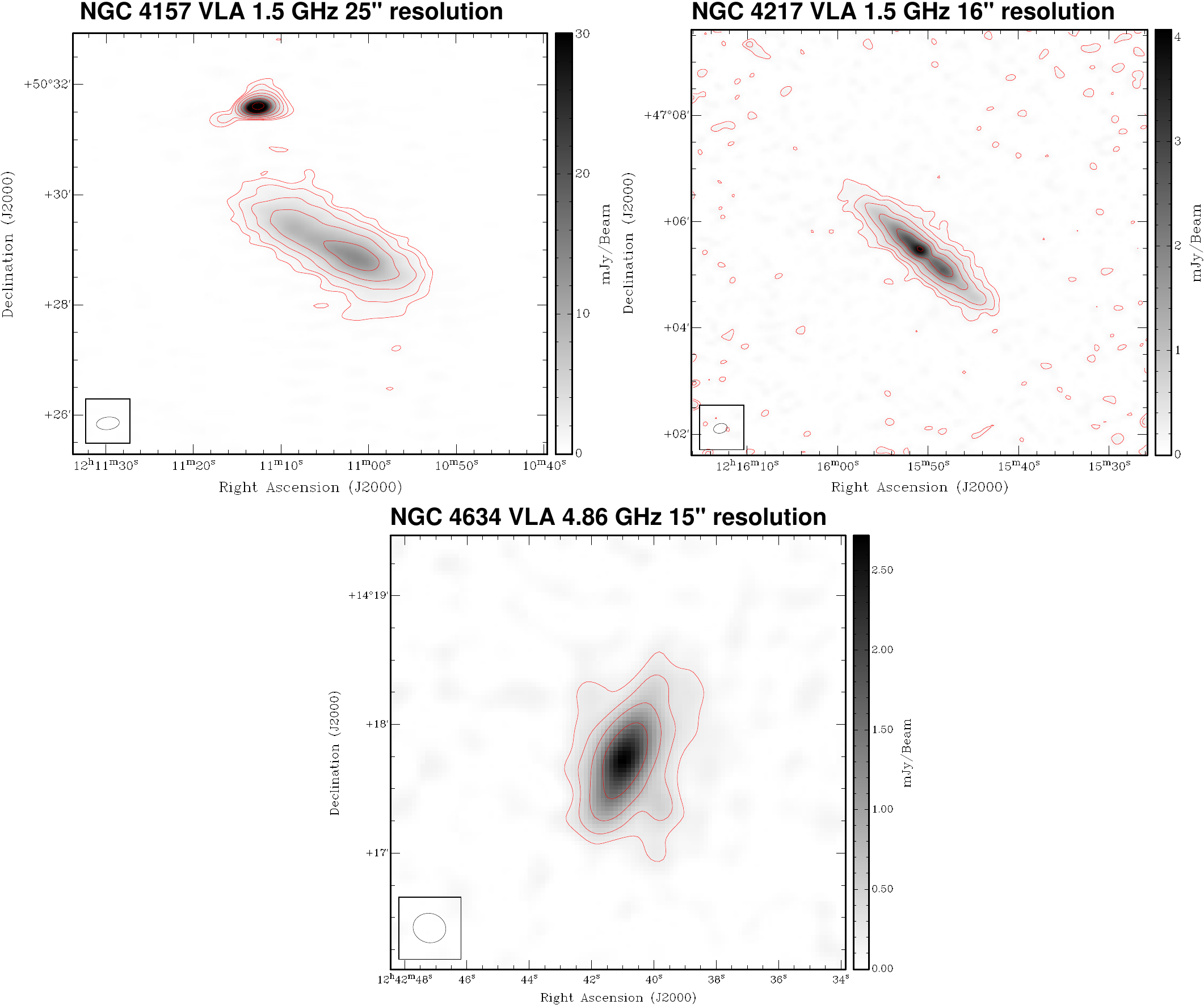}
  \caption{VLA maps of NGC~4157 (\emph{top left}), NGC~4217 (\emph{top right}) and NGC~4634
    (\emph{bottom}). Contours show the radio continuum emission at levels of
  $3\sigma\cdot2^n$ ($n=0, 1, 2, \ldots$). The size of the synthesized
  beam is shown in the bottom left corner.}
\label{fig:maps1}
\end{figure*}

Furthermore, we show in Fig.~\ref{fig:n4631_rainer} two maps of NGC~4631 at
$1.35$ and $1.65$~GHz observed with the VLA in D-configuration (R.~Beck 2016, priv.\ comm.). The data were
observed in August 1996, with 12~h on-source (ID: AG486) and reduced in
standard fashion with {\small AIPS}.  The maps
have an angular resolution of 52~arcsec, so that we did not use them in the
analysis, but they also show the halo of this galaxy very well. Lastly, we obtained maps of three further edge-on galaxies observed with the
VLA (NGC~4157, 4217 and
4634), which we present in Fig.~\ref{fig:maps1}. We reduced the data as
described in Section~\ref{observations}, but since we had only one frequency available and
no spectral index map, we did not use them in the analysis. The maps of
NGC~4157 and 4217 were created by re-reducing archive data (IDs
AI23, AF85, AH457 and AS392 for NGC~4157 and ID AM573 for
NGC~4217). The map of NGC~4634 was created by using so far unpublished data
from the VLA (ID: AD538).

\section{Image atlas}
\label{sec:image_atlas}

In this appendix, we present maps and vertical
profiles for individual galaxies. Each page shows results for one galaxy.\footnote{The complete image atlas (12 images) is available in the online journal.} The
top row shows maps of the radio continuum emission at $L$ and $C$ band
($1.4$ and 5~GHz, except for NGC~5775 where instead of a $C$-band map we show an $X$-band map at $8.5$~GHz) and
the thermal radio continuum emission at $5$~GHz (again, with the exception of NGC~5775, where we show the emission at $X$ band instead). The second row shows thermal fraction at
$C$ band (again except for NGC~5775 where this is at $X$ band), the
non-thermal radio
spectral index between $L$ and $C$ band (for NGC~5775 between $L$ and $X$
band) and the corresponding error of the non-thermal radio spectral index. In the radio maps,
contours show the radio continuum emission at levels of  (3, 6,
12, 24, 48, 96) $\times\sigma$, where $\sigma$ is the rms noise of the map. For the
thermal radio continuum map we use the same contours as of the radio continuum map of either $C$
or $X$ band. In the first and second row, the size of the synthesized beam is shown in the bottom left
corner and all maps are rotated so that the major axis is horizontal. In panels (c)--(f), a mask has been applied (Section~\ref{sec:masking}).

The third row shows vertical profiles of the non-thermal radio continuum intensities, vertical non-thermal intensity model profiles at one frequency and vertical profiles of the
non-thermal radio spectral index. Panels (g) show exponential (solid) and Gaussian
fits (dashed lines) to the data, panels (h) show model intensity profiles
and panels (i) show radio spectral index profiles. The fourth row shows the
reduced $\chi^2$ of the cosmic-ray transport model in the northern halo,
in the southern halo and the vertical model profile of the magnetic field strength. In
panels (j) and (k) contours show $\chi^2_{\rm int,min}+1$ for the intensity
fitting (solid lines) and $\chi^2_{\rm spix,min}+1$ for spectral index fitting (dashed lines). The colour-scale shows the total reduced $\chi^2$ (sum of intensity and spectral index
fitting). The best-fitting models (with uncertainties) lie in the intersection
of the two contours and are marked by yellow stars. In panels (l), the grey-shaded
area indicates the uncertainty of the magnetic field model.

\begin{figure*}
  \includegraphics[width=1.0\hsize]{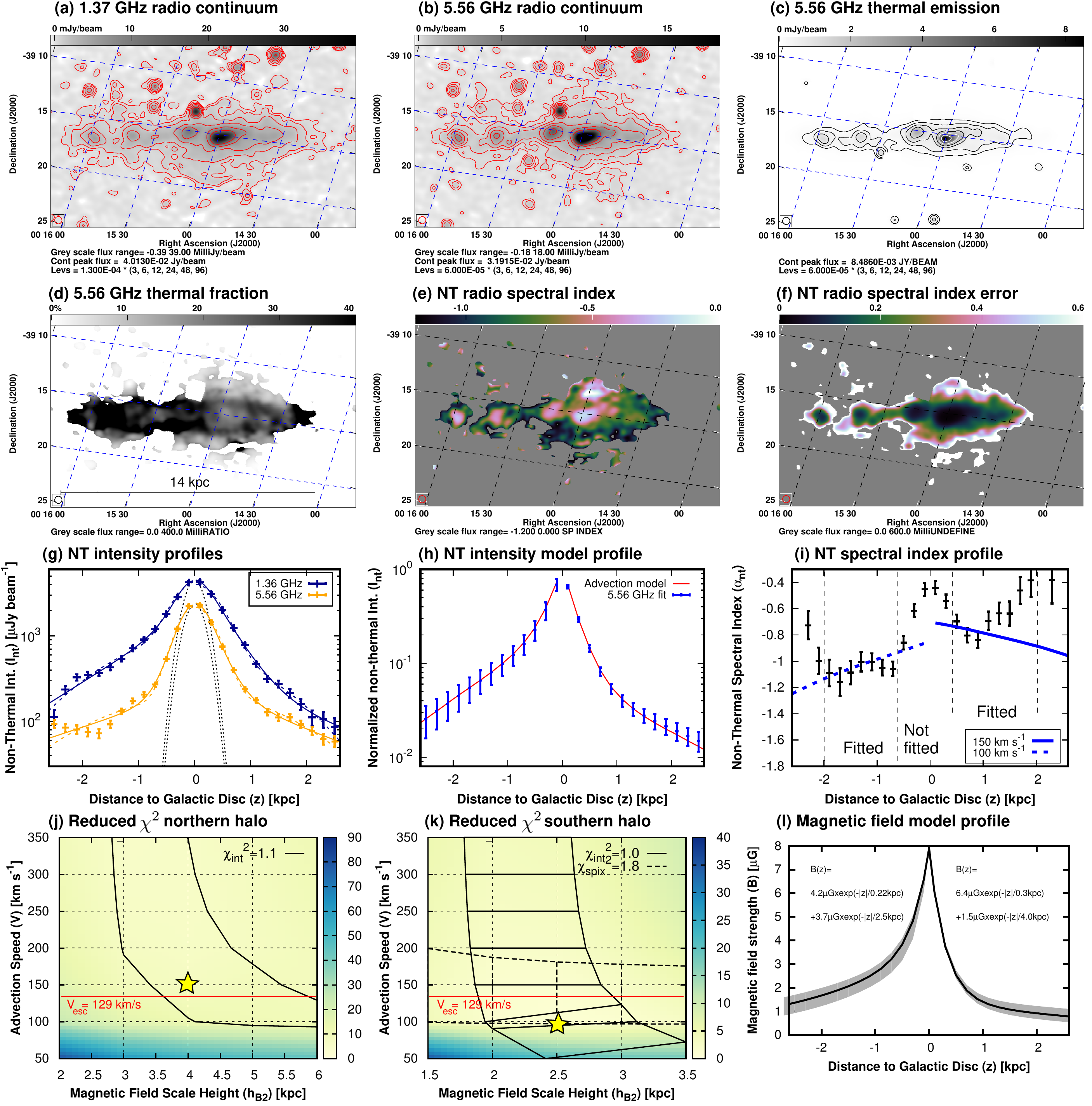}
  \caption{NGC~55. (a) radio continuum emission at $1.37$~GHz. (b) radio continuum emission at $5.56$~GHz. (c) thermal radio continuum emission at
    $5.56$~GHz. Contours in panels (a) and (b) are at (3, 6, 12, 24, 48 and 96)
    $\times \sigma$, where $\sigma$ is the rms map noise. In panel (c), the same contour levels as in (b) are used. (d) thermal
  fraction at $5.56$~GHz, where the grey-scale ranges from 0 to 40 per cent. (e) non-thermal radio spectral index between $1.37$
  and $5.56$~GHz, where the colour scale ranges from $-1.2$ to $0$. (f) error of the non-thermal radio spectral index, where the colour scale ranges from 0 to $0.6$. Panels (a)--(f) are rotated so that the major axis ($PA=108\degr$)
is horizontal and the synthesized beam is shown in the bottom
left corner. (g) vertical non-thermal intensity profiles at both
frequencies, where solid lines show two-component exponential fits and dashed
lines two-component Gaussian fits. (h) normalized non-thermal vertical intensity model profile at
    $5.56$~GHz with best-fitting advection model. (i) vertical non-thermal radio spectral index profiles
with best-fitting advection model. (j) reduced $\chi^2$ in the northern halo as
function of advection speed and magnetic field scale height in the thick
disc. (k) same as (j) but in the southern halo. The red lines in (j) and (k) show the escape velocity near the
  midplane. (l) vertical magnetic field model profile.\vspace{4pt}\hspace{\textwidth}(The complete image atlas  (12 images) is available in the online journal)}
\label{fig:n55}
\end{figure*}

\begin{figure*}
  \includegraphics[width=1.0\hsize]{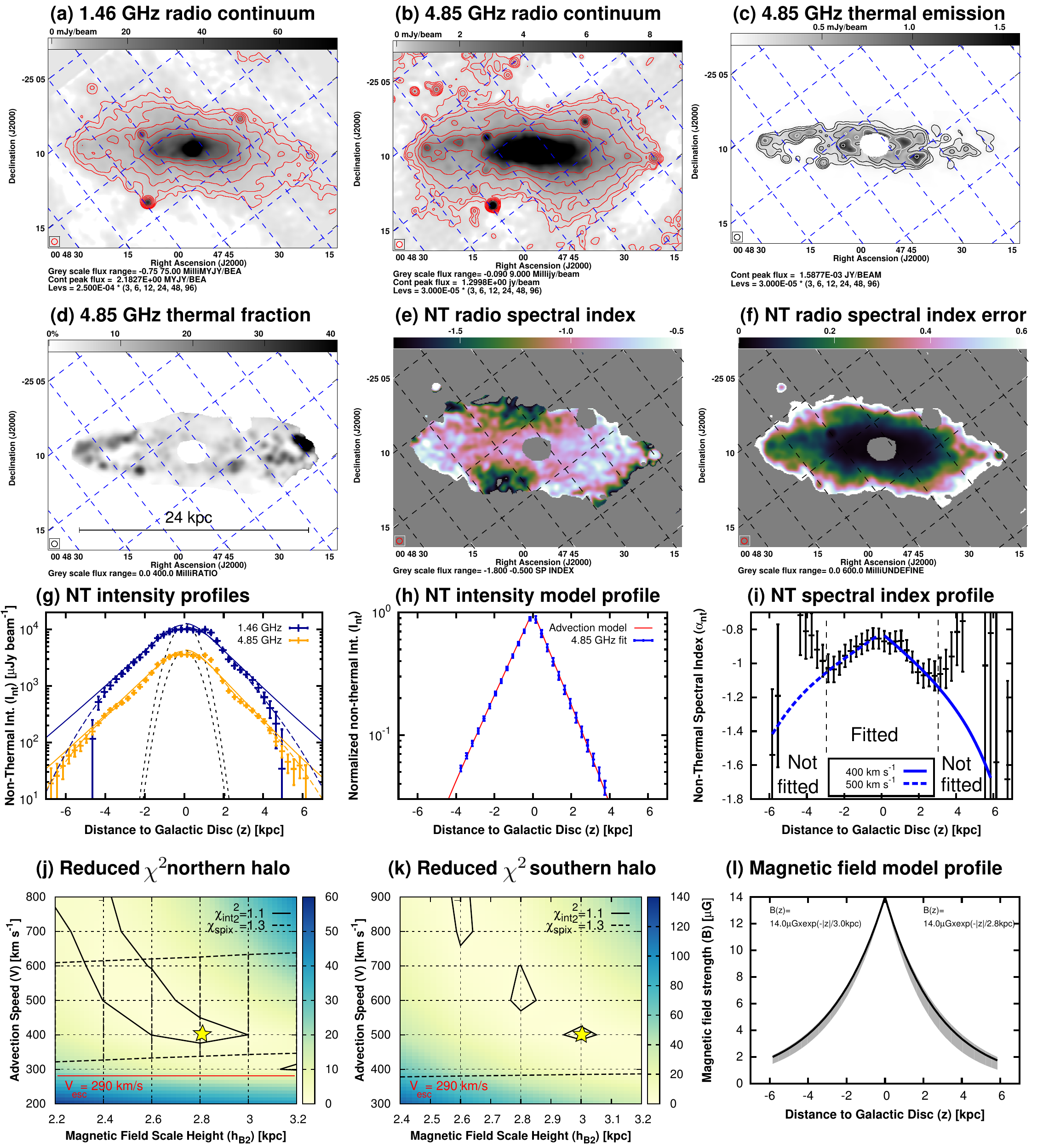}
  \caption{NGC~253. (a) radio continuum emission at $1.46$~GHz. (b) radio continuum emission at $4.85$~GHz. (c) thermal radio continuum emission at
    $4.85$~GHz. Contours in panels (a)--(c) are at (3, 6, 12, 24, 48 and 96)
    $\times \sigma$, where $\sigma$ is the rms map noise. In panel (c), the same contour levels as in (b) are used. (d) thermal
  fraction at $4.85$~GHz, where the grey-scale ranges from 0 to 40 per cent. (e) non-thermal radio spectral index between $1.46$
  and $4.85$~GHz, where the colour scale ranges from $-1.8$ to $-0.5$.  (f) error of the non-thermal radio spectral index, where the colour scale ranges from 0 to $0.6$. Panels (a)--(f) are rotated so that the major axis ($PA=52\degr$)
is horizontal and the synthesized beam is shown in the bottom
left corner. (g) vertical non-thermal intensity profiles at both
frequencies, where solid lines show two-component exponential fits and dashed
lines two-component Gaussian fits. (h) normalized vertical non-thermal intensity model profile at
    $4.85$~GHz with best-fitting advection model. (i) vertical non-thermal radio spectral index profile
with best-fitting advection model. (j) reduced $\chi^2$ in the northern halo as
function of advection speed and magnetic field scale height in the thick
disc. (k) same as (j) but in the southern halo. The red line in (j) shows the escape velocity near the
  midplane. (l) vertical magnetic field model profile.}
\label{fig:n253}
\end{figure*}

\begin{figure*}
  \includegraphics[width=1.0\hsize]{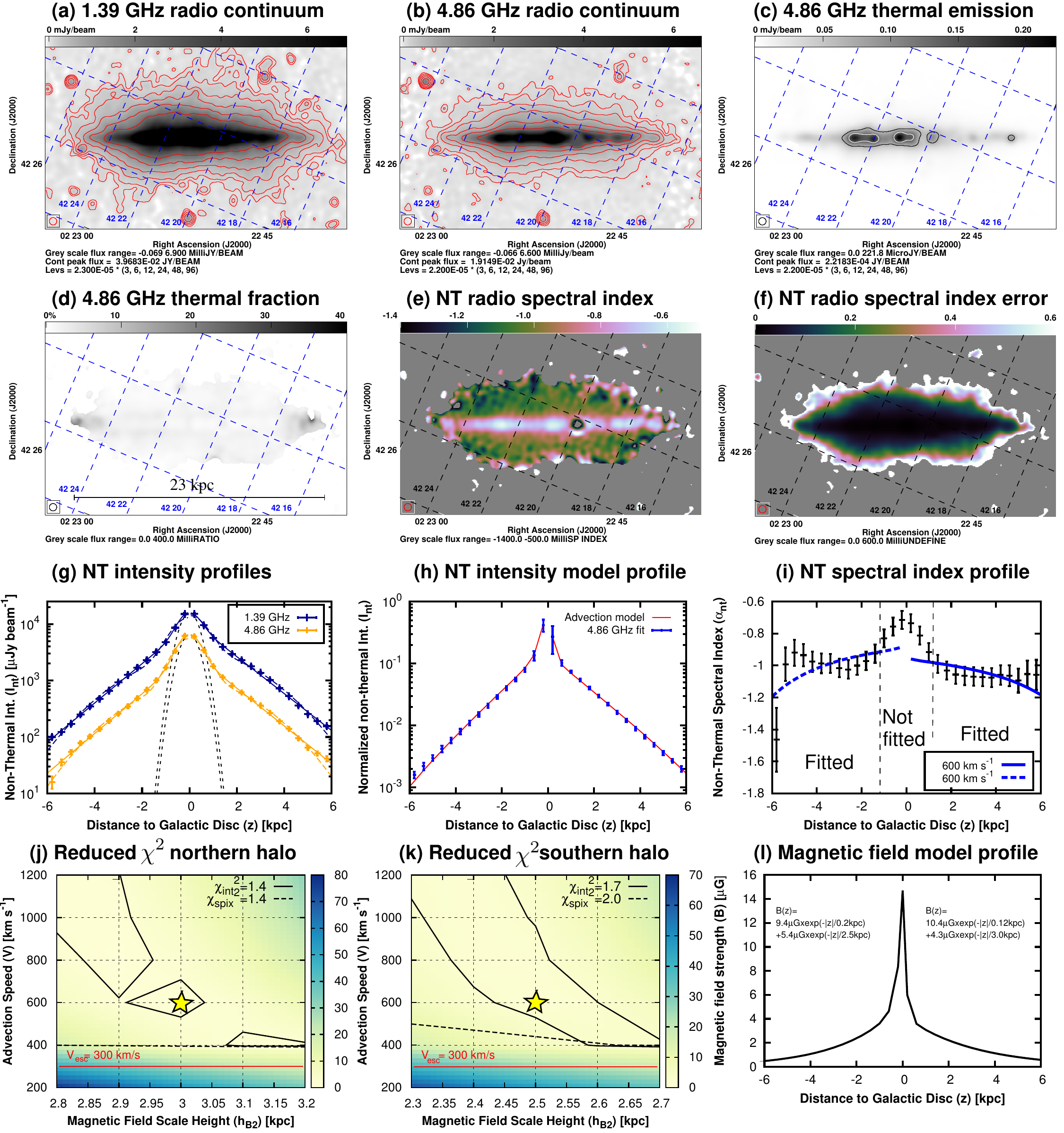}
  \caption{NGC~891. (a) radio continuum emission at $1.39$~GHz. (b) radio continuum emission at $4.86$~GHz. (c) thermal radio continuum emission at
    $4.86$~GHz. Contours in panels (a)--(c) are at (3, 6, 12, 24, 48 and 96)
    $\times \sigma$, where $\sigma$ is the rms map noise. In panel (c), the same contour levels as in (b) are used. (d) thermal
  fraction at $4.86$~GHz, where the grey-scale ranges from 0 to 40 per cent. (e) non-thermal radio spectral index between $1.39$
  and $4.86$~GHz, where the colour scale ranges from $-1.4$ to $-0.5$. (f) error of the non-thermal radio spectral index, where the colour scale ranges from 0 to $0.6$. Panels (a)--(f) are rotated so that the major axis ($PA=23\degr$)
is horizontal and the synthesized beam is shown in the bottom
left corner. (g) vertical non-thermal intensity profiles at both
frequencies, where solid lines show two-component exponential fits and dashed
lines two-component Gaussian fits. (h) normalized vertical non-thermal intensity model profile at
    $4.86$~GHz with best-fitting advection model. (i) vertical non-thermal radio spectral index profile
with best-fitting advection model. (j) reduced $\chi^2$ in the northern halo as
function of advection speed and magnetic field scale height in the thick
disc. (k) same as (j) but in the southern halo. The red lines in (j) and (k) show the escape velocity near the
  midplane. (l) vertical magnetic field model profile.}
\label{fig:n891}
\end{figure*}

\begin{figure*}
  \includegraphics[width=1.0\hsize]{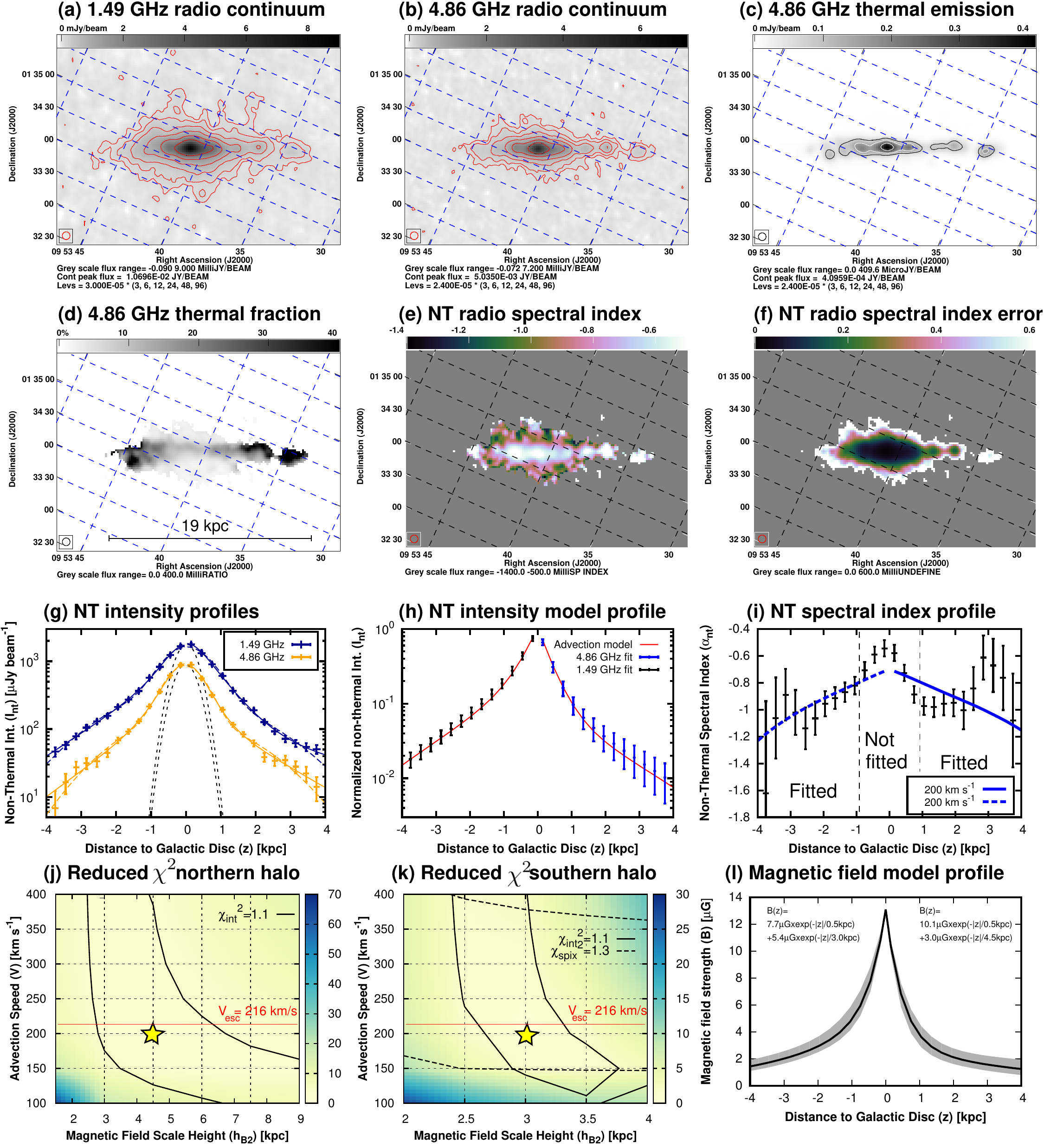}
  \caption{NGC~3044. (a) radio continuum emission at $1.49$~GHz. (b) radio continuum emission at $4.86$~GHz. (c) thermal radio continuum emission at
    $4.86$~GHz. Contours in panels (a)--(c) are at (3, 6, 12, 24, 48 and 96)
    $\times \sigma$, where $\sigma$ is the rms map noise. In panel (c), the same contour levels as in (b) are used. (d) thermal
  fraction at $4.86$~GHz, where the grey-scale ranges from 0 to 40 per cent. (e) non-thermal radio spectral index between $1.49$
  and $4.86$~GHz, where the colour-scale ranges from $-1.4$ to $-0.5$. The spectral index error is $0.4$
in regions of low intensities and decreases to $0.1$ in regions of high intensities. (f) error of the non-thermal radio spectral index, where the colour scale ranges from 0 to $0.6$. Panels (a)--(f) are rotated so that the major axis ($PA=114\degr$)
is horizontal and the synthesized beam is shown in the bottom
left corner. (g) vertical non-thermal intensity profiles at both
frequencies, where solid lines show two-component exponential fits and dashed
lines two-component Gaussian fits. (h) normalized vertical non-thermal intensity model profile at
    $1.49$ and $4.86$~GHz with best-fitting advection model. (i) vertical non-thermal radio spectral index profile
with best-fitting advection model. (j) reduced $\chi^2$ in the northern halo as
function of advection speed and magnetic field scale height in the thick
disc. (k) same as (j) but in the southern halo. The red lines in (j) and (k) show the escape velocity near the
  midplane. (l) vertical magnetic field model profile.}
\label{fig:n3044}
\end{figure*}

\begin{figure*}
  \includegraphics[width=1.0\hsize]{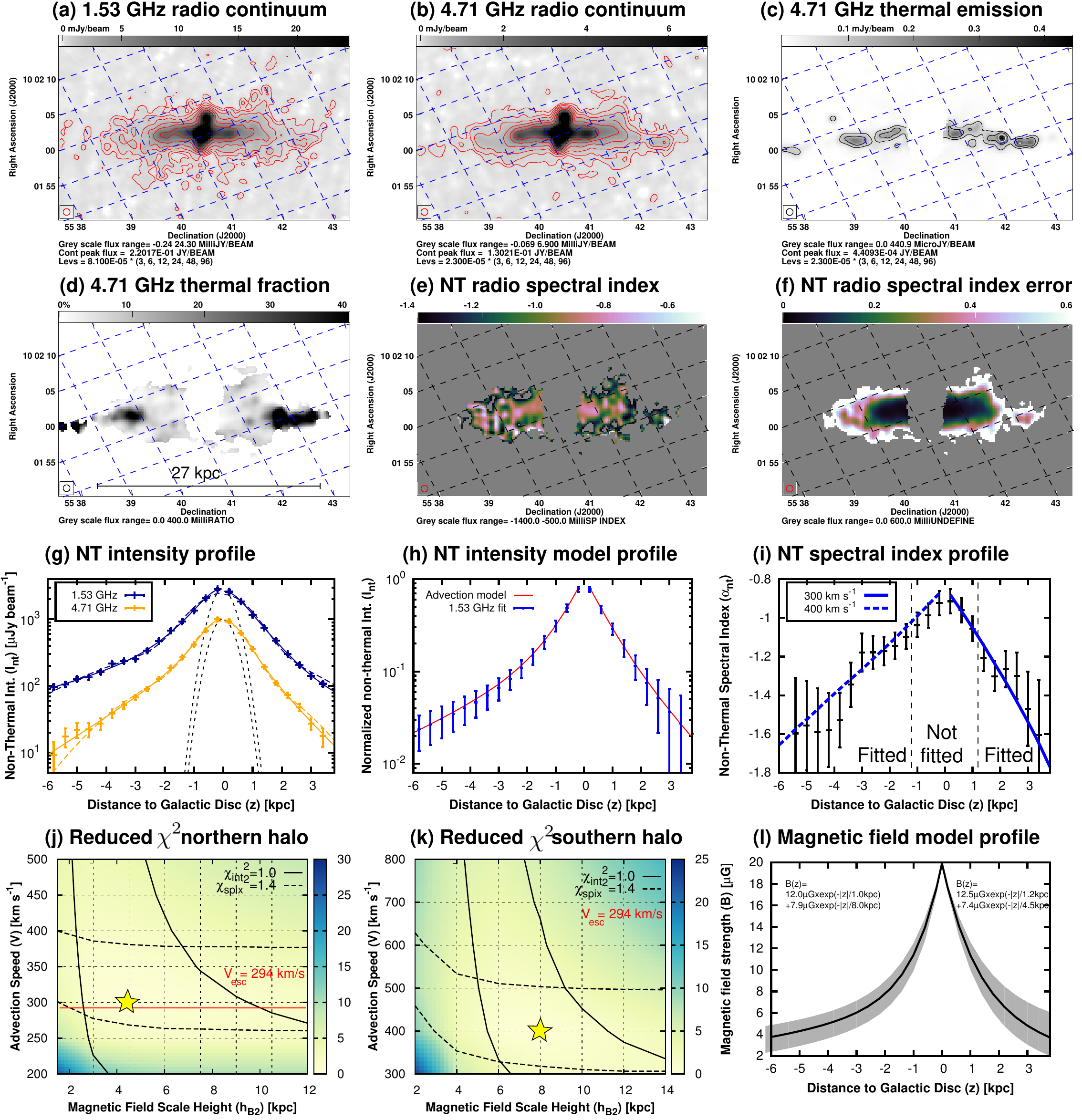}
  \caption{NGC~3079. (a) radio continuum emission at $1.53$~GHz. (b) radio continuum emission at $4.71$~GHz. (c) thermal radio continuum emission at
    $4.71$~GHz. Contours in panels (a) and (b) are at (3, 6, 12, 24, 48 and 96)
    $\times \sigma$, where $\sigma$ is the rms map noise. In panel (c), the same contour levels as in (b) are used. (d) thermal
  fraction at $4.71$~GHz, where the grey-scale ranges from 0 to 40 per cent. (e) non-thermal radio spectral index between $1.53$
  and $4.71$~GHz, where the colour-scale ranges from $-1.4$ to $-0.5$. (f) error of the non-thermal radio spectral index, where the colour scale ranges from 0 to $0.6$. Panels (a)--(f) are rotated so that the major axis ($PA=167\degr$)
is horizontal and the synthesized beam is shown in the bottom
left corner. (g) vertical non-thermal intensity profiles at both
frequencies, where solid lines show two-component exponential fits and dashed
lines two-component Gaussian fits. (h) normalized vertical non-thermal intensity model
profile at
 $1.53$~GHz with best-fitting advection model. (i) vertical non-thermal radio spectral index profile
with best-fitting advection model. (j) reduced $\chi^2$ in the northern halo as
function of advection speed and magnetic field scale height in the thick
disc. (k) same as (j) but in the southern halo. The red line in (j) shows the escape velocity near the
  midplane. (l) vertical magnetic field model profile.}
\label{fig:n3079}
\end{figure*}

\begin{figure*}
  \includegraphics[width=1.0\hsize]{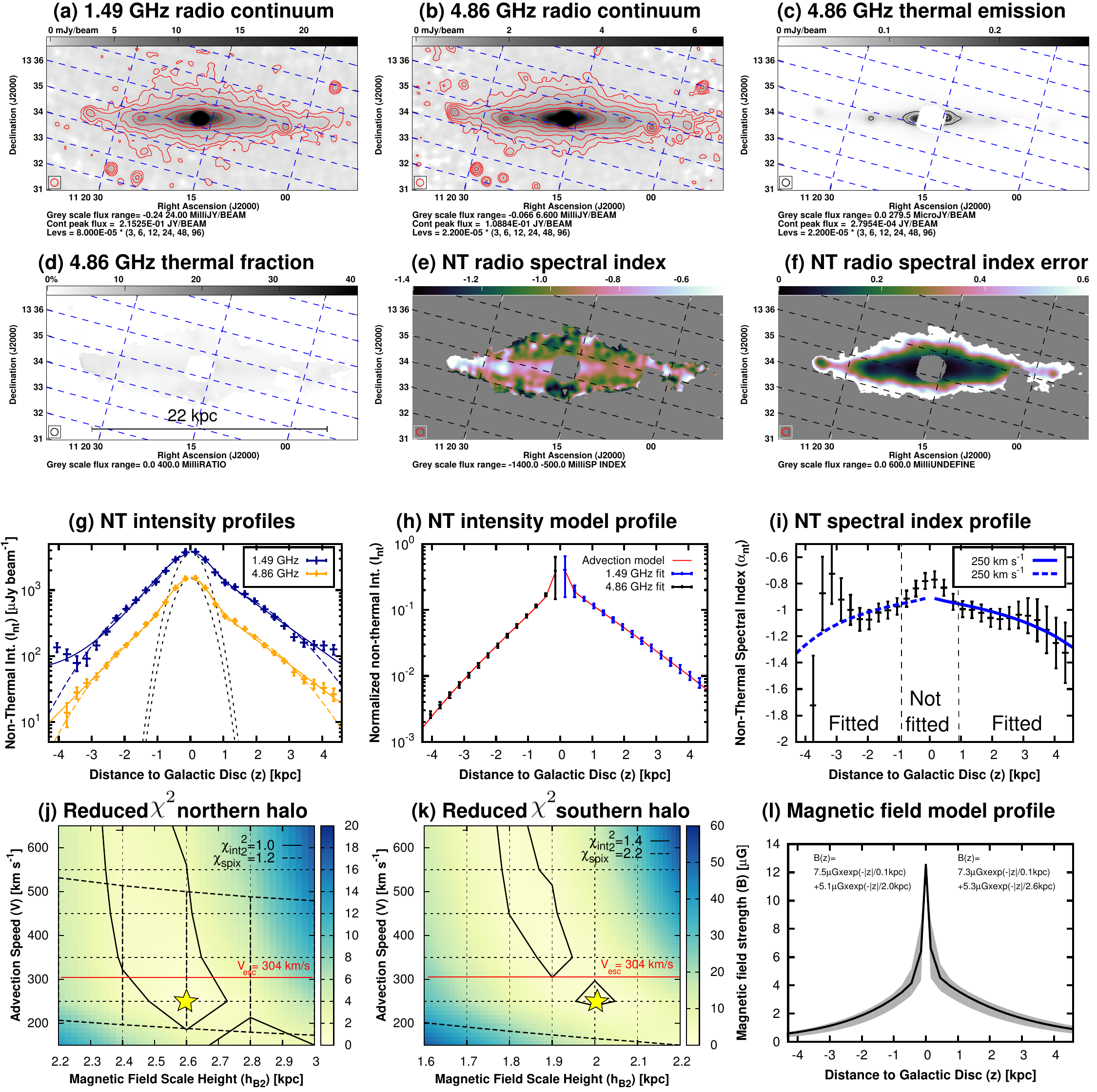}
  \caption{NGC~3628. (a) radio continuum emission at $1.49$~GHz. (b) radio continuum emission at $4.86$~GHz. (c) thermal radio continuum emission at
    $4.86$~GHz. Contours in panels (a)--(c) are at (3, 6, 12, 24, 48 and 96)
    $\times \sigma$, where $\sigma$ is the rms map noise. In panel (c) the rms
  noise of the map shown in (b) is used to ease the comparison. (d) thermal
  fraction at $4.86$~GHz, where the grey-scale ranges from 0 to 40 per cent. (e) non-thermal radio spectral index between $1.49$
  and $4.86$~GHz, where the colour-scale ranges from $-1.4$ to $-0.5$. (f) error of the non-thermal radio spectral index, where the colour scale ranges from 0 to $0.6$. Panels (a)--(f) are rotated so that the major axis ($PA=105\degr$)
is horizontal and the synthesized beam is shown in the bottom
left corner. (g) vertical non-thermal intensity profiles at both
frequencies, where solid lines show two-component exponential fits and dashed
lines two-component Gaussian fits. (h) normalized vertical non-thermal intensity model
profile at
 $1.49$ and $4.86$~GHz with best-fitting advection model. (i) vertical non-thermal radio spectral index profile
with best-fitting advection model. (j) reduced $\chi^2$ in the northern halo as
function of advection speed and magnetic field scale height in the thick
disc. (k) same as (j) but in the southern halo. The red lines in (j) and (k) show the escape velocity near the
  midplane. (l) vertical magnetic field model profile.}
\label{fig:n3628}
\end{figure*}

\begin{figure*}
  \includegraphics[width=1.0\hsize]{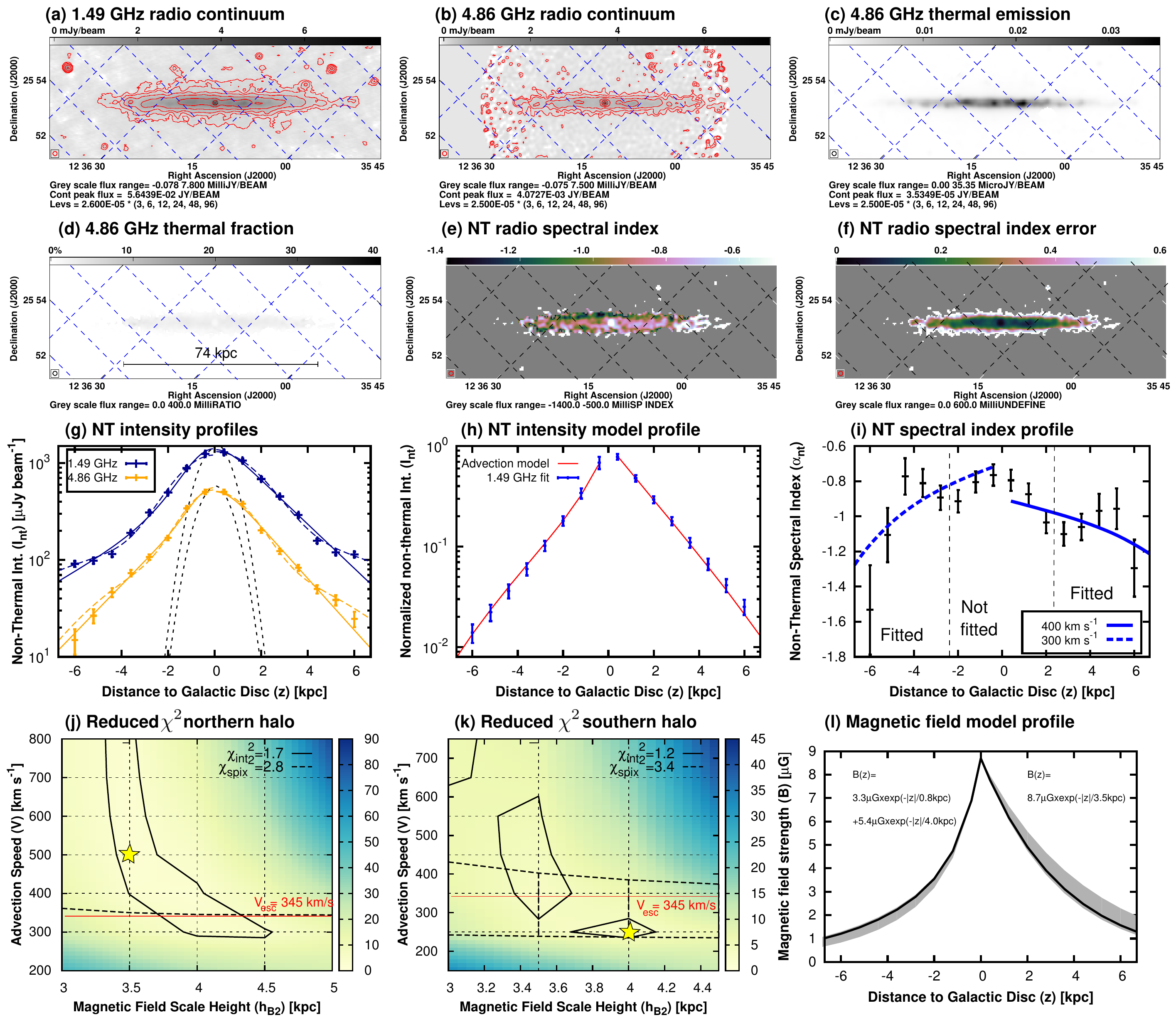}
  \caption{NGC~4565. (a) radio continuum emission at $1.49$~GHz. (b) radio continuum emission at $4.86$~GHz. The increased noise level at the map edge is due to the correction for the primary beam attenuation of the VLA. (c) thermal radio continuum emission at
    $4.86$~GHz. Contours in panels (a) and (b) are at (3, 6, 12, 24, 48 and 96)
    $\times \sigma$, where $\sigma$ is the rms map noise. In panel (c), the same contour levels as in (b) are used. (d) thermal
  fraction at $4.86$~GHz, where the grey-scale ranges from 0 to 40 per cent. (e) non-thermal radio spectral index between $1.49$
  and $4.86$~GHz, where the colour-scale ranges from $-1.4$ to $-0.5$. The spectral index error is $0.4$
in regions of low intensities and decreases to $0.1$ in regions of high
intensities. (f) error of the non-thermal radio spectral index, where the colour scale ranges from 0 to $0.6$. Panels (a)--(f) are rotated so that the major axis ($PA=135\fdg 5$)
is horizontal and the synthesized beam is shown in the bottom
left corner. (g) vertical non-thermal intensity profiles at both
frequencies, where solid lines show two-component exponential fits and dashed
lines two-component Gaussian fits. (h) normalized vertical non-thermal intensity model
profile at
 $1.49$~GHz with best-fitting advection model. (i) vertical non-thermal radio spectral index profile
with best-fitting advection model. (j) reduced $\chi^2$ in the northern halo as
function of advection speed and magnetic field scale height in the thick
disc. (k) same as (j) but in the southern halo. The red lines in (j) and (k) show the escape velocity near the
  midplane. (l) vertical magnetic field model profile.}
\label{fig:n4565}
\end{figure*}

\begin{figure*}
  \includegraphics[width=1.0\hsize]{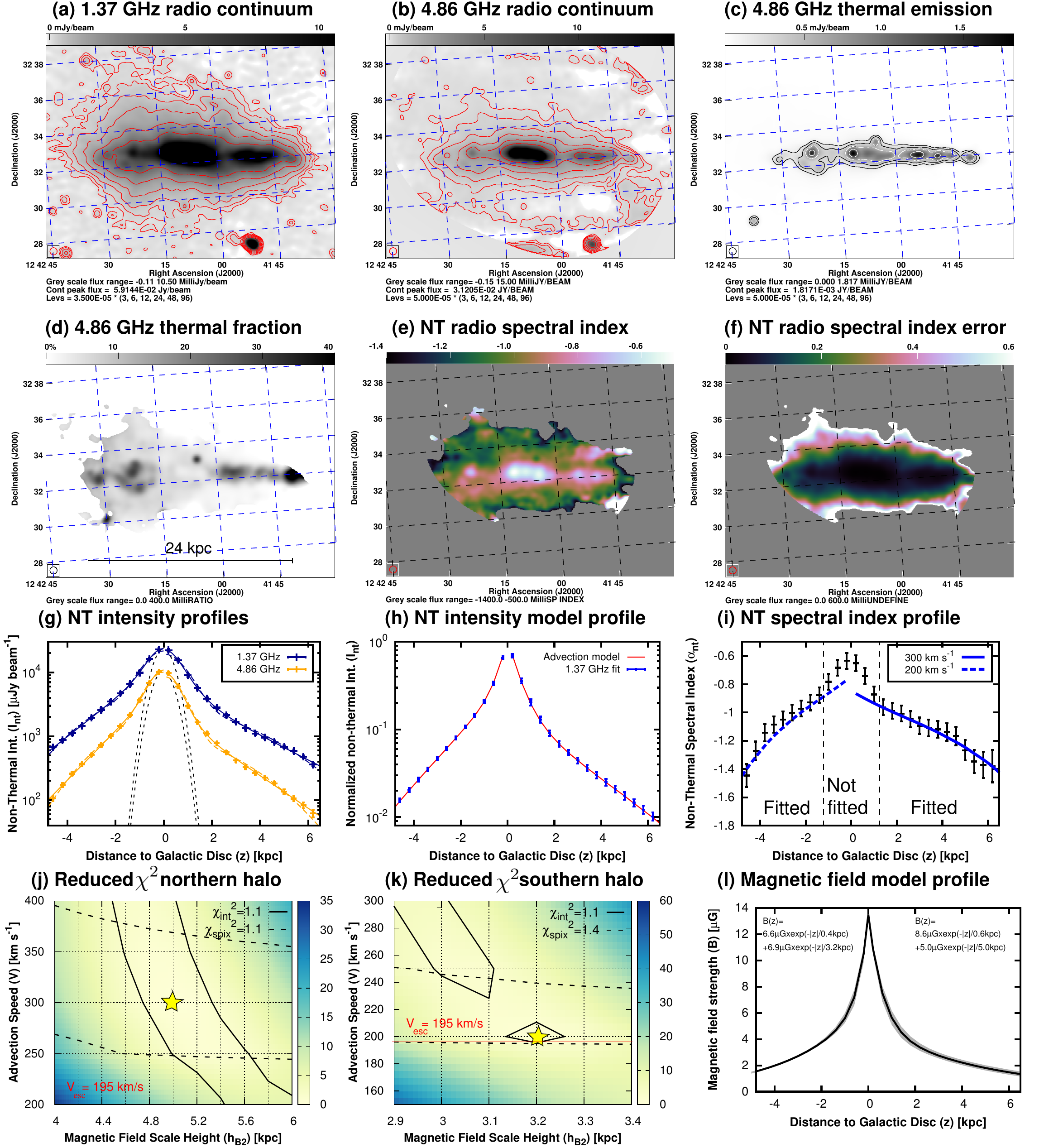}
  \caption{NGC~4631. (a) radio continuum emission at $1.37$~GHz. (b) radio continuum emission at $4.86$~GHz. The high noise level near the edge of the map stems from the correction for the primary beam attenuation of the VLA. (c) thermal radio continuum emission at
    $4.86$~GHz. Contours in panels (a)--(c) are at (3, 6, 12, 24, 48 and 96)
    $\times \sigma$, where $\sigma$ is the rms map noise. In panel (c), the same contour levels as in (b) are used to ease the comparison. (d) thermal
  fraction at $4.86$~GHz, where the grey-scale ranges from 0 to 40 per cent. (e) non-thermal radio spectral index between $1.37$
  and $4.86$~GHz, where the colour scale ranges from $-1.4$ to $-0.5$. (f) error of the non-thermal radio spectral index, where the colour scale ranges from 0 to $0.6$. Panels (a)--(f) are rotated so that the major axis ($PA=86\degr$)
is horizontal and the synthesized beam is shown in the bottom
left corner. (g) vertical non-thermal intensity profiles at both
frequencies, where solid lines show two-component exponential fits and dashed
lines two-component Gaussian fits. (h) normalized vertical non-thermal intensity model
profile at
 $1.37$~GHz with best-fitting advection model. (i) vertical non-thermal radio spectral index profile
with best-fitting advection model. (j) reduced $\chi^2$ in the northern halo as
function of advection speed and magnetic field scale height in the thick
disc. (k) same as (j) but in the southern halo. The red line in (k) shows the escape velocity near the
  midplane. (l) vertical magnetic field model profile.}
\label{fig:n4631}
\end{figure*}

\begin{figure*}
  \includegraphics[width=1.0\hsize]{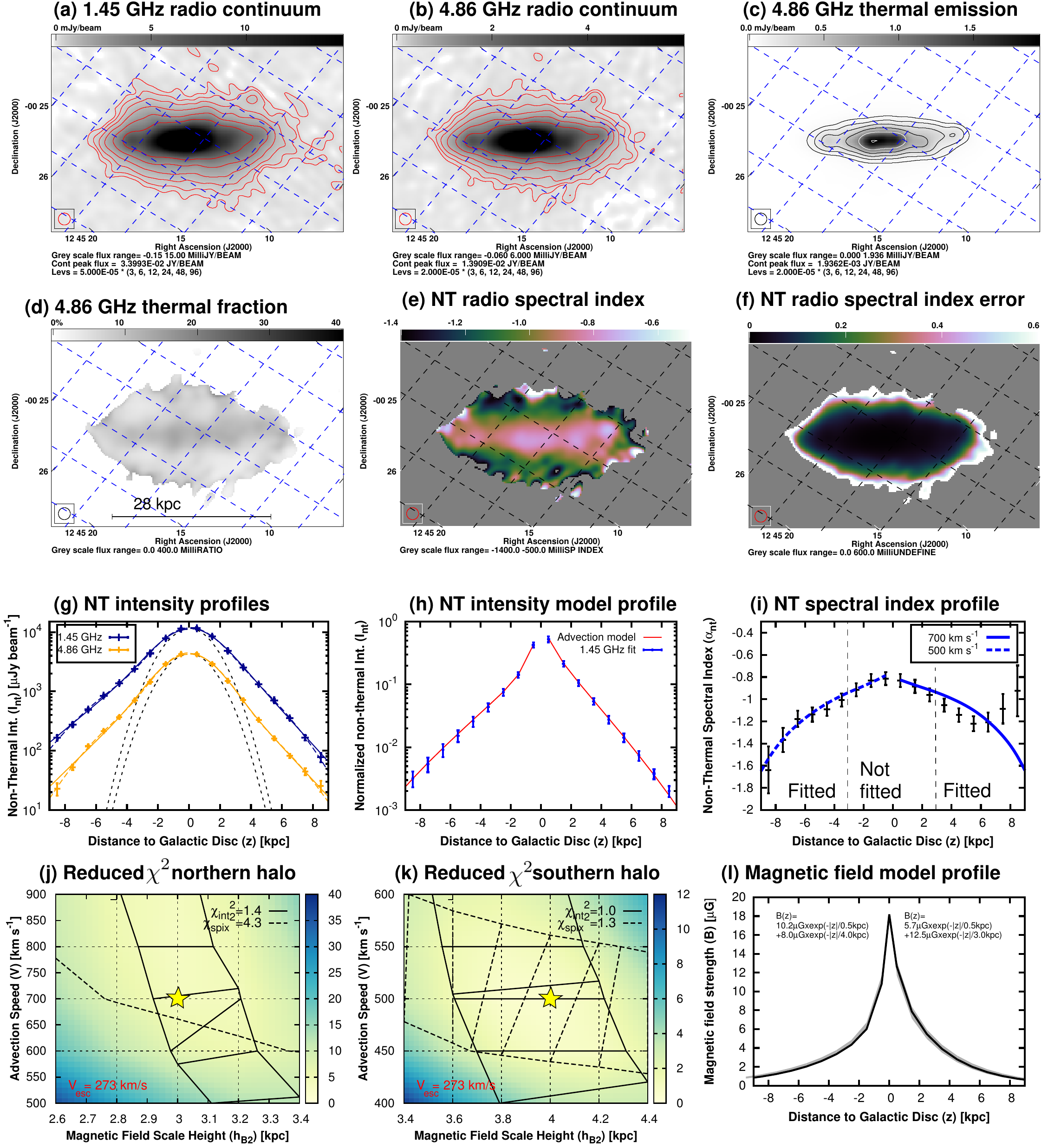}
  \caption{NGC~4666. (a) radio continuum emission at $1.45$~GHz. (b) radio continuum emission at $4.86$~GHz. (c) thermal radio continuum emission at
    $4.86$~GHz. Contours in panels (a) and (b) are at (3, 6, 12, 24, 48 and 96)
    $\times \sigma$, where $\sigma$ is the rms map noise. In panel (c), the same contour levels as in (b) are used. (d) thermal
  fraction at $4.86$~GHz, where the grey scale ranges from 0 to 40 per cent. (e) non-thermal radio spectral index between $1.45$
  and $4.86$~GHz, where the colour scale ranges from $-1.4$ to $-0.5$. (f) error of the non-thermal radio spectral index, where the colour scale ranges from 0 to $0.6$. Panels (a)--(f) are rotated so that the major axis ($PA=40\degr$)
is horizontal and the synthesized beam is shown in the bottom
left corner. (g) vertical non-thermal intensity profiles at both
frequencies, where solid lines show two-component exponential fits and dashed
lines two-component Gaussian fits. (h) normalized vertical non-thermal intensity model
profile at
 $1.45$~GHz with best-fitting advection model. (i) vertical non-thermal radio spectral index profile
with best-fitting advection model. (j) reduced $\chi^2$ in the northern halo as
function of advection speed and magnetic field scale height in the thick
disc. (k) same as (j) but in the southern halo. (l) vertical magnetic field model profile.}
\label{fig:n4666}
\end{figure*}

\begin{figure*}
  \includegraphics[width=1.0\hsize]{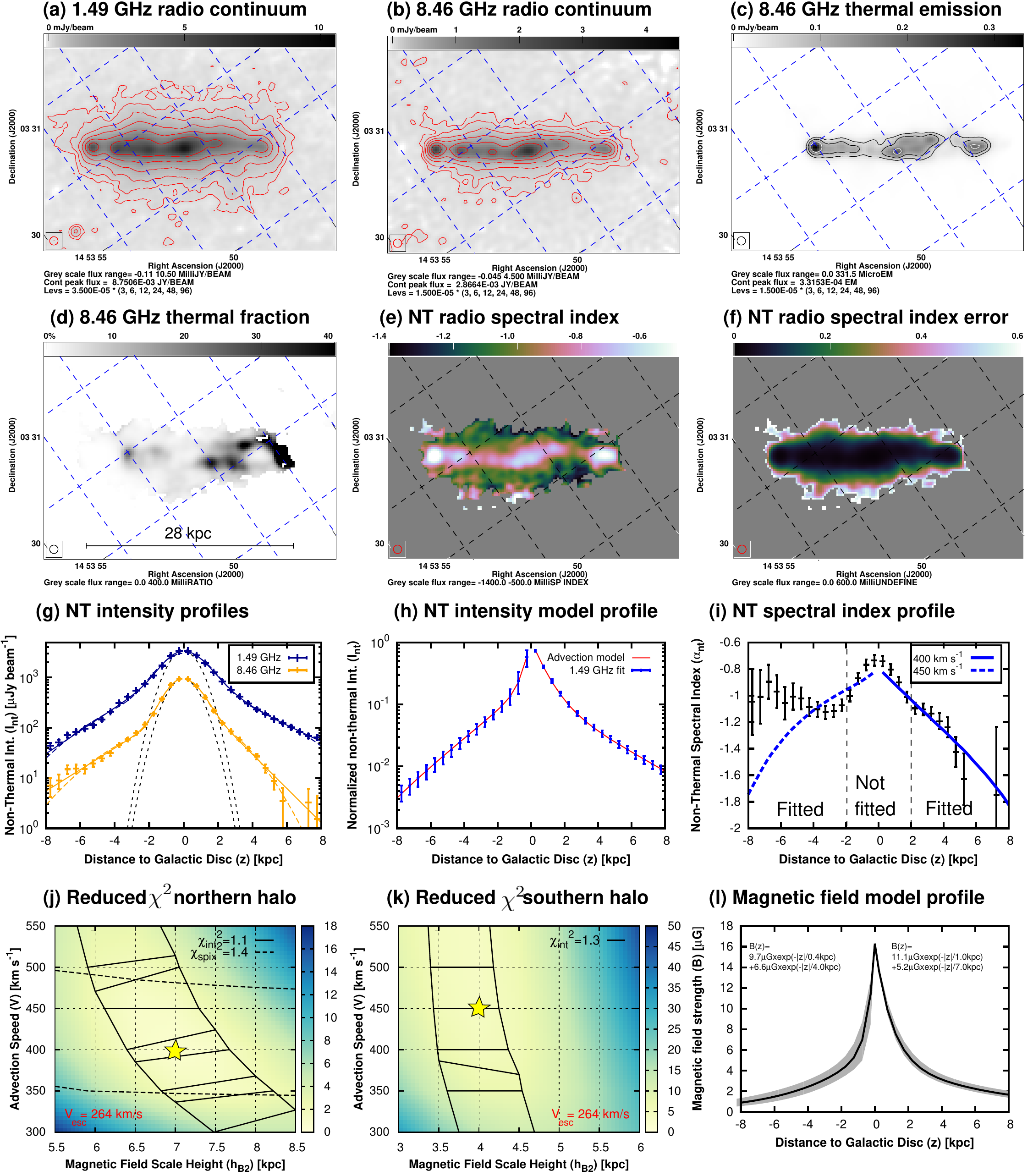}
  \caption{NGC~5775. (a) radio continuum emission at $1.49$~GHz. (b) radio continuum emission at $8.46$~GHz. (c) thermal radio continuum emission at
    $8.46$~GHz. Contours in panels (a) and (b) are at (3, 6, 12, 24, 48 and 96)
    $\times \sigma$, where $\sigma$ is the rms map noise. In panel (c), the same contour levels as in (b) are used. (d) thermal
  fraction at $8.46$~GHz, where the grey-scale ranges from 0 to 40 per cent. (e) non-thermal radio spectral index between $1.49$
  and $8.46$~GHz, where the colour-scale ranges from $-1.4$ to $-0.5$. (f) error of the non-thermal radio spectral index, where the colour scale ranges from 0 to $0.6$. Panels (a)--(f) are rotated so that the major axis ($PA=145\degr$)
is horizontal and the synthesized beam is shown in the bottom
left corner. (g) vertical non-thermal intensity profiles at both
frequencies, where solid lines show two-component exponential fits and dashed
lines two-component Gaussian fits. (h) normalized vertical non-thermal intensity model
profile at
 $1.49$~GHz with best-fitting advection model. (i) vertical non-thermal radio spectral index profile
with best-fitting advection model. (j) reduced $\chi^2$ in the northern halo as
function of advection speed and magnetic field scale height in the thick
disc. (k) same as (j) but in the southern halo. (l) vertical magnetic field model profile.}
\label{fig:n5775}
\end{figure*}

\begin{figure*}
  \includegraphics[width=1.0\hsize]{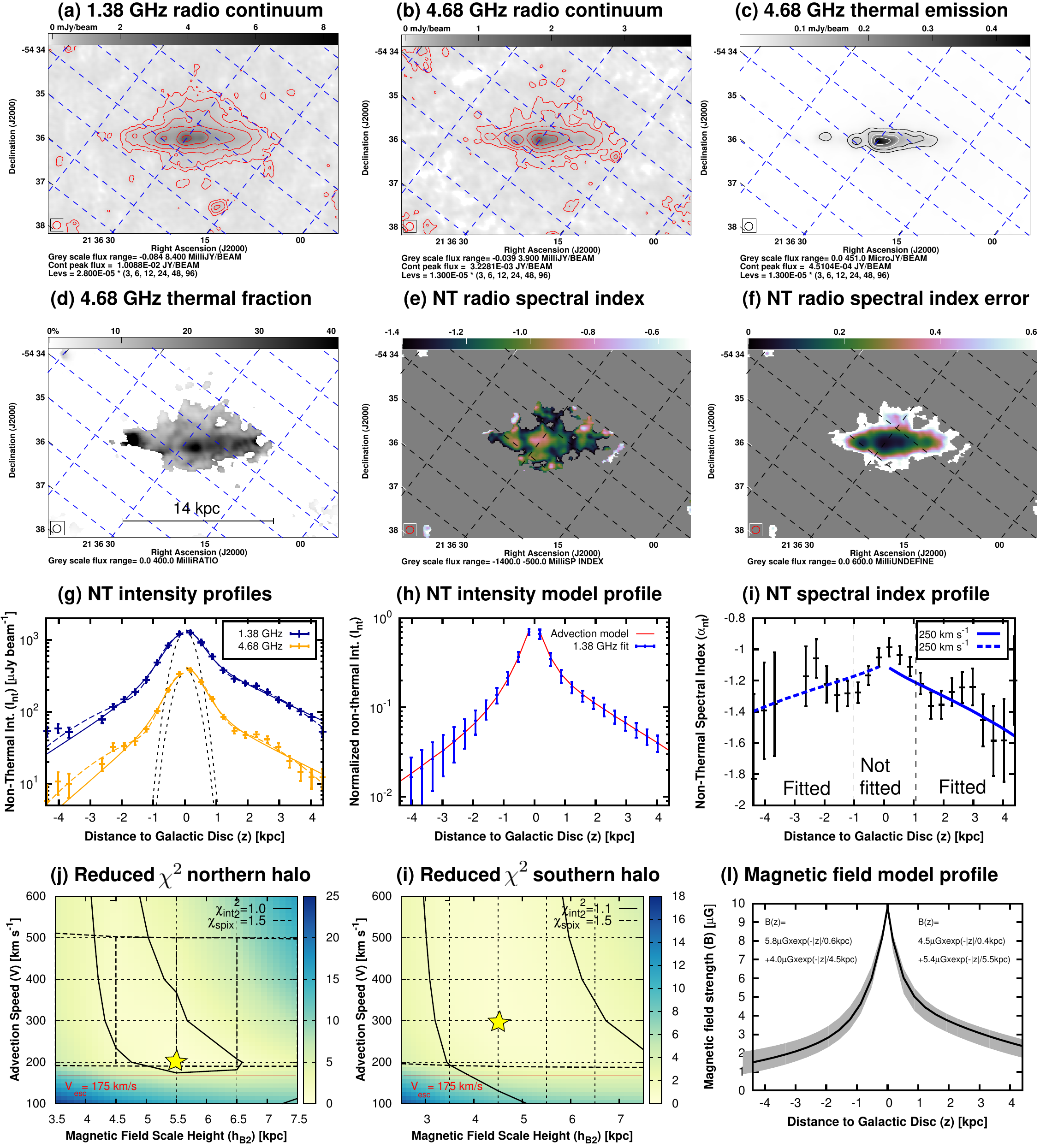}
  \caption{NGC~7090. (a) radio continuum emission at $1.38$~GHz. (b) radio continuum emission at $4.68$~GHz. (c) thermal radio continuum emission at
    $4.68$~GHz. Contours in panels (a) and (b) are at (3, 6, 12, 24, 48 and 96)
    $\times \sigma$, where $\sigma$ is the rms map noise. In panel (c), the same contour levels as in (b) are used. (d) thermal
  fraction at $4.68$~GHz, where the grey-scale ranges from 0 to 40 per cent. (e) non-thermal radio spectral index between $1.38$
  and $4.68$~GHz, where the colour-scale ranges from $-1.4$ to $-0.5$. (f) error of the non-thermal radio spectral index, where the colour scale ranges from 0 to $0.6$. Panels (a)--(f) are rotated so that the major axis ($PA=128\degr$)
is horizontal and the synthesized beam is shown in the bottom
left corner. (g) vertical non-thermal intensity profiles at both
frequencies, where solid lines show two-component exponential fits and dashed
lines two-component Gaussian fits. (h) normalized vertical non-thermal intensity model
profile at
  $1.38$~GHz with best-fitting advection model. (i) vertical non-thermal radio spectral index profile
with best-fitting advection model. (j) reduced $\chi^2$ in the northern halo as
function of advection speed and magnetic field scale height in the thick
disc. (k) same as (j) but in the southern halo. The red lines (j) and (k) show the escape velocity near the
  midplane. (l) vertical magnetic field model profile.}
\label{fig:n7090}
\end{figure*}

\begin{figure*}
  \includegraphics[width=1.0\hsize]{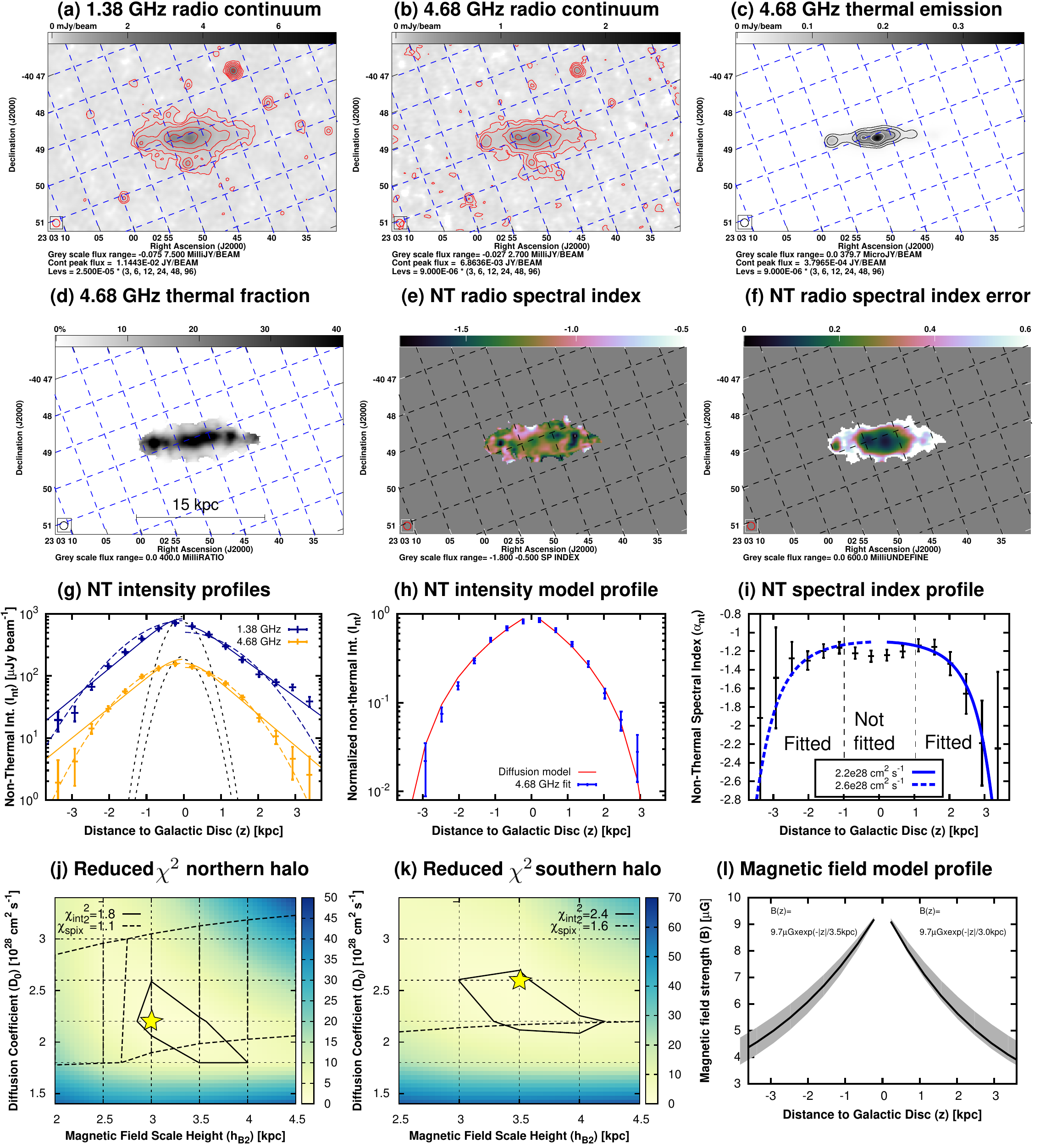}
  \caption{NGC~7462. (a) radio continuum emission at $1.38$~GHz. (b) radio continuum emission at $4.68$~GHz. (c) thermal radio continuum emission at
    $4.68$~GHz. Contours in panels (a) and (b) are at (3, 6, 12, 24, 48 and 96)
    $\times \sigma$, where $\sigma$ is the rms map noise. In panel (c), the same contour levels as in (b) are used. (d) thermal
  fraction at $4.68$~GHz, where the grey-scale ranges from 0 to 40 per cent. (e) non-thermal radio spectral index between $1.38$
  and $4.68$~GHz, where the colour-scale ranges from $-1.8$ to $-0.5$. The spectral index error is $0.4$
in regions of low intensities and decreases to $0.1$ in regions of high intensities. (f) error of the non-thermal radio spectral index, where the colour scale ranges from 0 to $0.6$. Panels (a)--(f) are rotated so that the major axis ($PA=73\degr$)
is horizontal and the synthesized beam is shown in the bottom
left corner. (g) vertical non-thermal intensity profiles at both
frequencies, where solid lines show two-component exponential fits and dashed
lines two-component Gaussian fits. (h) normalized vertical non-thermal intensity model
profile at
 $1.38$~GHz with best-fitting diffusion model. (i) vertical non-thermal radio spectral index profile
with best-fitting diffusion model. (j) reduced $\chi^2$ in the northern halo as
function of diffusion coefficient and magnetic field scale height in the thick
disc. (k) same as (j) but in the southern halo. (l) vertical model magnetic field profile.}
\label{fig:n7462}
\end{figure*}

\bsp	
\label{lastpage}

\end{document}